\documentclass[12pt,cite,epsfig]{article}
\usepackage{times,epsfig}
\newcommand{\nc}{\newcommand}
\nc{\bb}{\bibitem}
\nc{\be}{\begin{equation}}
\nc{\ee}{\end{equation}}
\nc{\pa}{\partial}
\nc{\parsym} {\stackrel{\leftrightarrow}{\pa}}
\nc{\ra}{\rightarrow}
\nc{\la}{\leftarrow}
\nc{\etp}{{\eta^\prime}}
\nc{\omg}{\omega}
\nc{\ggam}{\gamma \gamma}
\nc{\gam}{\gamma }
\nc{\bea}{\begin{eqnarray}}
\nc{\eea}{\end{eqnarray}}
\nc{\beas}{\begin{eqnarray*}}
\nc{\eeas}{\end{eqnarray*}}
\nc{\non}{\nonumber}
\nc{\second}{{\prime\prime}}
\def\hhht{\rule[ 0.mm]{0.mm}{6.mm}}

\def\hhhbb{\rule[-1.mm]{0.mm}{8.mm}}
\def\hhhc{\rule[-3.mm]{0.mm}{3.mm}}
\def\hhhd{\rule[-3.mm]{0.mm}{2.mm}}
\def\hhhe{\rule[-3.mm]{0.mm}{4.mm}}
\def\hhhu{\rule[-3.mm]{0.mm}{12.mm}}
\def\hhhv{\rule[-3.mm]{0.mm}{9.mm}}
\def\hhhq{\rule[-5.mm]{0.mm}{12.mm}}

\begin{document}
\begin{titlepage}
\vbox{~~~ \\
                                   \null \hfill LPNHE 2011--01\\
			           \null \hfill DESY 11-095\\
				   \null \hfill HU-EP-11/26\\
\title{Upgraded Breaking Of The HLS Model~:\\
A Full Solution to the $\tau-e^+e^-$  and $\phi$ Decay  Issues\\
And Its Consequences On $g-2$ VMD Estimates
   }
\author{
M.~Benayoun$^a$, P.~David$^a$, L.~DelBuono$^a$, F.~Jegerlehner$^{b,c}$ \\
\small{$^a$ LPNHE des Universit\'es Paris VI et Paris VII, IN2P3/CNRS, F-75252 Paris, France }\\
\small{ $^b$ Humboldt--Universit\"at zu Berlin, Institut f\"ur Physik, Newtonstrasse 15, D--12489 Berlin,
Germany }\\
\small{ $^c$ Deutsches  Elektronen--Synchrotron (DESY), Platanenallee 6, D--15738 Zeuthen, Germany}
}
\date{\today}
\maketitle
\begin{abstract}
The  muon anomalous magnetic moment $a_\mu$ and  the
hadronic vacuum polarization are examined using data analyzed within the framework
of a suitably broken HLS model. The analysis relies on all available scan data samples
and leaves provisionally aside the existing ISR data. 
Our HLS model based global fit approach allows for a better check of consistency between data sets and we investigate
how results depend on different strategies which may be followed.
Relying on global fit qualities, 
we find several acceptable solutions leading to ambiguities in the reconstructed value 
for $(a_\mu)_{th}$. Among these, the most conservative solution is
 $a_\mu^{\rm had,LO}[{\rm HLS \ improved}]=687.72(4.63) \times 10^{-10}$
 and $(a_\mu)_{th}=11\,659\,175.37(5.31)\times 10^{-10}$ corresponding to 
 a  $4.1 \sigma$ significance  for the difference
$\Delta a_\mu=(a_\mu)_{exp}-(a_\mu)_{th}$.
It is also shown that the various contributions accessible through the model
yield uniformly a factor 2 improvement of their uncertainty.
The breaking procedure implemented in the HLS model is an extension of the 
former procedure based on a mechanism defined by Bando, Kugo and Yamawaki
(BKY) . This yields a quite satisfactory
simultaneous description of  most $e^+e^-$ annihilation channels up to and including the $\phi$ meson
($\pi^+\pi^-$, $\pi^0\gamma$, $\eta\gamma$, $\pi^+\pi^-\pi^0$, $K^+K^-$, $K^0 \overline{K}^0$)
and of a set of 10 (mostly radiative) decay widths of light mesons. It also allows to achieve
the proof of consistency between the $e^+e^- \to \pi^+\pi^-$ annihilation and the
$\tau^\pm \ra \pi^\pm\pi^0 \nu$ decay and gives a solution to the reported problem
concerning the measured partial width ratio $\Gamma(\phi \to K^+K^-)/\Gamma(\phi \to K^0 \overline{K}^0)$.
Prospects for improving the VMD based estimates of $a_\mu$ are emphasized.

\end{abstract}
}
\end{titlepage}

\section{Introduction}
\indent \indent
The muon anomalous magnetic moment $a_\mu$ is a physics piece of information which has been measured with
the remarkable accuracy of $6.3 \times 10^{-10}$  \cite{BNL,BNL2}. From a theoretical point of view,
$a_\mu$ is the sum of several contributions; the most prominent  contributions can be predicted
with a very high accuracy by the Standard Model. This covers the QED contribution which presently 
reaches an accuracy better than $1.6 \times 10^{-12}$ \cite{Passera06} or the 
electroweak contribution where the
precision is now $1.8 \times 10^{-11}$ \cite{Fred09}. The light--by--light contribution 
to $a_\mu$  is more complicated to estimate and is currently known 
with an accuracy of $2.6 \times 10^{-10}$ \cite{LBL}.  

Another important  contribution to $a_\mu$ is the  hadronic vacuum polarization (HVP).
Perturbative QCD allows to compute a part of this with an accuracy of the order $ 10^{-11}$;
this covers the high energy tail and the perturbative window
between the $J/\psi$ and $\Upsilon$ resonance regions.
For the region below  this threshold, one is in the non--perturbative region of
QCD where estimates of the hadronic VP cannot so far be directly derived from QCD, relying 
on first principles only. However, this may change in a
 future. Indeed, some recent progress  in Lattice QCD 
\cite{Aubin2006,DellaMorte2010,Feng2011} gives hope that 
reliable calculations of the HVP are now 
in reach in the next years. They would be an important complement to the 
standard approaches, as well as to the approach
presented here.

One is, therefore, left with estimates numerically derived from experimental data.
Indeed, it has been proved long ago that the contribution of an intermediate hadronic state
$H_i$ to $a_\mu$ is related with the  annihilation cross section  $\sigma_{H_i}(s) \equiv 
\sigma(e^+ e^- \ra H_i)$ by ~: 
$$ a_\mu(H_i) = \frac{1}{4 \pi^3} \int_{s_{{H_i}}}^{s_{cut}} ds K(s) \sigma_{H_i}(s)$$
where $K(s)$ is a known kernel \cite{Fred09} enhancing the weight of the low  $s$ region, close
to the threshold $s_{H_i}$ of the final state $H_i$.
Then, the total non--pertubative HVP can be estimated by $a_\mu(H)=\sum a_\mu(H_i)$, where
the sum extends over all final states $H_i$ which can be reached in $e^+ e^- $ annihilations.

The accuracy of $a_\mu(H_i)$ is, of course, tightly related with the accuracy
of the experimental data set used to perform numerically the integration shown above.  
When different data sets are available for a given annihilation channel $H_i$, a combination
of the corresponding $a_\mu(H_i)$'s  is performed by weighting each estimate with
the reported uncertainties affecting each data set,  using standard statistical methods
(see \cite{FredAndSimon}, for instance). Possible mismatches between the various estimates
are accounted for by methods like the S--factor technics of the Particle Data Group \cite{RPP2010}.
In this approach, of course, the accuracy of each $a_\mu(H_i)$ is solely determined 
by all the 
measurements covering the channel $H_i$ $only$, without any regard to the other channels
$H_j$ ($j\ne i$).

 This method succeeds in providing very precise values for the relevant contributions. Summing up
 the non--perturbative HVP estimated this way with the rest, one obtains an estimate
 of  $a_\mu$ quite comparable to the BNL average measurement \cite{BNL,BNL2}. 
  However, the prediction based on $e^+ e^-$ annihilation data
or $\tau$ decay data  \cite{Fred3,Davier2007,DavierHoecker,DavierHoecker3,Fred1,Fred09,Fred11}
exhibits a long--standing discrepancy; the exact value of this discrepancy has gone
several times back and forth, depending on whether one trusts the $\tau$  data based
analyses or the scan $e^+ e^-$ annihilation data, which are obviously more directly related with
$ a_\mu(H)$.  With the advent of the high statistics data samples collected using the Initial 
State Radiation (ISR) method \cite{KLOE08,BaBar,KLOE10}, a precise value for this -- possible --
discrepancy has become harder to define unambiguously.

In order to get a firm conclusion concerning the numerical difference between the measured and
calculated values of  the muon anomalous magnetic moment $\Delta a_\mu=(a_\mu)_{exp}-(a_\mu)_{th}$,
one should first understand why $\tau$ based and $e^+ e^-$ based analyses differ; one should also 
understand the differences between scan data and ISR data and possibly the differences 
between the various available ISR data samples, as the KLOE samples \cite{KLOE08,KLOE10} and the
BaBar sample \cite{BaBar} seem to lead to somewhat conflicting results.

Anyway, while all proposed values for  $(a_\mu)_{th}$ differ from the average for 
$(a_\mu)_{exp}$ , the theoretical uncertainties
start to be comparable to the experimental one. Therefore,
it becomes interesting  to look for a method able to reduce the uncertainty on  $(a_\mu)_{th}$ 
by simply using the existing data. It is also an important issue to have a framework where the 
properties of each data set can be examined.

 \vspace{0.5cm}

In order to cover the low energy regime of strong interactions, the most common approach is to use
effective Lagrangians which preserve the symmetry properties of QCD. At very  
low energies,  
Chiral Perturbation Theory (ChPT) represents such a framework. However, the realm covered
by the usual ChPT is very limited (not much greater than the $\eta$ mass); Resonance 
Chiral Perturbation Theory (R$\chi$PT) permits to go much deeper inside the resonance region; 
it thus defines a framework  suited to study the non--perturbative   hadronic VP (HVP). 

It was soon recognized \cite{Ecker1} that the coupling constants occuring at order $p^4$ in ChPT 
were saturated by low lying meson resonances of various kinds (vector, axial, scalar, pseudoscalar)
as soon as they can contribute. This emphasized the role of the fundamental vector meson nonet 
and confirmed the relevance  of the Vector Meson  Dominance (VMD) concept in low energy physics. 
Soon after, \cite{Ecker2} 
proved that the Hidden Local Symmetry (HLS) model \cite{HLSOrigin,HLSRef} and 
the Resonance Chiral Perturbation Theory (R$\chi$PT) 
were equivalent. Therefore, one may think that the HLS model provides a convenient and 
constraining QCD inspired framework for an improved determination of the HVP. It is, therefore,
quite legitimate to check wether the HLS model allows  a better determination of the  
 HVP  than the usual method sketched above. 
 
The basic HLS model has an important limitation for HVP studies~: The vector resonances 
entering the model are only those embodied in the lowest mass vector meson nonet. This 
certainly limits upwards the relevant energy range to $\simeq 1.05$ GeV, {\it i.e.} slightly
above the $\phi(1020)$ meson mass; going beyond while staying within the standard HLS framework
certainly entails uncontrolled uncertainties due to the contribution of higher mass vector meson nonets.

However, relying on the standard method, one can
estimate the contribution of  the region $\sqrt{s} \in [m_{\pi^0}, m_\phi]$ 
to 83.3\% of the total HVP and show that its uncertainty is also
a large fraction of the total HVP uncertainty~: 
 $\simeq 4 \times 10^{-10}$ when using only scan data or
$\simeq 2.7 \times 10^{-10}$ when using also the recent ISR data samples. For comparison,
the uncertainty provided by the region above $\simeq 1.05$ GeV is $\simeq 4 \times 10^{-10}$.
Therefore, any significant improvement on the knowledge of $(a_\mu)_{th}$ in
the region $\sqrt{s} \le 1.05$ GeV is certainly valuable.

 \vspace{0.5cm}

The (basic) HLS model provides a framework where the interrelations between the
various observed decay channels are made explicit. The point is that the use of an adequate model allows 
for a global fit strategy. All available cross-section data are used to constrain the model parameters,
which in turn allows us to predict physical amplitudes.
Therefore,
if the model provides a statistically acceptable common solution to some
set ${\cal H} \equiv \{H_i\}$ of different processes\footnote{These can be cross sections 
or various kinds of meson partial widths, or also decay spectra.
Indeed, any piece of information able to constrain the model parameters is valuable.}, 
each covered by one or several data sets,
the fit results can serve to reconstruct reliably the   $a_\mu(H_i)$ ($H_i \in {\cal H}$).

Indeed, if a global fit of the set ${\cal H} $ of the various data samples is
successfully performed, then the parameter values and their error covariance
matrix summarize reliably all the knowledge of the set ${\cal H} $, including
the {\it physics  correlations} among them. Then, all
cross sections contained in  ${\cal H} $  can be estimated with an information
improved by having taken into account all the underlying physics correlations.

With the present  formulation, of the HLS model,  the various $a_\mu(H_i)$ can be 
reliably and accurately determined up to $\simeq 1.05$ GeV, just including the $\phi$ 
resonance region. 
All the rest should presently be estimated by the methods usual in this field.

One can substantiate the benefits drawn for using such a global model~: 

\begin{itemize}
\item As the model is global, it implies algebraic relations between 
the various channels it encompasses. Therefore, the accuracy of the estimate for 
$a_\mu(H_i)$ is determined by the statistics available for any channel $H_i$ $and$ 
also by the statistics associated with all the other channels
contained in ${\cal H} $. 

For instance, the accuracy for $a_\mu(\pi^+ \pi^-)$ is certainly determined by the available statistics for 
$e^+ e^- \to \pi^+ \pi^-$  but all other data, acting  as $constraints$,  
also contribute to the accuracy for $a_\mu(\pi^+ \pi^-)$. This is the role
of the  $e^+ e^- \to \eta \gamma$ or $e^+ e^- \to \pi^0 \gamma$ annihilation data,
but also those of the decay width for $\phi \ra \eta^\prime
\gamma$ or of the dipion spectrum in the $\tau \ra \pi \pi \nu$ decay, etc.

Conversely, the accuracy for $a_\mu(\pi^0 \gamma )$, for instance, is not
only governed by the statistics available for  $e^+ e^- \to \pi^0 \gamma$, but also
by those provided by the $e^+ e^- \to \pi^+ \pi^-$ or  $e^+ e^- \to \eta \gamma$ data,  etc. 

Therefore, the improvement expected from a global model should affect 
$simultaneously$ all
the channels contained in ${\cal H}$ and contributing to $a_\mu$.

\item As the breaking procedure is global, it affects simultaneously all physics 
channels related with each other by the Lagrangian model. A successfull  global fit thus
implies that it is validated by the fit quality of the largest possible set of data samples.
This high degree of consistency  indicates that the breaking 
model\footnote{We mean that the breaking procedure 
we define is certainly a model, but it is not intended to solve only one issue in isolation,
like the consistency between $e^+ e^- \to \pi^+ \pi^-$ and $\tau \ra \pi \pi \nu$,
without any regard to the rest of the correlated physics. Stated otherwise, it is
validated only if its consequences for the other related physics channels are
accepted by the corresponding data.} is not simply {\it had hoc}.

\item Any data set is certainly subject to specific systematics; however,
taking into account that the study we plan
relies on 45  different data sets covering 6 different 
annihilation channels, 10 partial width decays (taken  from the Review of 
Particle Properties \cite{RPP2010}) and some decay spectra\footnote{Actually,
it affects the dipion spectrum in the decay $ \tau \ra \pi \pi \nu $ and in 
the anomalous $\eta  / \eta^\prime \ra \pi \pi \gamma$ decays, among others.},
one may consider the effects of correlated
systematics reasonably well smeared out. Indeed, one may consider unlikely
that the systematics affecting as many different objects can pile up.

\end{itemize}

 Basically, what is proposed is to introduce the theoretical prejudice
 represented by one formulation of the VMD concept in order to constrain 
 the data beyond genuine statistical consistency of the  various data samples
 referring to the $same$ physics channel. It has 
 already been  shown \cite{ExtMod2}
  that theoretical (VMD) relationships among various channels are highly constraining.
  The present work plans to better explore such a framework with a much improved modelling.

 Conceptually, the idea to include some theoretical prejudice in order
 to reduce the uncertainties on $a_\mu$ is not completely new. A method
 to complement the  $e^+e^- \ra \pi^+ \pi^- $ data with
  the constraints of analyticity, unitarity and chiral symmetry 
  has been initiated by \cite{Leutwyler02,Colangelo04,Colangelo06} with
  the aim of improving the $\pi^+ \pi^- $ contribution to $a_\mu$,
  but this has not been finalized.

  \vspace{0.5cm}

For the present study, we have found appropriate to discard the data collected using
the Initial State Radiation (ISR) method \cite{KLOE08,BaBar,KLOE10}; indeed, because
of the complicated structure of their systematics, they almost certainly call
for a more complicated statistical treatment than the usual  $e^+ e^-$ scan data.
 The use of ISR data will be addressed in a 
forthcoming publication.

 \vspace{0.5cm}

The HLS model \cite{HLSOrigin,HLSRef} complemented with its
anomalous sector \cite{FKTUY} 
provides a framework able to encompass
a large realm of low energy physics. This anomalous sector
 will be referred to hereafter as FKTUY sector.
The non--anomalous sector allows
to cover most  $e^+ e^-$ annihilation channels  and some $\tau$ decays.
Thanks to its anomalous sector, the same framework also includes
the radiative decays of light flavor mesons with couplings of the 
form\footnote{In the following, we may
denote by $V$ and $P$ any of respectively the vector or 
the pseudoscalar light flavor 
mesons. This does not rise ambiguities.} 
$VP\gamma$ and $P\gamma \gamma$ and also several anomalous annihilation channels.
Actually, up to the $\phi$ meson mass, the only identified channel
which remains outside the scope of the HLS model is the $e^+ e^- \ra \pi^0 \omg $
annihilation channel, due to the large effect of high mass vector resonances \cite{GLi,Arbuzov}
presently not included in the HLS model. 

However, in order to use the HLS model beyond rather qualitative studies
and yield precise descriptions of experimental data, symmetry
breaking procedures have to be implemented. A simple mechanism 
breaking the SU(3) flavor symmetry  \cite{BKY} has been 
introduced, followed by several useful variants \cite{BGP,BGPbis,Heath}. 
Nonet symmetry breaking in the pseudoscalar 
sector has also been introduced by means of determinant terms  \cite{tHooft}.
This breaking procedure  has been shown to describe  precisely the radiative decays
of light mesons  \cite{rad,box} and to meet \cite{WZWChPT} all 
expectations of Chiral Perturbation Theory. 

In order to account for the reported mismatch between the pion form factor
in $e^+ e^-$  annihilation  and in the $\tau$ decay, it has been proposed \cite{taupaper}
to take into account loop effects. Indeed kaon loops 
produce a mixing of the neutral vector mesons which is a 
consequence for the $K^0-K^\pm$ mass splitting. These turn out to modify
effectively the vector meson mass term by identified $s$--dependent 
terms. 

Introducing the
physical vector fields which correspond to the eigenstates
of the  loop modified vector meson mass matrix, provides a mixing
mechanism of the triplet $\rho^0-\omg-\phi$ system. In this change of
fields the charged vector mesons remain unchanged. With this  
$s$--dependent mixing of neutral vector mesons, the fit residuals to the
pion form factor in  $e^+ e^-$ annihilations and in $\tau$ decays
did not exhibit any longer any mass dependence \cite{taupaper};  
thus this mechanism provides an important part of the solution
to the so--called $e^+ e^-$--$\tau$ puzzle\footnote{
A similar result has been obtained in \cite{Fred11} relying rather
on a  $\rho^0-\gamma$ mixing mechanism; it should be interesting to
study a more general $V-\gamma$ mechanism supplementing the $\rho^0-\omg-\phi$ 
mixing scheme.}. 

However, this solution is only partial. Indeed, if the dipion spectrum lineshape in the
decay of the $\tau$ lepton is clearly predicted \cite{taupaper,ExtMod2} from $e^+ e^-$ data, 
there is  still some problem with its absolute magnitude. 
This issue has been found to be cured
by allowing {\bf i/} a mass ($\delta m$) and a coupling ($\delta g$) difference  between the neutral 
and charged $\rho$ mesons, {\bf ii/} a rescaling of the $\tau$ dipion spectra consistent with the
reported uncertainties on the absolute scales of the various measured spectra \cite{Aleph,Belle,Cleo}.
The results returned by fits did not lead to a significant mass difference\footnote{
\label{cottingham}
The mass difference following from fit corresponds to $\delta m=0.68 \pm 0.40$ MeV is in accord
with what is expected for the electromagnetic mass difference \cite{Cottingham,Harari} 
of the $\rho$ mesons \cite{LeutwPriv} $\delta m \simeq 0.81$ MeV. } but, instead,
$\delta g$ and the fitted scales of the experimental spectra were found highly 
significant \cite{ExtMod2}.

However, the numerical values of these parameters  (never more than a few
percent) suggest that some unaccounted for isospin breaking effects have not yet 
been included. 

On the other hand, the HLS model  supplemented with the SU(3)/U(3)
breaking reminded above  accounts successfully -- and simultaneously --
for the measured
cross sections in the $e^+ e^- \ra \pi^+\pi^-$, $e^+ e^- \ra \pi^0\gamma$,
$e^+ e^- \ra \eta\gamma$, $e^+ e^- \ra \pi^+\pi^-\pi^0$ annihilation channels
and for the additional set of  9 decay widths, especially the radiative decays of the form $VP\gamma$ or 
$P\gamma\gamma$, needed in order to constrain more tightly the model.
This has been proved in  \cite{ExtMod1}. However, as it stands, the HLS model 
fails to account for the annihilation channels $e^+ e^- \ra K^+K^-$ and 
$e^+ e^- \ra K^0 \overline{K}^0$ simultaneously. This is obviously related to
the puzzling issue thoroughly discussed in \cite{BGPter} concerning
the branching fraction ratio $\phi \to K^+ K^- / \phi \to K^0 \overline{K}^0$.
The reported disagreement with theoretical expectations
is found significant and amounts to a few percent. This also allows thinking 
that some isospin breaking effects are not yet fully accounted for.

\vspace{0.5cm}

In the present paper, we define a symmetry breaking procedure which is
nothing but an extension of the BKY mechanism referred to above, but including now
breaking in the non--strange sectors. This mechanism is only an upgrade of the
BKY mechanism and applies likewise to the two
different sectors (the so--called $ {\cal L}_A$ and $ {\cal L}_V$
sectors) of  the non--anomalous HLS Lagrangian. We show that the  $\tau$ scale issue
is solved by breaking the $ {\cal L}_V$ Lagrangian piece while the
$\phi \to K^+ K^- / \phi \to K^0 \overline{K}^0$ puzzle yields its solution
from applying the same mechanism to the $ {\cal L}_A$ Lagrangian piece.
Stated otherwise, within the framework of the HLS model broken in this way,
the $e^+ e^-$--$\tau$ and the $\phi \to K  \overline{K} $ puzzles
appear as twin phenomena yielding parent explanations.

Actually, equipped with this upgraded  breaking mechanism, the HLS model provides
a satisfactory description of all the physics information listed above, including
now both $e^+ e^- \ra K  \overline{K} $ annihilations.

 Having discarded the 3 existing ISR data samples, {\it a priori}
 45  different data sets of scan data are relevant for our 
 present analysis. At each step of our analysis, we have checked
 the consistency of the various data samples with each other 
 by relying, as strictly as possible, on the information provided by the 
 various groups without any further assumption.
 We have found that 2 among them have a behavior not in agreement
 with what can be expected from the rest (43 data sets). One could have
 attempted to use them by weighting their contribution to the global $\chi^2$
 (a sort of S--factor); however, for now, we have preferred discarding them.
 Therefore our analysis relies on 43 data sets -- mostly produced by
 the CMD--2 and SND Collaborations -- and 10 accepted partial width
 information, which represents already an unusually large set of data consistently
 examined and satisfactorily understood.
 
\vspace{0.5cm}

 The paper is organized as follows. In Section \ref{sect1}, we briefly outine 
 the basics of the HLS model and its various sectors. In Section \ref{sect2},
 we define  the upgraded breaking procedure which is a trivial extension to
 the $u$ and $d$ sectors of the BKY breaking scheme as redefined 
 in \cite{Heath}-- the so--called "new scheme". Section \ref{brkLa} and Section \ref{brkLv}
 examine the consequence for the modified BKY breaking scheme on the two
 different  parts (${\cal L}_A$ and ${\cal L}_V$ ) of the non--anomalous
 HLS Lagrangian. In Section \ref{brkHLS} we first remind  the loop
 mixing scheme \cite{taupaper,ExtMod1} of the vector mesons and, next, construct  the pion form factor 
 in $e^+e^-$ annihilation and in the decay of the $\tau$ lepton. The condition
 $F_\pi(0)=1$ has some consequences for how  parametrizing 
 the Breit--Wigner amplitudes  should be done for narrow objects like the $\omg$ and $\phi$ mesons. 
Other topics are also examined~: the direct $\omg \pi \pi$ coupling and the 
$\phi \ra K \overline{K}$ couplings. The anomalous sector is examined in Section \ref{brkFKTUY}
where we also provide  the expressions for the $e^+ e^- \to \pi^0 \gamma$,
$e^+ e^- \to \eta \gamma$ and $e^+ e^- \to \pi^+ \pi^- \pi^0$ cross sections.
The expression for the various couplings of the form $P \gamma  \gamma $
and $V P \gamma $ are also derived; these are important ingredients for
the set of radiative decays included into the HLS framework.

We have found it appropriate to summarize the main features of the HLS model under 
the upgraded breaking scheme which underlies the present study; this
is the matter of Section \ref{summary}. Section \ref{DataSets} is 
devoted to list the different data sets  available for each physics
channel; in this Section, our fitting method, previously defined and
used in \cite{taupaper,ExtMod1,ExtMod2}, is reminded.

At this point, we are in position to confront our model 
and the data. Section \ref{3pions} examines the fit properties of the 
available $e^+ e^- \to \pi^+ \pi^-\pi^0$ data and Section \ref{KKbar}
reports on the simultaneous analysis of the $e^+ e^- \ra K^+K^-$ and 
$e^+ e^- \ra K^0 \overline{K}^0$ annihilation data. The analysis
of the  $e^+ e^- \ra K  \overline{K} $ channels allows us to show how
the problem raised by both $\phi \ra K  \overline{K}$ decay widths is solved
within the new release of the broken HLS model.

Section \ref{fit_tau}
provides our analysis of the dipion spectrum in the $\tau$ decay
in conjunction with $all$ $e^+ e^-$ data. It is shown therein
that  $e^+ e^-$ data and $\tau$  data are fully reconciled;
the precise mechanism solving this issue, somewhat unexpected, is exhibited. 

  The short Section \ref{omgdecay} is devoted to examining the
  exact structure of the $\omg \pi \pi$ coupling and compare
  with similar results of other authors \cite{Maltman1996,Maltman2009}.
  Similarly, another short Section \ref{PSproperties} examines
  in some detail some properties of the $\pi^0-\eta-\eta^\prime$
  mixing; it is shown here that the conclusions derived in
  \cite{WZWChPT} about the mixing angles $\theta_0$ and $\theta_8$
  introduced by \cite{leutwb,leutw} are confirmed, together with
  their relationship with the traditional singlet--octet 
  mixing angle  $\theta_P$. In Section \ref{Kubis}, one examines
  the fitted values of the parameters involved in the
  absolute scale of the FKTUY anomalous Lagrangian pieces and compare with
  existing estimates; this leads to the conclusion that
  the usual assumption $c_3=c_4$ is consistent with data.
  
  Section \ref{gmoins2} is devoted to study in detail the consequences
  for our HLS model  determination of the non--perturbative 
  part of the photon hadronic vacuum polarization. This is found to yield much
  reduced uncertainties compared to estimates derived by the direct averaging of data.
  
  The consequence for
  $g-2$ are also examined with the conclusion that the theoretical
  prediction differs from the BNL measurement \cite{BNL2}. The significance
  of this difference is shown to stay in between $4.07 \sigma$ and $4.33 \sigma$.  
 This looks an important improvement, as  we are still not using  the ISR data.
  
  Finally Section \ref{Conclusion} provides a summary of our
  conclusions and the perspectives. A large part of the
  formulae have been pushed inside several Appendices in order to 
  ease as much as possible the reading of the main text.

\section{The HLS Lagrangian}
\label{sect1}
\indent \indent The Hidden Local Symmetry Model (HLS) has been 
presented  in full detail in \cite{HLSOrigin} and, more recently,  
in \cite{HLSRef}. One can also find brief accounts  in
\cite{Heath1998,Heath}.

Beside its non--anomalous sector, which allows to address 
most $e^+ e^-$ annihilation channels and some $\tau$ decays up to about the $\phi$ meson
mass \cite{taupaper,ExtMod1},  the HLS Model also contains 
an anomalous (FKTUY) sector \cite{FKTUY} which provides couplings of the form
$VVP$, $VPPP$,  $\gamma PPP$,$VP\gamma$ or  $P\gamma \gamma$
among light flavor mesons. These are the key  in order to incorporate within
the HLS framework the radiative decays of the form $VP\gamma$ or  
$P \ra \gamma \gamma$, or decays importantly influenced by the box
anomaly like $\eta/\eta^\prime \ra \pi^+ \pi^- \gamma$ (see \cite{Abele,box}
for instance). It has been shown  that, while implementing
 (U(1)) nonet symmetry and  SU(3) symmetry breakings, one reaches
a remarkable agreement with data \cite{rad,box}.

 The anomalous pieces of the HLS Model are also the key tool when dealing with 
 annihilation
processes like $e^+ e^- \ra \pi^0 \gamma$, $e^+ e^- \ra \eta \gamma$
or $e^+ e^- \ra \pi^0 \pi^+\pi^-$ as successfully shown in \cite{ExtMod1}.
 
 In order to be self--contained, 
and without going into unnecessary detail, let us briefly remind the salient 
features of the HLS Model relevant for the present purpose.  

One defines the $\xi$ fields by~:
\be
\xi_{R,L} = \displaystyle  
\exp{[i \displaystyle \sigma/f_\sigma]} \exp{[\pm i \displaystyle P/f_\pi]}
\label{eq1}  
\ee
where the scalar field $\sigma$ is usually eliminated by means of a suitable
gauge choice  \cite{HLSOrigin} (the so--called unitary gauge). However the decay 
constant $f_\sigma$ still survives through
the ratio $a=f_\sigma^2/f_\pi^2$ which is a basic (free) ingredient
of the HLS Model. The standard VMD Lagrangian corresponds to having $a=2$.
The pseudoscalar field matrix $P$~:
\be
P= P_8 +P_0=\frac{1}{\sqrt{2}}
  \left( \begin{array}{ccc}
          \displaystyle  \frac{1}{\sqrt{2}}\pi_3+\frac{1}{\sqrt{6}}\eta_8+
            \frac{1}{\sqrt{3}}\eta_0&\displaystyle \pi^+ & \displaystyle  K^+ \\
            \displaystyle \pi^-  & \displaystyle -\frac{1}{\sqrt{2}}\pi_3+\frac{1}{\sqrt{6}}\eta_8
            +\frac{1}{\sqrt{3}}\eta_0  &  \displaystyle K^0 \\
           \displaystyle  K^-             &  \displaystyle \overline{K}^0  &\displaystyle 
             -\sqrt{\frac{2}{3}}\eta_8 +\frac{1}{\sqrt{3}}\eta_0 \\
         \end{array} 
  \right),
\label{eq2}  
\ee
contains singlet ($P_0$) and octet ($P_8$) terms. By $\pi_3$ we denote  the bare neutral pion field; 
the traditional naming $\pi^0$ will be devoted
to the fully renormalized neutral pion field. On the other hand, the usual $\eta$ and $\eta^\prime$
meson fields are (essentially) combinations of the $\eta_8$ and $\eta_0$ fields shown in Eq. (\ref{eq2}).
 
The HLS Lagrangian is defined by~:
 \be
\left \{
 \begin{array}{lll}
 {\cal L}_{HLS}= &{\cal L}_A + a {\cal L}_V & ~~\\[0.5cm]
 {\cal L}_A = &\displaystyle  -\frac{f_\pi^2}{4} {\rm Tr} [(D_\mu\xi_L\xi_L^\dagger -D_\mu\xi_R\xi_R^\dagger)^2] &
 \displaystyle \equiv -\frac{f_\pi^2}{4} {\rm Tr} [L-R]^2\\[0.5cm]
 {\cal L}_V = &\displaystyle  -\frac{f_\pi^2}{4} {\rm Tr} [(D_\mu\xi_L\xi_L^\dagger +D_\mu\xi_R\xi_R^\dagger)^2] &
 \displaystyle \equiv -\frac{f_\pi^2}{4} {\rm Tr} [L+R]^2\\[0.5cm]
 \end{array} 
 \right .
\label{eq3}  
\ee
where the covariant derivatives are given by~:
 \be
\left \{
\begin{array}{ccc}
D_\mu \xi_L  = \displaystyle  \pa_\mu \xi_L -i g V_\mu \xi_L +i \xi_L {\cal L}_\mu\\[0.5cm]
D_\mu \xi_R  = \displaystyle  \pa_\mu \xi_R -i g V_\mu \xi_R +i \xi_R {\cal R}_\mu
 \end{array} 
 \right .
\label{eq4}  
\ee
with~:
\be
\left \{
\begin{array}{l}
{\cal L}_\mu =   \displaystyle  e Q A_\mu + \frac{g_2}{\cos{\theta_W}} (T_z -\sin^2{\theta_W})Z_\mu 
+\frac{g_2}{\sqrt{2}} (W^+_\mu T_+ + W^-_\mu T_-)\\[0.5cm]
{\cal R}_\mu =   \displaystyle e Q A_\mu - \frac{g_2}{\cos{\theta_W}} \sin^2{\theta_W} Z_\mu 
 \end{array}  
 \right .
\label{eq5}  
\ee
exhibiting the $Z$, $W^\pm$ boson fields together with the photon field $A_\mu$.
The vector field matrix is given by~:
\be
V=\frac{1}{\sqrt{2}}
  \left( \begin{array}{ccc}
   \displaystyle (\rho^I+\omega^I)/\sqrt{2}  & \displaystyle \rho^+             &  \displaystyle K^{*+} \\[0.5cm]
    \displaystyle  \rho^-    & \displaystyle  (-\rho^I+\omega^I)/\sqrt{2}    &  \displaystyle  K^{*0} \\[0.5cm]
      \displaystyle        K^{*-}           & \displaystyle  \overline{K}^{*0}  &  \displaystyle  \phi^I   
         \end{array}
  \right) 
\label{eq6}
\ee

The quark charge matrix $Q$ is standard and the matrix $T_+=[T_-]^\dagger$ is constructed out
of matrix elements of the Cabibbo--Kobayashi--Maskawa matrix \cite{HLSRef,taupaper}. One should
note that the neutral charge entries of the vector field matrix 
$V$ are expressed in terms of the so--called ideal fields ($\rho^I$, $\omega^I$ and $\phi^I $).

In the expressions above, one observes the electric charge $e$, the universal vector coupling $g$
and the weak coupling $g_2$ (related with the Fermi constant by $g_2=2 m_W \sqrt{G_F\sqrt{2}}$).
As the influence of the $Z$ boson field is quite negligible in the physics we address, the
Weinberg angle $\theta_W$  plays no role.

We do not present here the anomalous sectors which can be found in the original HLS literature
\cite{HLSOrigin,FKTUY,HLSRef}. A summarized version, well suited to the present purpose, can be found
in  \cite{ExtMod1}) and will not be repeated.

The non--anomalous Lagrangian ${\cal L}_{HLS}$ at lowest order in field
derivatives can be found expanded in
\cite{Heath,Heath1998}. Its $\tau$ sector is explicitly given in
 \cite{taupaper,ExtMod1}.

The  HLS Lagrangian  fulfills a $U(N_f) \times U(N_f)$ symmetry rather than
$SU(N_f) \times  SU(N_f)$. The additional axial U(1) symmetry has several undesirable
features \cite{tHooft,feldmann_2}, especially a ninth light pseudoscalar meson. 
This symmetry can easily be reduced by adding appropriate terms to the effective 
Lagrangian. Defining  \cite{HLSOrigin} the chiral field 
$U\equiv \xi_L^\dag \xi_R=\exp{2 i P/f_\pi}$, this reduction is achieved by adding
 determinant terms\cite{tHooft} to the HLS Lagrangian. After this operation, 
 one gets \cite{WZWChPT}~:
  
\be
{\cal L}={\cal L}_{HLS}+{\cal L}_{tHooft}={\cal L}_{HLS}+\frac{\mu^2}{2} \eta_0^2  +
\frac{1}{2}\lambda \partial_\mu \eta_0 \partial^\mu \eta_0
\label{eq7}
\ee
where $\mu$ has obviously a mass dimension and $\lambda$ is dimensionless. In the following,
the additional Lagrangian piece will not be modified while breaking symmetries. Actually,
in the present work, one is only concerned by the perturbation
of the pseudoscalar meson kinetic energy.

\section{The BKY--BOC Breaking of the HLS Lagrangian}
\label{sect2}
\indent \indent The HLS Lagrangian above is certainly
an interesting and attractive framework. 
However, without introducing suitable mechanisms for symmetry breaking effects,
one cannot account for the experimental data at the level of
precision required by their accuracy. There is no unique 
way to implement such a mechanism within the HLS model and, actually, several SU(3) breaking 
schemes exist. The basic SU(3) symmetry breaking scheme has been proposed by Bando, Kugo and 
Yamawaki (BKY) \cite{BKY}. It has, however, some undesirable properties which have motivated
its modification. A first acceptable modification has been proposed by Bramon, Grau and 
Pancheri \cite{BGP,BGPbis} and another one in \cite{Heath}, where
the  various schemes  have been critically
examined.  Following this study, we prefer using in the following  the so--called 
"new scheme" variant defined in  \cite{Heath}; when referring to the BKY mechanism
throughout this paper, we always mean the "new scheme" variant just mentioned.
It will be referred to as either BKY or BKY--BOC. 

This breaking mechanism  (BKY--BOC) has been examined in detail  and
 its predictions -- relying on fits to experimental data --  have been found to meet
 the corresponding  ChPT expectations \cite{WZWChPT} at first order in the breaking parameters.  
It has also  been extensively used in several successful studies performed on radiative decays
of light mesons\cite{rad} and on $e^+e^-$ annihilation cross sections \cite{ExtMod1}. 
Up to now, the BKY mechanism was limited to SU(3) symmetry breaking effects; the issue now is to 
examine its extension to isospin symmetry breaking.

Briefly speaking, our variant of the BKY mechanism \cite{Heath}
turns out to define the broken non--anomalous HLS Lagrangian pieces by~:
 
\be
\left \{
 \begin{array}{ccc}
  {\cal L}_A  & \displaystyle \equiv -\frac{f_\pi^2}{4} {\rm Tr} [(L-R)X_A]^2\\[0.5cm]
 {\cal L}_V  & \displaystyle \equiv -\frac{f_\pi^2}{4} {\rm Tr} [(L+R)X_V]^2\\[0.5cm]
 \end{array} 
 \right .
\label{eq8}  
\ee
\noindent where $X_A$ and $X_V$ are matrices carrying the SU(3) symmetry breaking associated
with, respectively, ${\cal L}_A$ and ${\cal L}_V$. These are written as~:
 \be
\left \{
 \begin{array}{ll}
 X_A=& {\rm Diag}(1,1,z_A)\\[0.5cm]
 X_V=& {\rm Diag}(1,1,z_V)
 \end{array} 
 \right .
\label{eq9}	
\ee
\noindent  and departures of $z_A$ and $z_V$ from 1 account for  SU(3) symmetry breaking effects
in the ${\cal L}_A$ and ${\cal L}_V$ Lagrangian pieces\footnote{In
the following $X_A$ and $X_V$ are named breaking matrices; this convenient naming should not hide
that the true breaking matrices are rather $X_A-1$ and $X_V-1$.}. 
A priori, these two parameters are
unrelated and should be treated as independent of each other.

In order to extend to  isospin symmetry breaking, we propose to generalize Eqs. (\ref{eq9}) 
above to~:
 \be
\left \{
 \begin{array}{ll}
 X_A=& {\rm Diag}(q_A,y_A,z_A)\\[0.5cm]
 X_V=& {\rm Diag}(q_V,y_V,z_V)~~~~~.
 \end{array} 
 \right .
\label{eq10}	
\ee

As isospin symmetry breaking is expected milder than SU(3) breaking,
 the additional breaking parameters are obviously expected to fulfill~:
 \be
 \begin{array}{ll}
 |q_A-1|,~|y_A-1| <<  |z_A-1|~~,&~|q_V-1|,~|y_V-1| <<  |z_V-1|
 \end{array} 
\label{eq11}	
\ee 
In previous fits, performed with only SU(3) symmetry breaking, we got (see for instance
\cite{taupaper,ExtMod1}) $|z_A-1|,~|z_V-1|\simeq 0.5$. Such ways  of extending the 
BKY breaking mechanism have been already proposed within
similar contexts \cite{Harada1995,Hashimoto}.

We find appropriate to define~:
 \be
 \left \{
 \begin{array}{ll}
 q_{A/V} = \displaystyle 1 +  \epsilon_{A/V}^u=1 +  \frac{\Sigma_{A/V}+\Delta_{A/V}}{2}\\[0.5cm]
 y_{A/V} = \displaystyle 1 +  \epsilon_{A/V}^d=1 +  \frac{\Sigma_{A/V}-\Delta_{A/V}}{2}
 \end{array}
 \right . 
\label{eq12}	
\ee 
exhibiting the sum and difference of $\epsilon_{A/V}^u$ and $\epsilon_{A/V}^d$.
Indeed, the expressions for most physical couplings are simpler in terms of
these rather than in terms of  $\epsilon_{A/V}^u$ and $\epsilon_{A/V}^d$.

As clear from Equations (\ref{eq8}), the BKY breaking of the HLS Lagrangian
exhibits  a global character. It does not correspond to some systematic
way of including specific breaking terms of given kind or order  as done within ChPT.
As the numerical values of the breaking parameters are phenomenologically
derived from fits to a large set of experimental data,  they account $globally$
for several effects of different origin without any way to
disentangle the various contributions. This remark is especially relevant for
the breaking parameters corresponding to the $u$ and $d$ entries of  $X_A$ and $X_V$
which are small enough that several competing effects can mix together\footnote{This may include 
effects due to the quark mass breaking and to electromagnetic corrections.
It may also absorb corrections to hadronic vertices which can hardly
be derived from an effective  Lagrangian.}; because of their relatively large magnitude, the
SU(3) breaking effects can be more easily identified \cite{WZWChPT,rad}.

\section{Breaking the ${\cal L}_A$ Lagrangian Piece}
\label{brkLa}
 \indent \indent  The pseudoscalar kinetic energy term of the full (broken) Lagrangian is
 carried by  ${\cal L}_{A}+{\cal L}_{tHooft}$. In terms of bare fields, it is~:
 \be
\begin{array}{ll}
{\cal L}_{Kin} =  \displaystyle q_A y_A \partial \pi^+ \cdot \partial \pi^-
+ \frac{q_A^2+y_A^2}{4}  \partial \pi_3 \cdot  \partial \pi_3 
+ q_A z_A  \partial K^+ \cdot  \partial K^- + y_A z_A  \partial K^0 \cdot  \partial \overline{K}^0
\\[0.5cm]
 +\displaystyle \left [ \frac{z_A^2}{3}+\frac{q_A^2+y_A^2}{12} \right ] \partial\eta_8\cdot\partial\eta_8
+\left [\frac{z_A^2+q_A^2+y_A^2}{6} + \frac{\lambda}{2}\right ]\partial \eta_0 \cdot \partial \eta_0
\\[0.5cm]
\displaystyle 
+\sqrt{2}\frac{(q_A^2+y_A^2)-2 z_A^2}{6} \partial\eta_8 \cdot \partial\eta_0
+\frac{q_A^2-y_A^2}{\sqrt{12} }\partial \pi^0  \cdot \partial\eta_8
+\frac{q_A^2-y_A^2}{\sqrt{6} }\partial \pi^0  \cdot \partial\eta_0
\end{array}
\label{eq13}
\ee
which is clearly non--canonical.
In order to restore the canonical structure, one should perform a change of fields. This is done
in two steps, as in \cite{WZWChPT}. 

\subsection{First Step PS Field Renormalization}
\indent \indent  
The first step renormalization turns out to define the (step one) renormalized pseudoscalar
field matrix $P^{R_1}$ in term of the bare one $P$ by~:
 \be
P^{R_1}=X_A^{1/2} P X_A^{1/2}
\label{eq14}
\ee
The charged pion and both kaon terms in this expression actually  undergo their 
final renormalization already at
this (first) step. Indeed, the axial currents are defined  as in \cite{WZWChPT} and are given by~:
 \be
 J_\mu^{a}=\displaystyle -2 f_\pi [{\rm Tr} [ T^a X_A \partial_\mu P X_A] 
 + \lambda \delta^{a,0} \partial_\mu  \eta_0]
\label{eq15}
\ee
in terms of the bare fields and of the Gell--Mann matrices $T^a$, normalized
such as ${\rm Tr}[T^a T^b]=\delta^{a~b}/2$. 
The $\pi^\pm$, $K^\pm$ and $K^0/\overline{K}^0$ decay constants
are defined by the relevant axial current matrix elements closed on the 
renormalized PS meson fields~:
$<0|J_\mu^{\pi^\pm/K}|\pi^\pm_{R}> = if_{\pi/K}q_\mu $. As one chooses the renormalized (charged)
pion decay constant to coincide with its experimental central value \cite{RPP2006} 
($f_{\pi^\pm}=92.42$
MeV), this turns out to impose $q_A y_A=1$. At leading order in breaking parameters, 
this implies $\Sigma_A=0$. Then, the breaking matrix $X_A$ writes~:
 \be
  X_A={\rm Diag}(1+\frac{\Delta_A}{2},1-\frac{\Delta_A}{2},z_A)
\label{eq16}	
\ee
depending on only two free parameters ($\Delta_A$ and $z_A$). On the other 
hand, the kaon decay constant is~:
 \be
\displaystyle 
f_{K^\pm}=\sqrt{z_A}(1+\frac{\Delta_A}{4})f_\pi
\label{eq17}
\ee
One thus yields a marginal change compared to the previous  BKY breaking scheme
\cite{Heath,WZWChPT} (dealing only with SU(3) symmetry)
as one got $[f_{K^\pm}/f_{\pi^\pm}]^2=z_A$.
Anticipating somewhat on our numerical results, let us mention that the fits  always return
 $\Delta_A \simeq (5. \div 6.) \%$, much larger than  expected from solely
 an effect of the light quark mass difference \cite{GassLeutw}; this will be further discussed
 in Subsection \ref{phiKKbar}. 

 As clear from Eq. (\ref{eq16}), the $X_A$ matrix resembles the usual
 quark mass breaking matrix. However,  the entry $z_A$ is  essentially related 
 with the ratio $[f_K/f_\pi]^2 \simeq 1.5$, while the corresponding entry
 in the quark mass breaking matrix is numerically $\simeq 20$. Therefore,
 the correspondence between these two matrices is only formal.

The following relationship defines  some bare PS fields
in terms of their (fully) renormalized partners~:
\be
\left \{
\begin{array}{lll}
\displaystyle 
\pi^\pm &= \displaystyle \pi^\pm_R\\[0.5cm]
K^{\pm}&=\displaystyle \frac{1}{\sqrt{z_A}}(1-\frac{\Delta_A}{4})
K^{\pm}_R\\[0.5cm]
K^{0}&=\displaystyle \frac{1}{\sqrt{z_A}}(1+\frac{\Delta_A}{4}) K^{0}_R
\end{array}
\right.
\label{eq18}
\ee
\indent \indent  
This produces changes going in opposite directions for the couplings 
involving the physical $K^{\pm}$ and $K^{0}$ mesons compared to their bare partners.
This has a clear influence on the cross section ratio $\sigma(e^+e^- \ra K^+K^-)/\sigma(e^+e^- 
\ra K^0 \overline{K}^0)$. On the other hand, one also gets at leading order in 
breaking parameters the following relationship between some bare PS fields and the (first step) 
renormalized PS fields~: 

\be
\left \{
\begin{array}{ll}
\pi_3 =  \displaystyle \pi^{R_1}_3
-\frac{\Delta_A}{2\sqrt{3}}\eta_8^{R_1}
-\frac{\Delta_A}{\sqrt{6}}\eta_0^{R_1}\\[0.5cm]
\eta_0 =  
\displaystyle -\frac{\Delta_A}{\sqrt{6}}\pi^{R_1}_3
+\frac{\sqrt{2}}{3}\frac{z_A -1}{z_A} \eta_8^{R_1}
+\frac{1}{3}\frac{2 z_A+1 }{z_A} \eta_0^{R_1}\\[0.5cm]
\eta_8 =  
\displaystyle -\frac{\Delta_A}{2\sqrt{3}}\pi^{R_1}_3
+\frac{1}{3}\frac{z_A  +2}{z_A} \eta_8^{R_1}
+\frac{\sqrt{2}}{3}\frac{z_A -1 }{z_A} \eta_0^{R_1}
\end{array}
\right.
\label{eq19}
\ee

\subsection{Second Step PS Field Renormalization}
\indent \indent  While propagating the field redefinition displayed into 
Eqs. (\ref{eq18}, \ref{eq19})
in the expression for ${\cal L}_{Kin}$ (Eq. (\ref{eq13})), one should neglect second 
(and higher)
 order terms in the breaking parameters  $\Delta_A$ and $\lambda$. Indeed,
 both of these  are expected
 small (of the order of a few percent at most); instead, as $|z_A-1|$ is rather large 
 ($z_A\simeq 1.5$), we do not proceed alike with the SU(3) breaking term. Doing so,
 in terms of the (redefined) fields, the only surviving non--canonical piece 
 ${\cal L}_{0,8}$ writes~: 
\be
\begin{array}{ll}
2 {\cal L}_{0,8}= \displaystyle
[\partial\eta^{R_1}_0]^2+ [\partial\eta^{R_1}_8]^2 + \frac{\lambda}{9} \left[
\left(2+\frac{1}{z_A}\right) \partial\eta^{R_1}_0 + \sqrt{2}
\left(1-\frac{1}{z_A}\right)\partial\eta^{R_1}_8
\right]^2
\end{array}
\label{eq20}
\ee
and is independent of $\Delta_A$. As we  get -- at leading order -- the same dependence
as before \cite{WZWChPT}, the diagonalization procedure for this term is known
(see Section 3.1 in \cite{WZWChPT}). Let us only recall the results in the present
notation set~:
\be
\left \{
\begin{array}{lll}
\displaystyle
\pi^{R_1}_3=\pi^{R}_3\\[0.5cm]
\displaystyle
\eta^{R_1}_8=\frac{1+v\cos^2{\beta}}{1+v}\eta^R_8
-\frac{v\sin{\beta}\cos{\beta}}{1+v}\eta^R_0\\[0.5cm]
\displaystyle
\eta^{R_1}_0=-\frac{v\sin{\beta}\cos{\beta}}{1+v}\eta^R_8
+ \frac{1+v\sin^2{\beta}}{1+v}\eta^R_0
\end{array}
\right.
\label{eq21}
\ee
where~:
\be
\left \{
\begin{array}{ccc}
\displaystyle
\cos{\beta}=\frac{2z_A+1}{\sqrt{3(2 z_A^2+1)}}~~,~~
\displaystyle
\sin{\beta}=\frac{\sqrt{2}(z_A-1)}{\sqrt{3(2 z_A^2+1)}}\\[0.5cm]
\displaystyle
v=\sqrt{1+\lambda \frac{(2 z_A^2+1)}{3z_A^2}} -1 
\simeq \frac{\lambda}{2}\frac{(2 z_A^2+1)}{3z_A^2}~~ .
\end{array}
\right.
\label{eq22}
\ee

It thus looks more appropriate to use $v(\simeq \lambda/2)$ rather than \cite{WZWChPT} 
$\lambda$  as a breaking parameter, as it allows to work with simpler expressions. 
$v$ is the first parameter in our model which exhibits the intricacy between
U(1) and SU(3) breakings ($\lambda$ and $z_A$).
The canonical PS  fields -- 
denoted by the superscript $R$ -- are finally defined by Eqs. 
(\ref{eq18},\ref{eq19},\ref{eq21}).

\subsection{The $\pi^0-\eta-\eta^\prime$ Mixing}
\label{PSmixing}
\indent \indent  The physically observed $\eta$ and $\eta^\prime$ are 
traditionally described as
mixtures of the singlet and octet PS fields $\eta^8$ and $\eta^0$ involving
one mixing angle named here  $\theta_P$. Some
authors, following \cite{feldmann_1,feldmann_2} prefer now using 
mixtures of the $u\bar{u}+d\bar{d}$ and $s\bar{s}$ wave functions. However,
as these two approaches are equivalent, we prefer sticking to the traditional
description.   

Since \cite{leutw,leutwb}, it is admitted that the most appropriate ChPT 
description of the $\eta-\eta^\prime$ mixing involves two decay constants ($F^0$ and
$F^8$) and two mixing angles ($\theta_0$ and $\theta_8$). However, \cite{WZWChPT}
has shown how,  within VMD, the usual octet--singlet mixing scheme connects with this
new approach. In this reference, it was also shown that, relying on experimental data, 
the broken HLS model favors $\theta_0=0$ with a very good accuracy; 
this led to a relation between  $\theta_8$ and $\theta_P$ numerically close
to $\theta_8 = 2 \theta_P$. Comparing accepted ChPT numerical values for $\theta_8$ -- like
those in \cite{leutw} --  with the one derived from $\theta_P$ (determined in VMD fits)
was found quite satisfactory. Moreover, it was shown that fits to experimental data
lead to an algebraic relation of the form $\theta_P=f(\lambda,z_A)$. We will check
whether this relation for $\theta_P$  still fits within the present form of our broken HLS 
model.

As in all previous studies in this series, one could have limited oneself to
considering  only the $\eta-\eta^\prime$ mixing, decoupling this from the $\pi^0$ 
sector. However, it is a classical matter of Chiral Perturbation Theory to
address the issue of  $\pi^0-\eta$ mixing, as this is related with the (light)
quark mass difference \cite{Holstein}. Therefore, it may look interesting to see
if such a phenomenon could be exhibited from the experimental data we deal 
with. In this case, there is no
reason not to address the issue of the relevance of a full $\pi^0-\eta-\eta^\prime$
mixing mechanism. We choose to parametrize this PS mixing using \cite{leutw96}~:
 \be
\left \{
\begin{array}{lll}
\displaystyle 
\pi_R^3= \pi^0-\epsilon ~\eta-\epsilon^\prime ~\eta^\prime\\[0.5cm]
\displaystyle 
\eta_R^8= \cos{\theta_P} (\eta+\epsilon ~\pi^0)+
\sin{\theta_P} (\eta^\prime+\epsilon^\prime ~\pi^0) \\[0.5cm]
\displaystyle 
\eta_R^0 = -\sin{\theta_P} (\eta+\epsilon ~\pi^0)+
\cos{\theta_P} (\eta^\prime+\epsilon^\prime ~\pi^0) 
\end{array}
\right.
\label{eq23}
\ee
where $\pi_R^3$, $\eta_R^8$ and $\eta_R^0$ are the already redefined fields
(see Eqs. (\ref{eq21}) above),  the physically observable mesons
being $\pi^0$, $\eta$ and $\eta^\prime$. In the smooth limit of vanishing
$\epsilon$ and $\epsilon^\prime$, one recovers the usual $\eta-\eta^\prime$
mixing pattern with one ($\theta_P$) mixing angle, while the pion field decouples. 
Even if one does not expect a large influence of  $\epsilon$ and $\epsilon^\prime$
in the full data set collection we consider, it does not harm to examine
their effects and, if relevant impose $\epsilon=\epsilon^\prime=0$ to the model. 

Finally, at leading order in breaking parameters, the pseudoscalar
meson kinetic energy term is canonical  when expressed in terms of the
fully renormalized fields (those carrying a $R$ subscript).

\subsection{About The $\theta_8$, $\theta_0$ and $\theta_P$ Mixing Angles}
\label{mixingAngl}  
\indent \indent  Let us define~: 
\be
\left \{
\begin{array}{lll}
\displaystyle
 g^{\pi^0}(v)=  ( 1 -2 v) \rightarrow~~(\simeq  1 -20\%)
\\[0.5cm]
\displaystyle
g^0(v)= 1-\frac{v}{3}~\frac{(2z_A+1)^2 }{(2 z_A^2+1)}
\rightarrow~~ (\simeq  1 -10\%)\\[0.5cm]
 \displaystyle
g^8(v)= 1-\frac{2v}{3}~\frac{(z_A-1)^2 }{(2 z_A^2+1)}
\rightarrow(\simeq  1 -0.3\%)
\end{array}
\right.
\label{eq24}
\ee
These functions tend to unity when the $U_A(1)$ symmetry is restored ($\lambda=0$).
The property $g^8(v) \simeq  1$ is the simplest way to justify the approximation
done in our previous works to parametrize nonet symmetry breaking by the parameter
$x$ (see, for instance, the discussion in \cite{WZWChPT}). 
Using these functions, one can derive from  Eqs (\ref{eq15}) the following axial 
currents~: 
\be
\left \{
\begin{array}{lll}
J^{\pi_3}_\mu=&\displaystyle f_\pi \left \{
\partial_\mu \pi_R^3
+\Delta_A ~ g^{\pi^0}(v)~
\left[
\frac{1}{2\sqrt{3}}\partial_\mu \eta_R^8+
\frac{1}{\sqrt{6}}\partial_\mu \eta_R^0
\right]
\right\} \\[0.5cm]
\displaystyle 
J^{\eta^0}_\mu=&\displaystyle 
f_\pi \left \{
\frac{\Delta_A}{\sqrt{6}}\partial_\mu \pi_R^3+
\frac{z_A+2}{3}g^0(v)~~ \partial_\mu \eta_R^0 
-\sqrt{2}\frac{z_A-1}{3}g^8(v)~~ 
\partial_\mu \eta_R^8
\right\} 
\\[0.5cm]
J^{\eta^8}_\mu=&\displaystyle 
f_\pi \left \{
\frac{\Delta_A}{2\sqrt{3}}\partial_\mu \pi_R^3
-\sqrt{2}\frac{z_A-1}{3}g^0(v) ~~\partial_\mu \eta_R^0
+\frac{2z_A+1}{3}g^8(v)~~
\partial_\mu \eta_R^8
\right\} 
\end{array}
\right.
\label{eq25}
\ee
The mixing angles $\theta_8$, $\theta_0$  \cite{leutw,leutwb} yield the following
expressions~:
\be
\left \{
\begin{array}{lll}
\displaystyle \tan{\theta_8}=~\frac{<0|\partial^\mu J^8_\mu|\eta^\prime>}
				  {<0|\partial^\mu J^8_\mu|\eta>}= \tan{(\theta_P-A)}
~~~,& \displaystyle ~~\tan{A}\equiv \sqrt{2} \frac{z_A-1}{2z_A+1}\frac{g^0(v)}{g^8(v)}
 \\[0.5cm]
 \displaystyle \tan{\theta_0}=-\frac{<0|\partial^\mu J^0_\mu|\eta>}
				  {<0|\partial^\mu J^0_\mu|\eta^\prime>}= \tan{(\theta_P+B)}
~~~,& \displaystyle ~~\tan{B}\equiv \sqrt{2} \frac{z_A-1}{z_A+2}\frac{g^8(v)}{g^0(v)}
\end{array}
\right.
\label{eq26}
\ee
One can easily check that $g^8(v)/g^0(v)\simeq 1-\lambda/2$. A property to check
from fits using the present form of the model is  whether $\theta_0$ is still consistent
with zero \cite{WZWChPT}. In this case,  $\theta_P$ is still no longer a free
parameter, but fully determined by $\lambda$ and $z_A$, {\it i.e.} by breaking
parameters and $\theta_P$ tends to zero when the symmetries are restored.

One should also  note that the usual ChPT definition of the $\pi^0$
decay constant ($<0|J^{\pi_3}_\mu| \pi^0>=iq_\mu f_{\pi^0}$) 
provides $f_{\pi^0}=f_{\pi^\pm}$, not influenced by our isospin breaking procedure. 
However, as will be seen shortly -- and as already emphasized in \cite{WZWChPT} for 
the decays $\eta/\eta^\prime \ra  \gamma \gamma$ -- this is not the quantity actually
involved in the decay $\pi^0 \ra  \gamma \gamma$.

\section{Breaking the ${\cal L}_V$ Lagrangian Piece}
\label{brkLv}
 \indent \indent  The  ${\cal L}_V$ Lagrangian, is defined by Eqs. (\ref{eq8}--\ref{eq12}).
It yields the following  vector meson mass term ($m^2 \equiv a g^2 f_\pi^2$)~:
 \be
{\cal L}_{mass} = \displaystyle \frac{m^2}{2}
\left [(1+ \Sigma_V) \rho_I^2 +
       (1+ \Sigma_V) \omg_I^2  + 2\Delta_V \rho_I \cdot \omg_I+
	z_V \phi_I^2  + 2 (1+ \Sigma_V) \rho^+ \cdot \rho^- \right ]
\label{eq27}
\ee
while keeping only the leading terms in the breaking parameters $\Sigma_V$ and $\Delta_V$
(the $K^*$ mass term has been dropped out). As can be seen, the canonical structure of
the mass term is broken by a $\rho_Î \cdot \omg_Î$ term. 

In order to restore the canonical form of the mass term, one should perform a field redefinition in only 
the $(\rho_Î -\omg_Î)$ sector. Interestingly, the requested transform is not a  rotation 
but~:
 \be
 \left (
 \begin{array}{ll}
 \rho_I \\[0.5cm]
 \omg_I
\end{array}
\right)=
\left (
\begin{array}{c}
\rho_{R_1} \\[0.5cm]
\omg_{R_1}
\end{array}
\right) - \Delta_V
\left(
\begin{array}{c}
h_V\omg_{R_1}\\[0.5cm]
(1-h_V)\rho_{R_1} 
\end{array}
\right) 
\label{eq28}
\ee
makes the work when  non--leading terms in $\Sigma_V$ and $\Delta_V$ are neglected
\footnote{Eq. (\ref{eq27}) can be diagonalized by a 45$^\circ$ rotation,
however, this solution is physically unacceptable as it has not the requested smooth
limit when $\Delta_V \ra 0$.}.
In terms of the $R_1$ renormalized fields, one gets~: 
 \be
{\cal L}_{mass} = \displaystyle \frac{m^2}{2}
\left [(1+ \Sigma_V) \rho_{R_1}^2 +
       (1+ \Sigma_V) \omg_{R_1}^2  + 
	z_V \phi_{R_1}^2  + 2 (1+ \Sigma_V) \rho^+ \cdot \rho^- \right ]
\label{eq29}
\ee
having renamed for convenience $\phi_I \equiv \phi_{R_1}$. A few remarks are 
worth being done~: 
\begin{itemize}
\item[(1)]
Beside the two breaking parameters $\Sigma_V$ and $\Delta_V$,
one gets a third free parameter $h_V$ which governs the  mixing of
$\rho_{R_1}$ and $\omg_{R_1}$.

Therefore, the exact content of isospin 1 ($\rho_I$) inside $\omg_{R_1}$
and of isospin 0  ($\omg_I$) inside $\rho_{R_1}$ should be extracted
from data.

\item[(2)]
The masses for $\rho_{R_1}$, $\omg_{R_1}$ and $\rho^\pm$
remain degenerated at leading order in the breaking parameters as   
the needed $R_1$ change of fields results in a vector meson mass term 
independent of $\Delta_V$.

Therefore, even if  one may legitimately  think that breaking isospin symmetry
inside ${\cal L}_V$ could result into a non zero (Lagrangian) mass 
difference \cite{ExtMod2} $\delta m^2$  between the charged 
and the neutral $\rho$ mesons, our breaking  procedure 
rules out such a possibility at leading order in breaking parameters. 
Actually, electromagnetic corrections \cite{Cottingham,Harari}, presently neglected,
generate such a mass difference. Such a term has been considered in \cite{ExtMod2}
but found numerically insignificant; preliminary studies within the present framework
leading to the same conclusion, we have given up considering explicitly a $\rho^0-\rho^\pm$
mass difference. 

\item[(3)] The field transform (\ref {eq29}) propagates to the vector meson kinetic
energy by generating a term of the form $\Delta_V \partial \rho^0_{R_1} \partial \omg_{R_1}$
which breaks the canonical structure of the  kinetic energy. This is a classical 
issue \cite{Sakurai}  known to imply
the occurence of wave--function renormalization factors \cite{Sakurai} which are 
absorbed into the effective couplings defined by the Lagrangian vertices. In our case,
they are certainly absorbed in our breaking parameters. This is exactly
the same issue which arises in the electroweak Standard Model with
the $\gamma-Z^0$ mixing. This has been investigated in detail within
$Z^0$ lineshape studies (at the one plus two--loop level) and by the LEP experiments.
The same issue also appears when treating the $\gamma-\rho^0$ mixing
and has been discussed in [16].
\end{itemize}

The second step renormalization of the vector meson fields, 
which accounts for loop effects \cite{taupaper,ExtMod1}, is considered below.

\section{The Fully Broken Non--Anomalous HLS Lagrangian}
\label{brkHLS}
\indent \indent  For definiteness, we name (abusively) from now on
"HLS Lagrangian" the full expression given in Eq. (\ref{eq7}), {\it i.e.}
including the determinant terms. The HLS Lagrangian is explicitly
provided  in Appendix \ref{AA}, dropping out for conciseness  all terms 
not relevant for the purpose of the present study.

At the present step -- which does not still include the
(second step) redefinition of the neutral vector fields \cite{taupaper,ExtMod1} --
several remarks are worth being done~:
\begin{itemize}
\item  
The vector meson masses occuring in the Lagrangian fulfill
$m_{\rho^0}^2=m_{\rho^\pm}^2=m_{\omg}^2$. Thus, no mass splitting is generated,
except for the $\phi$ meson.
\item  
The couplings $\rho \pi \pi$ undergo isospin breaking ($\Sigma_V$) but remain
strictly identical for the charged and neutral $\rho$ mesons. Instead, a direct $\omg \pi \pi$
coupling is generated; it is proportional to $(1-h_V) \Delta_V$. 

\item  
The $\rho^0-\gamma$ and $\rho^\pm -W^\pm$ transition 
amplitudes\footnote{Compare
$f_{\rho \gamma}$ and $f_{\rho W}$ as given by Eqs. (\ref{AA3}) and (\ref{AA4}).}
may slightly differ, as $h_V \Delta_V/3$ should not exceed a few percent level. 
\end{itemize} 

Therefore,  non--vanishing 
$\delta m^2=m_{\rho^0}^2-m_{\rho^\pm}^2$ and 
$\delta g_{\rho \pi \pi}=g_{\rho^0 \pi^+ \pi^-}-g_{\rho^\pm \pi^\pm \pi^0}$, as stated in \cite{ExtMod2},
are not derived by extending the $X_A/X_V$ breaking scheme  to include
isospin  symmetry breaking\footnote{As stated above, electromagnetic corrections
contribute to generate a non--vanishing $\delta m^2$ without, however, a significant
influence on the fit properties.}. 

Therefore, non--vanishing $\delta m^2$ and $\delta g_{\rho \pi \pi}$ are
not the way followed by the (broken) HLS model in order to account for the (slightly) different
normalizations of the pion form factor in $\tau$ decays and in $e^+ e^-$ 
annihilations. The actual mechanism at work  is emphasized below.

\subsection{Loop Mixing of Vector Mesons}
\indent \indent   As remarked in \cite{taupaper}, pseudoscalar loops
modify the vector mass matrix by $s$--dependent terms. In this way,
the $\rho$, $\omg$ and $\phi$ squared masses become $s$--dependent
through  contributions at real $s$ of analytic functions\footnote{
Actually, the anomalous FKTUY Lagrangian and the Yang--Mills  terms
contribute respectively with $VP$ and $VV$ loops; one can consider their
influence  absorbed in the subtraction polynomials
of the $PP$ loops \cite{taupaper}. }, namely  the
$K\overline{K}$ loops and, for the  $\rho$, also the charged pion loop.
Conceptually, this turns out to remark that the inverse vector meson
propagator written $D^{-1}_V(s)=s-m_V^2-\Pi(s)$  in order to exhibit the loop
effects, can be thought as $D^{-1}_V(s)=s-m_V^2(s)$, reflecting the running
character of the vector meson squared mass. 

More important, however, is that this $s$--dependent mass matrix becomes
non--diagonal, showing that, at one--loop order, the $\rho$, $\omg$ and $\phi$
(corresponding here to the $R_1$ renormalized vector fields) are no--longer
mass eigenstates. Mass eigenstates can easily be constructed by standard 
perturbative methods \cite{BF} as shown in  \cite{taupaper}; one observes that they 
become $s$--dependent.

This mass matrix  can be written~:
\be
M^2(s) = M^2_0(s) + \delta M^2(s)
\label{Eq30}
\ee 
where\footnote{Entries are ordered  respectively  $\rho_{R_1}$, 
$\omg_{R_1}$ and $\phi_{R_1}$.}~:
\be
M^2_0 (s)= {\rm Diag}(m_\rho^2 +\Pi_{\pi \pi}(s) ,m_\omg^2,m_\phi^2)
\label{Eq31}
\ee 
is treated as the unperturbed part of the squared mass matrix. The pion loop is
weighted by the square   of the $\rho_{R_1} \pi \pi$ coupling constant 
(see Eq.(\ref{AA2}) in Appendix A) and has been included  
in the $\rho_{R_1}$ entry as $\Pi_{\pi \pi}(s)$ is not really
small in the timelike region. 
Instead, as the $\omg_{R_1} \pi \pi$
coupling is first order in $\Delta_V$, the pion loop contribution
to the $\omg_{R_1}$ entry should be neglected ($\simeq {\cal O}(\Delta_V^2)$). 
The values for these (Higgs--Kibble) masses can be found in Eq. 
(\ref{AA3}); they fulfill $m_\rho=m_\omg$. On the other hand, $\delta M^2(s)$ is 
given by~:
\be
\hspace{-1.5cm}
\delta M^2(s) = 
\left [
\begin{array}{cccc}
\displaystyle 
\epsilon_\rho~, &
\displaystyle 
\epsilon_{\rho \omg}~, &
\displaystyle 
\epsilon_{\rho \phi}  \\[0.5cm]
\displaystyle 
\epsilon_{\rho \omg}~, &
\displaystyle 
\epsilon_\omg~, &
\displaystyle 
\epsilon_{\omg \phi}\\[0.5cm]
\epsilon_{\rho \phi} ~, & \epsilon_{\omg \phi} ~, &\epsilon_\phi
\end{array}
 \right] 
\label{Eq32}
\ee 
and contains only the perturbations generated by kaon loop effects.
The kaon loop transition from a vector meson $V$ to another one $V^\prime$
 has been denoted $\epsilon_{VV^\prime}$.

One should note \cite{taupaper} that $M^2$ is an analytic function of $s$ 
satisfying the (so--called) hermitian analyticity condition~: 
$M^2(s^*)=[M^2(s)]^\dag$.

The  entries of these matrices are appropriately parametrized in terms of~:
\be
\left \{
\begin{array}{lll}
\epsilon_1(s)= ~~\Pi_+(s)-\Pi_0(s) \\[0.5cm]
\epsilon_2(s)= ~~\Pi_+(s)+\Pi_0(s) \\[0.5cm]
\Pi_{\pi \pi}(s) = g_{\rho \pi \pi}^2 \Pi^\prime(s)~~,& \displaystyle 
\left (g_{\rho \pi \pi}=\frac{ag}{2} (1+\Sigma_V)\right )
\end{array}
\right .
\label{Eq33}
\ee
where $\Pi^\prime(s)$ denotes the amputated pion loop, while $\Pi_+(s)$
and $\Pi_0(s)$ denote, respectively, the amputated charged and neutral kaon loops;
their analytic expressions can be found in the Appendices of \cite{taupaper}.
$\epsilon_1(s)$ $\epsilon_2(s)$ do not contain symmetry breaking terms 
beyond the effects of the kaon mass splitting. 
The expressions for the entries in $\delta M^2(s)$ are given in Appendix B
and show this dependence explicitly (see Eqs. (\ref{BB1})). 

One can construct, as in \cite{taupaper}, the eigensolutions of $M^2$. These 
are the final (step two) renormalized vector fields denoted respectively by $\rho_R$, 
$\omg_R$ and $\phi_R$ and are related with their $R_1$ partners by~:
\be
\left (
\begin{array}{lll}
\rho_{R_1}\\[0.5cm]
\omg_{R_1}\\[0.5cm]
\phi_{R_1}
\end{array}
\right ) = 
\left (
\begin{array}{cll}
\displaystyle  \rho_R -\alpha \omg_R+\beta \phi_R\\[0.5cm]
\displaystyle  \omg_R +\alpha \rho_R+\gamma \phi_R\\[0.5cm]
\displaystyle \phi_R -\beta  \rho_R -\gamma \omg_R
\end{array}
\right ) 
\label{eq34}
\ee
where the $s$--dependent mixing angles are defined by~:
\be
\left \{
\begin{array}{ll}
\displaystyle 
\alpha(s)= ~~\frac{\epsilon_{\rho \omg}}{\lambda_\rho -\lambda_\omg}\\[0.5cm]
\displaystyle 
\beta(s)=-\frac{\epsilon_{\rho \phi}}{\lambda_\rho -\lambda_\phi} \\[0.5cm]
\displaystyle 
\gamma(s)=-\frac{\epsilon_{\omg \phi}}{\lambda_\omg -\lambda_\phi}
\end{array}
\right .
\label{eq35}
\ee
using the eigenvalues of $M^2$ (at first order)~:
\be
\begin{array}{lll}
\lambda_\rho(s)=m_\rho^2+\Pi_{\pi \pi}(s) + \epsilon_\rho(s)~~,~
\lambda_\omg(s)=m_\omg^2+ \epsilon_\omg(s)~~,~
 \lambda_\phi(s)= m_\phi^2+ \epsilon_\phi(s)
\end{array}
\label{eq36}
\ee
The $\epsilon_\rho(s)$, $\epsilon_\omg(s)$ and $\epsilon_\phi(s)$ quantities, 
defined in Eqs.(\ref{BB1}), only depend on the kaon loop
functions and on breaking parameters.

\subsection{The Pion Form Factor in $e^+e^-$ Annihilations and in $\tau$ Decays }
\label{pionFF}
\indent \indent 
The pion form factor in the $\tau^\pm$ decay to $\pi^\pm \pi^0 \nu_\tau$ can easily be
derived from the Lagrangian piece ${\cal L}_\tau$ given in Eq. (\ref{AA4})~:
\be
\displaystyle 
F_\pi^\tau(s) = \left[
1-\frac{a}{2}(1+\Sigma_V) \right]- \frac{ag}{2}(1+\Sigma_V)F_\rho^\tau(s) 
\frac{1}{D_\rho(s)}
\label{eq37}
\ee
where~:
\be
\left \{
\begin{array}{lll}
\displaystyle 
F_\rho^\tau(s) =f_\rho^\tau - \Pi_{W}(s)\\[0.5cm]
\displaystyle 
D_\rho(s)=s-m^2_\rho -\Pi_{\rho \rho}^\prime(s)\\[0.5cm]
\displaystyle 
f_\rho^\tau =a g  f_\pi^2 (1+\Sigma_V) ~~~,~~ m^2_\rho=a g^2 f_\pi^2(1+\Sigma_V)
\end{array}
\right .
\label{eq38}
\ee
and the loop functions are~:
\be
\left \{
\begin{array}{lll}
\displaystyle \Pi_{W}(s)=\frac{ag}{2}(1+\Sigma_V)
\left [ (1-\frac{a}{2}(1+\Sigma_V)) \ell_\pi(s) + 
\frac{1}{2 z_A^2}(z_A-\frac{a}{2}(1+\Sigma_V)) \ell_{K}(s) 
\right ]+ P_W(s) \\[0.5cm]
\displaystyle \Pi_{\rho \rho}^\prime(s) = [\frac{ag}{2}(1+\Sigma_V)]^2  \left [ 
\ell_\pi(s) + \frac{1}{2 z_A^2} \ell_{K}(s)~
\right ] + P_\rho(s)
\end{array}
\right .
\label{eq39}
\ee
where $\ell_\pi(s)$ and $ \ell_{K}(s) $ denote respectively the amputated charged
pion and kaon loops, $P_W(s)$ and $P_\rho(s)$ being subtraction polynomials.
In order to fulfill current conservation, these polynomials should
vanish at $s=0$.
Here, as in former studies \cite{taupaper,ExtMod1,ExtMod2}, identifying
$P^\pm P^\mp$ and $P^\pm P^0$ loops has been found numerically justified.

If one compares with the corresponding formulae in \cite{ExtMod2} (Subsection
2.1.1), one sees that $\delta m^2$ and $\delta g$ -- supposed to reflect 
different properties of the charged and neutral $\rho$ mesons -- have been deleted.
As the loop functions  vanish at $s=0$, one clearly has $F_\pi^\tau(0)=1$.

The pion form factor in $e^+e^-$ annihilations is not as simply derived. One needs
first to propagate the transformation in Eq. (\ref{eq34}) into the Lagrangian
Eq. (\ref{AA2}) and collect all contributions to, respectively, $\rho_R$,
$\omg_R$ and $\phi_R$. In this way, the $V-\gamma$ couplings associated
with the fully renormalized vector fields become~:
\be
\left \{
\begin{array}{lll}
\displaystyle 
f_\rho^\gamma=a g  f_\pi^2 (1+\Sigma_V+\frac{h_V \Delta_V}{3}+\frac{\alpha(s)}{3}
+\frac{\sqrt{2}\beta(s)}{3} z_V)\\[0.5cm]
\displaystyle 
f_\omg^\gamma=\frac{a g  f_\pi^2}{3} (1+\Sigma_V+3 (1-h_V) \Delta_V-3\alpha(s)
+\sqrt{2}\gamma(s) z_V)\\[0.5cm]
\displaystyle 
f_\phi^\gamma=\frac{a g  f_\pi^2}{3} (-\sqrt{2}z_V + 3\beta(s)+\gamma(s))
 \end{array}
\right .
\label{eq40}
\ee
including the mixing angle contributions. Using the Lagrangian pieces given
in Eqs.(\ref{CC1}), one can construct easily the pion form factor~:
\be
\displaystyle
F_\pi^e(s) = \left [ 1-\frac{a}{2}(1+\Sigma_V+\frac{h_V \Delta_V}{3})\right] - 
F_{\rho \gamma}^e(s) \frac{g_{\rho \pi \pi}}{D_\rho(s)}
- F_{\omega \gamma}^e(s) \frac{g_{\omega \pi \pi}}{D_\omega(s)}
- F_{\phi \gamma}^e(s) \frac{g_{\phi \pi \pi}}{D_\phi(s)}
\label{eq41}
\ee
where~:
\be
\begin{array}{llll}
\displaystyle g_{\rho \pi \pi} = \frac{a g}{2}(1+\Sigma_V) ~,&
\displaystyle g_{\omega \pi \pi} = \frac{a g}{2} 
[(1-h_V)\Delta_V-\alpha(s)]~,&
\displaystyle g_{\phi \pi \pi} = \frac{a g}{2}  \beta(s)
\displaystyle   
\end{array}
\label{eq42}
\ee

\begin{table}[ph]
\begin{center}
\begin{tabular}{|| c  | c  || c ||}
\hline
\hline
\hhhc
\hhhd ~~~~ & \hhhv $F_\pi^\tau(s)$   &  \hhhv $F_\pi^e(s)$ (I=1) \\
\hline
\hline
Non--Resonant Term \hhhu &  $\displaystyle \left[ 1-\frac{a}{2}(1+\Sigma_V) \right] $ 	
		& $ \displaystyle  \left[1-\frac{a}{2}(1+\Sigma_V+\frac{h_V\Delta_V}{3})\right] $
		\\[0.5cm]
\hline
$\rho$  Mass Squared \hhhbb& $\displaystyle  a g^2 f_\pi^2(1+\Sigma_V) $ 
			&   $\displaystyle  a g^2 f_\pi^2(1+\Sigma_V) $ \\[0.5cm]
\hline
$\pi \pi$ Coupling $\displaystyle g_{\rho \pi \pi}$  \hhhv &  $\displaystyle  \frac{a g}{2}(1+\Sigma_V)$ 
				&  $\displaystyle  \frac{a g}{2}(1+\Sigma_V)$ \\[0.5cm]
\hline
 Amplitudes $f_{\rho}^\gamma$ \& $f_{\rho}^\tau$ 
\hhhq&  $  \displaystyle a g  f_\pi^2 (1+\Sigma_V)$	& 
        $  \displaystyle a g  f_\pi^2 (1+\Sigma_V+\frac{h_V \Delta_V}{3}+\frac{\alpha(s)}{3}
+\frac{\sqrt{2}\beta(s)}{3} z_V) $\\[0.5cm]
\hline
$  \displaystyle \frac{f_{\rho}^\gamma}{f_{\rho}^\tau}$ \hhhq& 
\multicolumn{2}{|c|}{ $  \displaystyle 1 +\frac{h_V \Delta_V}{3}+
\frac{\alpha(s)}{3}+\frac{\sqrt{2}\beta(s)}{3} z_V $ 
}\\[0.5cm]
\hline
Renormalization  factor of  \hhhv & ~~& ~~\\[-0.5cm]
$K \overline{K}$ couplings \hhhv & $\displaystyle \frac{1}{z_A}$
&$\displaystyle \frac{1}{z_A} (1\mp  \frac{\Delta_A}{2})$\\[0.4cm]
\hline
\hline
\end{tabular}
\end{center}
\caption {
\label{T1}
Comparison of the  pion form factor information in  $\tau$ decay and in $e^+e^-$ annihilation.
Second column lists only isospin 1 related information. In the last entry of the rightmost
data column,  the upper sign refers to $K^+K^-$ pairs, the lower to $K^0 \overline{K}^0$.}
\end{table}

The loop corrected $V-\gamma$ transitions amplitudes $F_{V\gamma}^e(s)$ are defined by~:
\be
\displaystyle F_{V\gamma}^e(s)= f_V^\gamma - \Pi_{V\gamma}(s)~~,~~
(V=\rho^0_R,~\omg_R,~\phi_R)~~,
\label{eq43}
\ee
with the $s$--dependent loop terms $\Pi_{V\gamma}(s)$ being defined in  Appendix C.
All $\Pi_{V\gamma}(s)$ are requested to vanish at $s=0$ because of current conservation.
The inverse $\rho$ propagator $D_\rho(s)$ is defined by (see Eq. (\ref{eq36}))~:
\be
D_ {\rho^0}(s)=s -\lambda_\rho(s) =s-m^2_\rho -\Pi_{\rho \rho}(s)\\[0.5cm]
\label{eq44}
\ee
\indent \indent As the $\rho$ self--mass $\Pi_{\rho \rho}(s)$  vanishes at $s=0$,
one certainly has $D_ {\rho^0}(0)=-m^2_\rho$. Concerning the $\omg$ and $\phi$
mesons, one can correspondingly write their inverse propagators as~:
\be
\displaystyle  
D_V(s)=s-m^2_V -\Pi_{V V}(s)~~,~ (m^2_\omg=m^2_\rho,~ m^2_\phi=z_V m^2_\rho)
\label{eq45}
\ee
and one can legitimately assume their self--energies to also vanish at $s=0$.  
Then, $D_ {\omg}(0)=-m^2_\rho$ and $D_ {\phi}(0)=-z_V m^2_\rho$ should certainly
be fulfilled. However (most of) the $\omg$ self--energy cannot be computed in
closed form and the 3--pion part of $\phi$ self--energy too. Therefore, convenient
forms for their propagators should be considered.
This issue is readdressed just below for both mesons.

At this step, it is of concern to compare the properties of the isospin 1 part
of $F_\pi^e(s)$ with $F_\pi^\tau(s)$. The most important pieces of information
are listed in Table \ref{T1}. The difference displayed for the non--resonant
term is tiny. One can see that there is no mass difference between the 
charged and neutral $\rho$ mesons, nor different couplings to a pion pair.
Instead, most of the difference is actually
carried out by the transition amplitudes (see the fifth data line in Table \ref{T1}) 
which are significantly $s$--dependent, as can be inferred from Figures 6 and
7 in \cite{ExtMod1}. 

Finally, it is interesting to note that the renormalization factor
introduced in couplings involving a kaon pair plays in opposite directions 
for charged and neutral kaon pairs.

\subsection{The $\omg \pi \pi$ Direct Coupling and the Condition $F_\pi^e(0)=1$}
\label{omgpipi}
\indent \indent
As can be seen from Eqs. (\ref{eq42}), the  fully broken  HLS model  reveals a total
 coupling of the $\omg$ to a pion pair given by~:

$$\displaystyle g_{\omega \pi \pi} = \frac{a g}{2} 
[(1-h_V)\Delta_V-\alpha(s)]~.$$

This expression illustrates that  the  $\omega \pi \pi$ coupling in our model 
is {\it a priori} a superposition  of a direct isospin breaking term and of
another piece generated by vector meson mixing through kaon loops. This kind of sharing has been
emphasized several times \cite{Maltman1996,Maltman2009}. The full data set we use should give
the most precise and motivated estimate  for these two pieces as this is
still presently controversial   \cite{Maltman1996,Maltman2009,Maltman1997}. 

\vspace{0.5cm}

On the other hand, the parametrization of the $\omg$ contribution to the pion form
factor may pose a conceptual problem related with  the condition $F_\pi^e(0)=1$ which
is worth addressing.

The pseudoscalar meson loops which enter the $V V^\prime$ transition amplitudes
(see Eqs. (\ref{Eq32}), (\ref{eq35}) and (\ref{BB1}))  behave
as ${\cal O}(s)$ near the origin.  The  running vector meson masses (see Eqs. (\ref{eq36}))
are such that $\lambda_\rho(s)-\lambda_\omg(s)$ vanishes at the origin, while
the two other differences which come into Eqs. (\ref{eq36}) tend to a
non--zero constant. Therefore, {\it ab initio}, the mixing angles 
are expected to fulfill~:
\be
\displaystyle 
\begin{array}{llll}
\beta(0)=\gamma(0)=0 ~~,~\alpha(0)  \ne 0
\end{array}
\label{eq46}
\ee

Even if clear in the previous publications using the loop mixing mechanism (Figure 7 
in \cite{taupaper} or  Figure 6 in  \cite{ExtMod1} clearly
show that $\alpha(0) \simeq -5\%$ ), this was not explicitly pointed out. 
Therefore, the $s$--dependent  $\omg \pi \pi$ coupling generated by
loop mixing\footnote{Actually, the non--identically vanishing $\epsilon_1(s)$ 
function providing the vector meson mixing via loops is also generated 
by isospin symmetry breaking, however in the pseudoscalar sector.
} does not vanish at the origin. This has some consequences.

Indeed, using Eqs. (\ref{eq40}) and  (\ref{eq42}), and the vanishing properties of the
functions $\Pi_{V\gamma}(s)$, $\Pi_{V V}(s)$ and $\beta(s)$ at the origin, one gets~:
\be
\displaystyle
F_\pi^e(0) = 1 + \frac{a}{6} 
\left[\frac{\alpha(0)}{3} -(1-h_V) \Delta_V  \right]
\left[1+ \frac{m_\rho^2}{D_\omg(0)}\right]
\label{eq47}
\ee
when keeping only the first--order terms in breaking parameters.
 
 As already discussed at the end of the previous Subsection, it
 is motivated to  assume the $\omg$ self--energy $\Pi_{\omg\omg}(s)$ vanishing 
 at the origin. Moreover, this allows to stay consistent with the  so--called 
  "Node theorem"  \cite{RhoOmg0,Klingl}.
  Then, the inverse propagator $D_\omg(s)=s-m_\rho^2 -\Pi_{\omg\omg}(s)$ 
 fulfills $D_\omg(0)=-m_\rho^2$. This provides the vanishing of the last bracket
in the formula above and, thus,   $F_\pi^e(0)=1$, 
 whatever the values for $h_V$, $\Delta_V$ and $\alpha(0)$.

However, in most applications, for objects carrying such a narrow width 
as the $\omg$ and $\phi$ mesons, one generally uses approximate inverse propagators,
e.g. either\footnote{Within the 
ongoing discussion, phenomenological  values --e.g. not derived
from the broken HLS model parameters values--
for vector mesons masses and widths, are denoted with a tilde symbol
in order to avoid confusion.}~:
 
 $$ D_V(s)=s- \tilde{m}_V^2 +i \tilde{m}_V  \tilde{\Gamma}_V~~{\rm (BW_a)~~~  or}
 ~~~ D_V(s)=s- \tilde{m}_V^2 +i\sqrt{s} \tilde{\Gamma}_V~~{\rm (BW_b)}.$$
  with values for $\tilde{m}_V$ and $\tilde{\Gamma}_V$ either taken from the 
  Review of Particle Properties or extracted from  one's fits.
 Then, with either of these Breit--Wigner lineshapes, the condition $F_\pi^e(0)=1$
 is not necessarily fulfilled. From our model results, this condition is even violated  
 at a few  percent level. However, it is easy to check that either of~:
   $$ D_\omg(s)=s- m_\rho^2 - \frac{s}{m_\rho^2}(\tilde{m}_\omg^2-m_\rho^2-i \tilde{m}_\omg \tilde{\Gamma}_\omg)~~
  ~~{\rm (BW _a}^\prime)$$
(remember that $m_\rho^2=m_\omg^2$) and~:
  $$ D_\omg(s)=s- m_\rho^2 - \frac{s}{m_\rho^2}(\tilde{m}_\omg^2-m_\rho^2- i \sqrt{s} \tilde{\Gamma}_\omg)
  ~~{\rm (BW _b}^\prime)$$
 \noindent certainly cures this disease. This turns out to parametrize the
$\omg$ self--energy $\Pi_{\omg\omg}(s)$ with an ansatz which satisfies
its vanishing at the origin.

It is worth stressing that using standard
Breit--Wigner lineshapes or their modified partners provides  practically unchanged
fit results. This is due to the fact that the $\rho$ and $\omg$ masses (with tilde or not)
are close to each other, and then, the factor $s/m_\rho^2$ is very well approximated by
 1 all along the sensitive region of the  $\omg$ peak.
 
In order to substantiate the possible changes, we have run our code using 
${\rm BW_a}$ and ${\rm BW _a}^\prime$ as inverse $\omg$ propagators.  As 
a typical  example of modification, one can compare
$\tilde{m}_\omg=782.44 \pm 0.06$ MeV and $\tilde{\Gamma}_\omg=8.46 \pm 0.09 $ MeV
while fitting with ${\rm BW_a}$, and 
$\tilde{m}_\omg=782.49 \pm 0.06$ MeV and $\tilde{\Gamma}_\omg=8.36 \pm 0.08 $ MeV
when using instead ${\rm BW _a}^\prime$.
For definiteness, in the fits presented in this paper, $ D_\omg(s)$
will be modified as just explained. As $\beta(0)=0$, the pion form factor
value at $s=0$ is not sensistive to how the $\phi$ propagator is approximated.

Even if our choice is motivated, others are certainly possible as exemplified  in 
\cite{DavierHoecker,Maltman2009}. Transposed to our model, the just mentioned choice 
would turn out to weight the full $\omg$
contribution to the pion form factor by a  factor $s/\tilde{m}_\omg^2$ or
$s/m_\rho^2$ which restores $F_\pi^e(0)=1$. The behavior of this choice
is identical to ours, basically  because  $\tilde{m}_\omg$ and $m_\rho$ 
are very close to each other.
 
\subsection{The Charged and Neutral Kaon Form Factors}
\indent \indent We give here the annihilation cross sections/form factors
within the extended BKY--BOC breaking of the HLS Lagrangian.
Cross sections and form factors are related through~:
\be 
\displaystyle
\sigma(e^+e^- \ra P \overline{P})=\frac{8 \pi \alpha_{em}^2}{3 s^{5/2}} q_P^3 |F^e_P(s)|^2
\label{eq48}
\ee
for any meson pair $P \overline{P}$. $q_P=\sqrt{s-4 m_P^2}/2$ is the $P$
momentum in the center--of--mass system. The kaon form factors are given by~:
\be 
\hspace{-2.3cm}
\left \{
\begin{array}{ll}
\displaystyle
F^e_{K_c}(s)= \left[1 -\frac{a}{6z_A} (2+z_V+2\Sigma_V+2 \Delta_V-\frac{\Delta_A}{2}(2+z_V) \right]
-\frac{g_{\rho K^+ K^-} F_{\rho\gamma}(s)}{D_\rho(s)}
-\frac{g_{\omg K^+ K^-} F_{\omg\gamma}(s)}{D_\omg(s)}
-\frac{g_{\phi K^+ K^-} F_{\phi\gamma}(s)}{D_\phi(s)}\\[0.5cm]
\displaystyle
F^e_{K_0}(s)= -\left[\frac{a}{6z_A} (1-z_V+\Sigma_V- \Delta_V+\frac{\Delta_A}{2}(1-z_V))\right]
-\frac{g_{\rho K^0  \overline{K}^0} F_{\rho\gamma}(s)}{D_\rho(s)}
-\frac{g_{\omg K^0  \overline{K}^0} F_{\omg\gamma}(s)}{D_\omg(s)}
-\frac{g_{\phi K^0  \overline{K}^0} F_{\phi\gamma}(s)}{D_\phi(s)}
\end{array}
\right.
\label{eq49}
\ee
where the $\gamma - V$ transition amplitudes $F_{V\gamma}$ have been already defined
(see Eq. (\ref{eq43})). The $VK\overline{K}$ couplings can be read off from the corresponding
Lagrangian pieces (Eqs. (\ref{CC2}) and (\ref{CC3})).

The kaon form factors  fulfill~:
\be 
\begin{array}{ll}
\displaystyle F^e_{K_c}(0)=1~~~,&~\displaystyle F^e_{K_0}(0)=0
\end{array}
\label{eq50}
\ee

However, it is easy to check that these conditions are both fulfilled, only if~:
\be 
\displaystyle \frac{m_\rho^2}{D_\omg(0)} =\frac{z_V m_\rho^2}{D_\phi(0)}=-1
\label{eq51}
\ee

Therefore a fixed width Breit--Wigner shape  for the  $\phi$  should be adapted  
as already discussed for the $\omg$.

\subsection{Parametrization of the $\phi$ Propagator}
\indent \indent
As for the pion form factor, in order to fulfill
$F^e_{K_c}(0)=1$ and $F^e_{K_0}(0)=0$, one should impose that
the $\omg$ and $\phi$ inverse propagators at  $s=0$ are equal in magnitude and opposite
in sign to their respective Lagrangian masses 
($m_\phi^2=z_V m_\omg^2=z_V m_\rho^2$). Here again, this turns out to
parametrize the full self--energy $\Pi_{\phi\phi}(s)$ by an ansatz 
vanishing at $s=0$. For the two--body loops, this is well known \cite{taupaper}; 
however, the three--body loop  is not known in closed form (as for $\Pi_{\omg\omg}(s)$).

However, in contrast with the case for $\omg$, using~:
\be 
\displaystyle D_\phi(s)=s- z_V m_\rho^2 - \frac{s}{z_V m_\rho^2} (\tilde{m}_\phi^2-z_V m_\rho^2-i 
\tilde{m}_\phi\tilde{\Gamma_\phi})~~  ~~{\rm (BW _a}^\prime)
\label{eq52}
\ee
for the $\phi$ inverse propagator, instead  of the usual (fixed width) ${\rm BW _a}$ form, should
be further commented, as $\sqrt{z_V} m_\rho$ significantly differs from $\tilde{m}_\phi$
normally fitted\footnote{More substantially, with appropriate fits,
one yields $\sqrt{z_V} m_\rho \simeq
925$ MeV, while a direct fit yields $\tilde{m}_\phi \simeq 1020$ MeV~!} 
(e.g. with ${\rm BW _a}$).

Even  if anticipating on our fit results, it is worth discussing this matter right
now. As far as cross sections are concerned, the two kinds of fits provide
almost identical results. In order to yield this result, almost all parameters
vary within errors except for\footnote{
In fits with $\rm BW _a$ for the $\phi$ meson, one gets $z_V=1.368\pm 0.005$,
while with ${\rm BW _a}^\prime$ the fit returns $z_V=1.472\pm 0.001$.
} $z_V$, which could have been expected. However, it will be shown that 
this change has a marginal influence on all information of physics importance.
Anyway, such kind of information is interesting as it provides a hint
on the model dependence of numerical results. Therefore,  it has been
of concern  to compare results obtained with either of  
${\rm BW _a}$ and ${\rm BW _a}^\prime$, when appropriate.
 
Before closing this Section, one may note that, at the   $\phi$ peak location
($\sqrt{s} \simeq 1020$ Mev), the modified Breit--Wigner lineshape provides~:
$$ z_V m_\rho^2 + \frac{s}{z_V m_\rho^2}(\tilde{m}_\phi^2-z_V m_\rho^2)
\simeq [(1.020)~{\rm MeV}]^2$$
which explains why the fit remains successful  when using  ${\rm BW _a}^\prime$
 and also why $z_V$ should change correspondingly, taking into account that $m_\rho^2$
 cannot much vary. The fit quality of the $e^+e^- \ra K \overline{K}$ cross sections will illustrate
 the validity of this parametrization of the $\phi$ propagator.
 
\subsection{The Coulomb Interaction Factor }
\indent \indent
Beyond modelling, there is an important issue  to discuss 
when dealing with the charged kaon form factor. In the decay
$\phi \ra K^+ K^-$, and more generally as close to the $K \overline{K}$ threshold, 
one has  to take into account the Coulomb interaction among
the emerging charged kaons. This has been first addressed in \cite{Gourdin}
and recently readdressed (and corrected) in \cite{BGPter}.
The net result of this effect is to multiply 
the charged kaon cross section by the Coulomb factor\footnote{
Actually, the full electromagnetic correction factor is more complicated,
but the main effect comes from the Coulomb factor. One assumes that the kaon data
which have been submitted to fit have been appropriately corrected
for soft photon corrections, which allows to cancel out the term
named $C_i$ in \cite{BGPter}.}~:
\be 
\begin{array}{ll}
\displaystyle
Z(s)=\left[1+\pi \alpha_{em} \frac{1+v^2}{2v} \right]~~~,~~& 
\displaystyle v=\frac{\sqrt{s-4 m_{K^\pm}^2}}{s}
\end{array}
\label{eq53}
\ee

In \cite{SNDKKb}, and later in \cite{CMD2KKb}, the cross section for charged kaons is multiplied
by $Z(s)/Z(m_\phi^2)$. This turns out to consider the Coulomb interaction as
a breaking mechanism which affects the charged kaon sector and not the neutral one;
as the corresponding $\phi $ branching fractions are fit independently, this should not
affect their results. One may just have to remark that  this turns out to incorporate the 
Coulomb effects inside the corresponding estimates for the $\phi \ra K^+ K^- $ branching
fraction. 
\subsection{About the $\phi \ra K^+ K^- / \phi \ra K^0 \overline{K}^0$ Ratio}
\indent \indent Up to well defined phase space factors generated by the kaon mass
splitting, the partial width ratio $\phi \ra K^+ K^- / \phi \ra K^0 \overline{K}^0$
is the square of the corresponding $s$-dependent effective coupling ratio. Neglecting for each coupling 
corrections terms of order greater than 1, one can derive from Eqs. (\ref{CC2}) and 
(\ref{CC3})~:
\be 
\displaystyle
\frac{g_{\phi K^+ K^-}} {g_{\phi K^0 \overline{K}^0}} =-
\frac{\sqrt{2} z_V -\beta(s) - \gamma(s)}
{\sqrt{2} z_V +\beta(s) - \gamma(s)} \left[ 1-\Delta_A  
\right] \simeq - \left[1-\Delta_A  \right]
\label{eq54}
\ee
where the last equation follows from remarking (see Figure 7 in \cite{ExtMod1})
that the mixing angle $\beta(s)$ -- defined by Eq. (\ref{eq35}) --
is negligibly small compared to $\sqrt{2} z_V$ in the $\phi$ 
mass region. Therefore, this mechanism proposes a way for this ratio to 
depart from unity. 

In their throughout study  of the $\phi \ra K^+ K^- / \phi \ra K^0 \overline{K}^0$ 
ratio, the authors of \cite{BGPter} examined this issue using several other
mechanisms than this one  and concluded
that none of them was able to accomodate a coupling constant ratio smaller than one
(in absolute magnitude). 
The global fit, based on the suitably broken HLS model, provides a new approach.
In this framework,  the determination of $\Delta_A$ is constrained 
 by both $e^+e^- \ra K \overline{K}$ annihilation 
cross sections {\it separately}, and by some more light meson anomalous decays, 
which also depend on  $\Delta_A$.

\section{The HLS Anomalous Sector}
\label{brkFKTUY}
\indent \indent In order to treat radiative decays, {\it i.e.} the $VP\gamma$ couplings,
and some important annihilation channels (namely  $e^+e^- \ra \pi^0 \gamma$, 
$e^+e^- \ra \eta \gamma$ and $e^+e^- \ra \pi^0\pi^+\pi^- $) within the HLS framework,
one needs to incorporate the appropriate Lagrangian pieces. 
These are given by the Wess--Zumino--Witten (WZW) terms \cite{WZ,Witten} which
traditionally account for the triangle ($AAP$) and box ($APPP$) anomalies, together with
the FKTUY Lagrangian pieces \cite{FKTUY,HLSRef}~:
\be 
\displaystyle
{\cal L}_{anom.} = {\cal L}_{WZW} +\sum_{i=1}^4 c_i{\cal L}_i
\label{eq55}
\ee
where the four $c_i$ are constants left unconstrained by theory \cite{FKTUY}.
A closer examination of the FKTUY Lagrangian allows 
to identify five different pieces -- listed in Appendix \ref{DD} -- and one
then remarks that the accessible physics is sensitive to the difference $c_1-c_2$
and not to each of them separately. One is then left {\it a priori} with
three unconstrained parameters \cite{HLSRef}. 

\vspace{0.5cm}

When no breaking is at work, the amplitudes for
the couplings\footnote{Here and in the following $P_0$
denotes either of the $\pi^0$, $\eta$ and $\eta^\prime$ mesons.}
 $P_0 \gamma \gamma$ and $P_0 \pi^+ \pi^- \gamma$  at the chiral point --
computed within the FKTUY--HLS framework\footnote{e.g. using 
Eq. (\ref{eq55}) $and$ the $V-\gamma$ transitions
provided by the non--anomalous HLS Lagrangian.}--
coincide with those directly
derived from the WZW piece in isolation \cite{ExtMod1}.  
Due to a sign error\footnote{We gratefully ackowledge B. Kubis   (HISKP, 
 Bonn University) for having kindly pointed out the issue.} 
in the FKTUY Lagrangian piece ${\cal L}_{AVP}$,
it was asserted in \cite{ExtMod1}  that the constraint $c_3=c_4$ was mandatory 
in order to recover this property. Actually, this property is automatically 
 satisfied \cite{HLSOrigin,HLSRef}. In addition, we have verified that
 this property is maintained within our fully broken HLS model.

However,  the condition $c_3=c_4$, which is fulfilled by VMD models \cite{HLSRef} is
 successful and only turns out to reduce the freedom in fits. Nevertheless,
 one has examined relaxing this condition and found that our fit results 
 are well compatible with the constraint $c_3=c_4$.
  
\subsection{Breaking the Anomalous HLS Lagrangian}
\indent \indent At this step, the anomalous HLS Lagrangian can be written~~:
\be 
\displaystyle
{\cal L}_{anom.} = {\cal L}_{AAP} +{\cal L}_{APPP}+{\cal L}_{VVP}+{\cal L}_{VPPP}
\label{eq56}
\ee
with pieces listed in Appendix \ref{DD}. As for the non--anomalous HLS Lagrangian,
each among these pieces may undergo specific symmetry breaking independently of 
each other. This may lead to plenty of free parameters as illustrated 
by M. Hashimoto \cite{Hashimoto} who implemented combined SU(3) and Isospin symmetry 
breakings in the anomalous sector.

A simpler mechanism has also been proposed for SU(3) breaking by Bramon, Grau and Pancheri
\cite{BGP,BGPbis}; however, this was insufficient to account for both 
$K^{*[\pm ,0]} \ra K^{[\pm,0]} \gamma$ decay widths. In \cite{rad,taupaper}
it was proposed to supplement it with a breaking of the vector field
matrix resembling a vector field redefinition. Quite unexpectedly, this
provides  a (successful) parametrization for the $K^*$ radiative partial widths
identical to those proposed by G. Morpurgo \cite{Morpurgo} within a completely different context.
Interestingly, this combined mechanism leaves totally unaffected the other sectors of the 
${\cal L}_{VVP}$ piece we deal with; this is well accepted by all data considered \cite{rad,taupaper}. 
This combined breaking mechanism has been studied
in detail  \cite{ExtMod1} for all pieces of ${\cal L}_{anom.}$ with similar
conclusions.

The  combined breaking mechanism, as presented in \cite{ExtMod1}, has been examined 
by combining SU(3) and 
Isospin symmetry breakings using the complete data set discussed below within
the minimization code underlying 
the present study. It was concluded that possible  Isospin symmetry breaking
effects -- not propagated from the field redefinitions provided by non--anomalous
HLS Lagrangian breaking -- provide invisible effects. It was then decided
to neglect this additional possible source of Isospin symmetry breaking, as
the parameter freedom it gives is found useless.

Therefore, for sake of clarity, one only quotes the specific forms for the decay 
amplitudes $K^{*[\pm ,0]} \ra K^{[\pm,0]} \gamma$, referring  the interested reader
to \cite{ExtMod1} for more information. 

As a summary, our dealing with the anomalous sector -- except for the limited
$K^*$ sector -- involves only 3 parameters~: $c_1-c_2$ and $c_3$ and $c_4$~; 
former studies \cite{ExtMod1,ExtMod2,box} remain valid, as the condition 
$c_4-c_3=0$ is well accepted by the data, as will be shown shortly.

\vspace{-0.5cm}

\subsection{Radiative Couplings}
\indent \indent 
For what concerns the radiative decays of light mesons and the 
$e^+e^- \ra P \gamma$ annihilation processes, one needs 
${\cal L}_{AAP}$ and an effective  piece named ${\cal L}_{AVP}^\prime$ defined below. 
 
In terms of the final renormalized pseudoscalar fields and assuming the 
$\pi^0-\eta -\eta^\prime$ mixing  defined in Section \ref{brkLa}, one can write~:
\be
\displaystyle
{\cal L}_{AAP}=-\frac{3 \alpha_{em}}{ \pi f_\pi} (1-c_4)\epsilon^{\alpha \beta \mu \nu}
\partial_\alpha A_\beta \partial_\mu A_\nu \left[
g_{\pi^0 \gamma \gamma} \frac{\pi^0}{6} + g_{\eta \gamma \gamma} \frac{\eta}{2\sqrt{3}}
 + g_{\eta^\prime \gamma \gamma} \frac{\eta^\prime}{2\sqrt{3}}
\right]
\label{eq57}
\ee
At leading order in breaking parameters, the coefficients $g_{P_0 \gamma \gamma}$
are given by\footnote{One could expand the $(1+v)^{-1}$ factor 
and keep only the contributions of orders 1 and $v$. However, in the present case, it does not
simplify the expressions.}~:
\be
\hspace{-2.0cm}
\left \{
\begin{array}{ll}
g_{\pi^0 \gamma \gamma}=  \displaystyle 1-\frac{5 \Delta_A}{6} +
 \frac{\epsilon}{\sqrt{3}}
\left \{\frac{5z_A-2}{3z_A}\cos{\theta_P}-\sqrt{2}\frac{5z_A+1}{3z_A}\sin{\theta_P}
\right\}+ \frac{\epsilon^\prime}{\sqrt{3}}\left \{\frac{5z_A-2}{3z_A}\sin{\theta_P}+
\sqrt{2}\frac{5z_A+1}{3z_A}\cos{\theta_P}\right\} \\[0.5cm]
g_{\eta \gamma \gamma} =  \displaystyle \frac{\cos{\theta_P}}{3} 
\left \{\frac{5z_A-2}{3z_A(1+v)}+v \frac{1+2z_A}{1+2z_A^2}-\frac{\Delta_A}{2}
\right\}-\sqrt{2} \frac{\sin{\theta_P}}{3} \left \{
\frac{5z_A+1}{3z_A(1+v)}+v \frac{1-z_A}{1+2z_A^2} -\frac{\Delta_A}{2}\right\}
-\frac{\epsilon}{\sqrt{3}}\\[0.5cm]
g_{\eta^\prime \gamma \gamma} =  \displaystyle \frac{\sin{\theta_P}}{3}
\left \{\frac{5z_A-2}{3z_A(1+v)}+v \frac{1+2z_A}{1+2z_A^2}
-\frac{\Delta_A}{2}
\right\} +\sqrt{2} \frac{\cos{\theta_P}}{3} \left \{
\frac{5z_A+1}{3z_A(1+v)}+v \frac{1-z_A}{1+2z_A^2}-\frac{\Delta_A}{2}\right\}-
\frac{\epsilon^\prime}{\sqrt{3}}
\end{array}
\right.
\label{eq58}
\ee
These clearly depend on the breaking parameters $\Delta_A$, $z_A$ and $v$ (the PS nonet symmetry breaking)
and on the
$\pi^0-\eta -\eta^\prime$ mixing scheme (see Eqs. (\ref{eq24})), especially 
on the singlet--octet mixing angle $\theta_P$. One should note that
$f_\pi/g_{\pi^0 \gamma \gamma}$ is another way to define the neutral pion decay constant.
The other equations also illustrate that the so--called octet and singlet decay constants
as derivable from there have little to do with the standardly defined ones, {\it i.e.}
from the currents in Eqs. (\ref{eq25}). This question has raised some confusion
which motivated the study in \cite{WZWChPT}.

In order to treat the $e^+e^- \ra \pi^0 \pi^+ \pi^-$ annihilation process  the part of the  
${\cal L}_{APPP}$ Lagrangian describing  the so--called box anomalies  is needed. 
This can be written~:
\be
\left \{
\begin{array}{ll}
\displaystyle
{\cal L}_{APPP}=  -iE \epsilon^{\mu \nu \alpha \beta} A_\mu 
\left[ g_{\pi^0 \pi^+\pi^-\gam} \partial_\nu \pi^0 +
g_{\eta \pi^+\pi^-\gam} \partial_\nu \eta +
g_{\eta^\prime \pi^+\pi^-\gam} \partial_\nu \eta^\prime
\right] \partial_\alpha  \pi^- \partial_\beta  \pi^+\\[0.5cm]
\displaystyle
E=-\frac{e }{ \pi^2 f_\pi^3} \left[ 1- \frac{3}{4}(c_1-c_2+c_4)
\right]
\end{array}
\right.
\label{eq59}
\ee
with~:
\be
\hspace{-1cm}
\left \{
\begin{array}{ll}
\displaystyle g_{\pi^0 \pi^+\pi^-\gam} =\frac{1}{4}
\left[ 1-\frac{\Delta_A}{2} + \frac{\cos{\theta_P}}{\sqrt{3}}
\left \{\epsilon+\sqrt{2}\epsilon^\prime\right \}
 -\frac{\sin{\theta_P}}{\sqrt{3}}
\left \{\sqrt{2}\epsilon-\epsilon^\prime\right \} \right]\\[0.5cm]
\displaystyle g_{\eta \pi^+\pi^-\gam} =\frac{\sqrt{3}}{12}
\left[\left \{  1+ 2 v z_A \frac{1-z_A}{2z_A^2+1} -\frac{\Delta_A}{2}\right \}\cos{\theta_P}
-\left \{ 1- v z_A\frac{2z_A+1}{2z_A^2+1} - \frac{\Delta_A}{2}\right \}
\sqrt{2}\sin{\theta_P}
\right] - \frac{\epsilon }{4}\\[0.5cm]
\displaystyle g_{\eta^\prime \pi^+\pi^-\gam} =\frac{\sqrt{3}}{12}\left[
\left \{ 1- v z_A\frac{2z_A+1}{2z_A^2+1} - \frac{\Delta_A}{2}\right \}
\sqrt{2}\cos{\theta_P}
+\left \{  1+ 2 v z_A \frac{1-z_A}{2z_A^2+1} -\frac{\Delta_A}{2}\right \}
\sin{\theta_P}
\right]- \frac{\epsilon^\prime }{4}
\end{array}
\right.
\label{eq60}
\ee

Eqs. (\ref{eq58}) and (\ref{eq60}) show how the triangle and box anomaly
amplitudes behave under isospin, SU(3) and PS nonet symmetry breakings. One should especially
note the intricacy of SU(3) and PS nonet symmetry breakings.

In order to derive the radiative decay couplings, an effective Lagrangian has been built 
up from
${\cal L}_{VVP}$  and the non--anomalous Lagrangian in the same way as in \cite{ExtMod1}. 
This can be written in terms of the renormalized $R_1$ fields~:
 \be
 \hspace{-1.5cm}
\left \{
\begin{array}{ll}
\displaystyle 
{\cal L}_{AVP}^\prime=G \epsilon^{\mu \nu \alpha \beta} F_{\mu \nu} \partial_\alpha A_\beta 
{\rm ~~with~~}  G=-\frac{ e g }{4 \pi^2 f_\pi} \frac{c_4+c_3}{2}\\[0.5cm]
\displaystyle 
GF_{\mu \nu}=\displaystyle \sum_{P=\pi^0,~\eta,~\eta^\prime} P \left [
g_{P \rho \gamma} ~\partial_\mu \rho_\nu^{R_1} + 
g_{P \omg \gamma} ~\partial_\mu \omg_\nu^{R_1} + 
g_{P \phi \gamma} ~\partial_\mu \phi_\nu^{R_1}\right] +
g_{\pi^\pm \rho^\mp \gamma} 
\left [ \pi^+ ~\partial_\mu \rho_\nu^- +
\pi^- ~\partial_\mu \rho_\nu^{+}\right]
\end{array}
\right.
\label{eq61}
\ee
The expression for the various coupling constants $g_{V P \gamma}$ can be 
found in Appendix \ref{EE}. In order
to derive the physical couplings, one should first apply the transformation given in Eq.(\ref{eq34})
and then collect the various contributions to each of the (neutral) $\rho_{R}$, $\omg_{R}$ and $\phi_{R}$.

Concerning the $AVP$ couplings, it is quite interesting to compare the expressions in Eqs.
(\ref{EE3})--(\ref{EE5}) with the corresponding ones in \cite{taupaper,WZWChPT}, derived using 
an  approximate expression for nonet symmetry breaking\footnote{In order to restore the condition
$c_3 \ne c_4$, one should simply make in  \cite{taupaper}
the replacement $c_3 \rightarrow (c_3+c_4)/2$.}
(the $x$ parameter in the quoted papers). Indeed, the three variants
by which nonet breaking occurs (see Eqs.(\ref{EE2}))
are close together and can reasonably well approximated by $x_{eff}=1-v\simeq 1-\lambda/2$.

\vspace{-0.5cm}

\subsection{Breaking the $VVP$ and $VPPP$ Anomalous Lagrangians}
\indent \indent The $VPPP$ anomalous Lagrangian is given by~:
 \be
\left \{
\begin{array}{ll}
\displaystyle 
{\cal L}_{VPPP}=-iD\epsilon^{\mu \nu \alpha \beta}  \left\{
\left [ g_{\rho \pi}^0 \partial_\nu \pi^0+g_{\rho \eta}^0 \partial_\nu \eta+
+g_{\rho \eta^\prime}^0 \partial_\nu \eta^\prime \right ] ~\rho^{R_1}_\mu\right.\\[0.5cm]
\displaystyle 
~~~+\left [ g_{\omg \pi}^0 \partial_\nu \pi^0+g_{\omg \eta}^0 \partial_\nu \eta+
+g_{\omg \eta^\prime}^0 \partial_\nu \eta^\prime \right ]~\omega_\mu^{R_1}
+ g_{\phi \pi}^0\partial_\nu \pi^0 ~\phi_\mu^{R_1} 
 \left\} \partial_\alpha\pi^-  \partial_\beta \pi^+ \right.\\[0.5cm]
\displaystyle 
{\rm ~~with~~} D=-\frac{ 3 g (c_1-c_2-c_3)}{4 \pi^2 f_\pi^3}
\end{array}
\right.
\label{eq62}
\ee
where one has limited oneself to display the $V P_0\pi^+\pi^- $ sector. The leading terms
of the couplings occuring in this expression are given in Appendix \ref{FF}.

\vspace{0.5cm}
\indent \indent The ${\cal L}_{VVP}$  Lagrangian piece plays an important role in
the annihilation process $e^+e^- \ra \pi^0 \pi^+ \pi^-$. Its relevant part 
is ~:
\be
\hspace{-1.cm}
\begin{array}{ll}
\displaystyle  
{\cal L}_{VVP}=&
\displaystyle  
\frac{C}{2} \epsilon^{\mu \nu \alpha \beta} \left \{
\left[  \partial_\mu \omg_\nu^{R_1} - (1-h_V) \Delta_V \partial_\mu \rho^{R_1}_\nu \right] 
\left[ \partial_\alpha \rho^+_\beta \pi^- +\partial_\alpha \rho^-_\beta \pi^+\right] +
+\partial_\mu \rho^{R_1}_\nu \partial_\alpha \omg^{R_1}_\beta \pi^0 \right . \\[0.5cm]
~~& \displaystyle \hspace{-1.cm}
\left.  
+  \left[\widetilde{g}_{\omg \pi^0}~
\partial_\mu \omg^{R_1}_\nu \partial_\alpha \omg^{R_1}_\beta + \widetilde{g}_{\rho \pi^0}~
 \partial_\mu \rho^{R_1}_\nu\partial_\alpha \rho^{R_1}_\beta+ \widetilde{g}_{\Phi\pi^0 }~
 \partial_\mu \Phi^{R_1}_\nu  \partial_\alpha \Phi^{R_1}_\beta 
\right] \pi^0  \right \}~~,~(C=-\frac{N_c g^2 c_3}{4 \pi^2 f_\pi})
\end{array}
\label{eq63}
\ee
where~:
\be
\hspace{-1.5cm}
 \left \{
\begin{array}{ll}
\displaystyle  
\widetilde{g}_{\omg \pi^0 }= -  \frac{\Delta_A}{4}    -h_V \Delta_V+
\epsilon ~\frac{\cos{\theta_P} -\sqrt{2}\sin{\theta_P}}{2\sqrt{3}}
+\epsilon^\prime ~\frac{\sqrt{2} \cos{\theta_P} +\sin{\theta_P}}{2\sqrt{3}}
\\[0.5cm]
\displaystyle  
\widetilde{g}_{\rho \pi^0}=-  \frac{\Delta_A}{4}   -(1-h_V )\Delta_V
+\epsilon ~\frac{\cos{\theta_P} -\sqrt{2}\sin{\theta_P}}{2\sqrt{3}}
+\epsilon^\prime ~\frac{\sqrt{2} \cos{\theta_P} +\sin{\theta_P}}{2\sqrt{3}}
\\[0.5cm]
\displaystyle  
\widetilde{g}_{\Phi \pi^0}=
-\epsilon ~\frac{\sqrt{2} \cos{\theta_P} +\sin{\theta_P}}{z_A\sqrt{6}}
+\epsilon_2 ~\frac{\cos{\theta_P} -\sqrt{2}\sin{\theta_P}}{z_A\sqrt{6}}
\end{array}
\right.
\label{eq64}
\ee
When going from $R_1$--renormalized  to the fully renormalized vector fields $R$, one has to take some
care with  attributing the $s$--dependence between the two neutral fields of each
monomial in the second line of Eq. (\ref{eq63}).  This should be tracked for each $R_1$ field
while applying Eq. (\ref{eq34}).


\subsection{The $e^+e^-\ra P_0 \gamma $ Annihilation Cross Sections}
\indent \indent Using the Lagrangian pieces given above, the transition amplitudes 
$\gamma^* \ra P \gamma$ can be written similarly to \cite{ExtMod1}~:
\be
T(\gamma^*  \ra P_0 \gamma)= i Y \left [ g~\frac{c_3+c_4}{2} K_{P_0}(s) - (1-c_4) L_{P_0} \right]
\epsilon^{\mu \nu \alpha \beta}
k_\mu \varepsilon_\nu(q) p_\alpha \varepsilon_\beta(p)~~~,~~ P_0=\pi^0,~\eta,~\eta^\prime
\label{eq65}
\ee
where $Y=-\alpha_{em} N_c/\pi f_\pi$ has been factored out. 
$q$ is the incoming photon momentum ($q^2=s$),
$p$ the outgoing photon momentum ($p^2=0$) and $N_c=3$. 
The pieces provided by ${\cal L}_{AAP}$ are\footnote{
The corresponding expressions  given in \cite{ExtMod1} carry
a missprint~: Each of the right--hand sides of Eqs. (41)  is missing a
factor of 2.}~:
\be 
\begin{array}{lll}
\displaystyle
L_{\pi^0}=\frac{g_{\pi^0 \gamma \gamma}}{3}~~~,~~
&\displaystyle
L_\eta=\frac{g_{\eta \gamma \gamma}}{\sqrt{3}}~~~,~~
&\displaystyle
L_\eta^\prime=\frac{g_{\eta^\prime \gamma \gamma}}{\sqrt{3}}
\end{array}
\label{eq66}
\ee
using the $g_{P_0 \gamma \gamma}$ couplings defined in Eqs. (\ref{eq58}),
where the  $(1-c_4)$ has been factored out.
The resonance contributions are gathered in $K_{P_0}(s)$~:
\be
K_{P_0}(s) =  \displaystyle \sum_{V_i= \rho^R, \omg^R,\phi^R}
\frac{H_{V_i}^{P_0}(s)  F_{V_i^R \gamma}(s)}{D_{V_i}(s)}~~~~,~~ P_0=\pi^0,~\eta,~\eta^\prime
\label{eq67}
\ee
where the $H_{V_i}^{P_0}(s)$ -- given in Appendix \ref{GG} -- are the resonance couplings to
$P_0 \gamma$ and the
$F_{V_i^R \gamma}(s)$ are the $V-\gamma$ transition amplitudes defined in Eq. (\ref{eq43}).
The $D_{V_i}(s)$ are the vector mesons inverse propagators already encountered. 
The cross sections can then be written~:
\be 
\left \{
\begin{array}{ll}
\displaystyle
\sigma(e^+e^- \ra P_0 \gamma)=
\frac{3 \pi \alpha_{em}^3}{8 \pi^2 f_\pi^2} 
\left[ \frac{s-m_{P_0}^2}{s} \right]^3|F^e_{P_0 \gamma}(s)|^2 \\[0.5cm]
\displaystyle F^e_{P_0 \gamma}(s)=g \frac{c_3+c_4}{2} K_{P_0}(s)-(1-c_4)L_{P_0}
\end{array}
\right.
\label{eq68}
\ee
\subsection{The $e^+e^-\ra \pi^0 \pi^+ \pi^- $ Annihilation Cross Section}
\indent \indent Following as closely as before the notations in \cite{ExtMod1}, the
amplitude for the $\gamma^* \ra \pi^+ \pi^- \pi^0$  is given by~:
 \be
T (\gamma^* \ra \pi^+ \pi^- \pi^0)= \displaystyle 
[T_{sym}(s) + T_{brk}(s) +T_{AVP}(s)] ~~\epsilon^{\mu \nu \alpha \beta}
\varepsilon_\mu(q) p^0_\nu p^+_\alpha  p^-_\beta
 \label{eq69}
\ee
where $\varepsilon_\mu(q)$ ($q^2=s$) is the (heavy) photon 
polarization vector. $T_{sym}$ is the symmetric part of the amplitude
(in terms of the $\rho \pi$ 'final' states), while $T_{brk}$ (denoted $T_\rho$  
in \cite{ExtMod1}) breaks this symmetry. We have found appropriate to introduce 
separately the contribution $T_{AVP}(s)$ to the full amplitude
generated by the ${\cal L}_{AVP}$ Lagrangian piece
(see Eqs. (\ref{DD2})); its first term is symmetric in terms of the $\rho \pi$ 'final' 
states. One has\footnote{
\label{cor3pion} The $N_5$ contribution was
wrongly omitted in the study \cite{ExtMod1}; the error was due
to having missed that the two occurences of the function $\gamma$ in the numerator in the 
last Equation (\ref{HH1}) come with two different arguments ($s_{+-}$ and $s$).
The authors of the study \cite{ExtMod1} apologize for this inconvenience. }~:
 \be 
\hspace{-2.5cm}
\left \{
\begin{array}{lll}
\displaystyle
T_{sym}(s)= & \displaystyle \frac{ie}{4 \pi^2f_\pi^3} \left[ 4 g_{\pi^0 \pi^+ \pi^-\gamma}
(1- \frac{3}{4}[c_1-c_2+c_4]) \right.\\[0.5cm]
~~~&\displaystyle \left.
-\frac{9}{4}g [c_1-c_2-c_3](N_1(s)+N_0(s))
+\frac{3}{2} m^2 g(1+\Sigma_V) c_3  N_1(s)N_2(s)
\right]\\[0.5cm]
\displaystyle
T_{brk}(s)= & \displaystyle \frac{ie}{4 \pi^2f_\pi^3} \left[ 
\frac{3}{2} m^2 g(1+\Sigma_V) c_3
\right] \left[\frac{F_{\rho \gamma}(s)}{D_{\rho^0}(s)} (N_3(s)+N_4(s))
+\frac{F_{\phi \gamma}(s)}{D_{\phi}(s)}N_5(s)
\right]\\[0.5cm]
T_{AVP}(s) = & \displaystyle -\frac{ie}{4 \pi^2f_\pi^3}\left[ \frac{c_4-c_3}{4}
m^2 (1+\Sigma_V)\right]
\left[ N_2(s)+3 N_3(s)+9 N_6(s)\right]
\end{array}
\right.
\label{eq70}
\ee
where all parameters and functions have been already defined, except for the $N_i(s)$
functions which are given and commented in Appendix \ref{HH}. One has kept as much as possible
the notations used in  \cite{ExtMod1} in order to exhibit the effects of our additional
isospin symmetry breaking effects by simple inspection.  Finally, $T_{AVP}(s)$ identically
vanishes when $c_4=c_3$.

The differential cross section writes~:
\be
\displaystyle
\frac{d^2 \sigma(e^+e^- \ra \pi^+ \pi^- \pi^0)}{dx ~dy}=
\frac{\alpha_{em}}{192 \pi^2} s^2 G(x,y) |T_{sym}(s)+T_{brk}(s)+T_{AVP}(s) |^2
\label{eq71}
\ee
 using the ($x$ and $y$) parametrization proposed by E. Kuraev and Z. Siligadze
 \cite{Kuraev} who provided the kernel function
 $G(x,y)$ reminded in Appendix \ref{HH}. Note also that each of
 $T_{sym}(s)$, $T_{brk}(s)$ and $T_{AVP}(s)$ also depend on $x$ and $y$.

\section{Ugraded Breaking of the HLS Model~: A Summary} 
\label{summary}
\indent \indent 
In the former studies performed along the present lines 
\cite{taupaper,ExtMod1,ExtMod2}, roughly speaking, one incorporated nonet 
symmetry and SU(3) symmetry breaking in the pseudoscalar (PS) sector. In the vector
meson sector, only  SU(3) symmetry breaking was considered.

However, some important effects can be already attributed to isospin breaking
effects in the PS sector. Indeed, it is  the non--vanishing character of the
mixing "angles" $\alpha(s)$ and $\beta(s)$  which induces $s$--dependent
$\rho-\omg$ and $\rho-\phi$ 
mixings at the one loop level. This non--vanishing of the $\alpha(s)$ and $\beta(s)$ functions
proceeds from the kaon mass splitting  which breaks the symmetry between the neutral and 
charged kaon loops and, then, allows to choose the analytic 
function $\epsilon_1(s)$ as non--identically vanishing.
Therefore, except for the $\omg-\phi$ system which would mix anyway at one loop,  the full 
loop mixing mechanism for vector mesons is the prominent consequence for this limited 
account of isospin breaking\footnote{Actually, as noted in previous 
works \cite{taupaper}, $VP$ and $VV$
loops contribute to the vector meson mixing; the  effect of these additional loops
can be considered as absorbed by the subtraction
polynomials of the kaon loops.}.

This quite limited breaking scheme, allows already for 
a good account \cite{taupaper,ExtMod1,ExtMod2} of the available data.
However, within the realm accessible to the HLS model, two experimental issues remain unsolved~:
\begin{itemize}
 \item[i)] The dipion spectrum lineshape in $\tau$ decays is consistent with expectations
from $e^+e^-$ annihilations \cite{taupaper,ExtMod1}, but not its absolute scale
\cite{ExtMod2}.
 \item[ii)] The partial width ratio $\Gamma(\phi\ra K^+ K^-)/\Gamma(\phi\ra K^0 \overline{K}^0)$
is found inconsistent with all reported expectations \cite{BGPter}. Obviously, 
this inconsistency propagates to the corresponding $e^+ e^-$ annihilation cross
sections.
\end{itemize}

The first topic has been shown to get a satisfactory -- but not perfect -- 
solution by allowing some difference 
between   $\rho^0$ and  $\rho^\pm$ meson properties to be fitted from data.
If the effect of a non--vanishing $\delta m^2=m_{\rho^0}^2-m_{\rho^\pm}^2$  
was found small, those generated by a non--vanishing
$\delta g =g_ {\rho^0 \pi^+ \pi^-} - g_ {\rho^\pm \pi^0 \pi^\pm}$                                                        
was found especially significant \cite{ExtMod2}.
Moreover,  some rescaling of the  $\tau$  spectra, consistent
with the reported experimental  scale uncertainties  remained
unavoidable.                                                                              
 
The second topic is experimentally addressed by considering  \cite{SNDKKb,CMD2KKb}
that the Coulomb interaction\footnote{The function $Z(s)$  in Eq.  (\ref{eq53}).
} plays as a symmetry breaking mechanism which modifies the SU(3) relationship
$g_{\phi K^+ K^-}=g_{\phi K^0 \overline{K}^0}~~$ between coupling constants to
$~~g_{\phi K^+ K^-}=g_{\phi K^0 \overline{K}^0}\sqrt{Z(m_\phi^2)}$. This approach,
which turns out to consider the Coulomb interaction as some breaking effect,
may look unsatisfactory; anyway, it does not fit with our breaking
scheme.
                                                                                                           
These two issues motivated an upgrade of the breaking scheme of the HLS model
in order to check whether an acceptable  solution can be derived. The extension
to isospin breaking of the BKY--BOC breaking mechanism is {\it a priori} an 
obvious candidate to examine. This has been done in the preceding Sections
with several interesting conclusions, which can be summarized as
follows~:

\begin{itemize}
 \item[j)] One does not find any signal for a mass or a coupling difference 
between the $\rho^0$ and  $\rho^\pm$ mesons\footnote{Electromagnetic effects
beyond the HLS model and the BKY breaking scheme may, of course, change
 a little bit this picture; however, the phenomenological consequences 
of letting free this mass difference
are known to be negligible \cite{ExtMod2} as reminded before. }. 
However the coupling difference
between $\rho-\gamma$ and $\rho-W$ might be enforced with respect to 
\cite{taupaper,ExtMod1,ExtMod2} if the breaking parameter product $h_V\Delta_V$ is 
found significantly non--zero (see Table \ref{T1}),
 \item[jj)] Everything goes as if the universal coupling $g$ remains
unchanged in the anomalous sector, while one observes that $g$ 
is effectively modified to  $g (1+\Sigma_V)$ for the whole non--anomalous sector. Therefore,
isospin breaking in the HLS model generates some mild disconnection
between anomalous and non--anomalous processes which needs to be explored.
 \item[jjj)] The partial width ratio 
$\Gamma(\phi\ra K^+ K^-)/\Gamma(\phi\ra K^0 \overline{K}^0)$                                                                                                        
is found subject to isospin breaking in a novel way compared with the various  possibilities
examined in  \cite{BGPter},                                                                                             
\end{itemize}

Topics {\bf j} and {\bf jj} are both important for scale issues. Indeed, by disconnecting
somewhat more than before the ratio of transition amplitudes $\rho-\gamma$ and $\rho-W$,
one allows the HLS model to get more freedom for the purpose to account  for 
scale issues. 
More important, both $\tau$ and  $e^+e^-$ physics share the same universal coupling
($g (1+\Sigma_V)$), but it is no longer common with the scale of the anomalous processes 
which
remains governed by $g$. Moreover, none among the anomalous couplings, all displayed in 
several of the Appendices,  exhibits a dependence upon $\Sigma_V$. Stated otherwise, the anomalous
couplings -- which fix the scales of the anomalous meson decay and annihilation processes --
no longer constrain the non--anomalous process scales as sharply as formerly assumed
\cite{taupaper,ExtMod1,ExtMod2}.

Concerning the topics {\bf ii} and {\bf jjj}, it should be stressed that the parameter
$\Delta_A$ governing the change of this ratio is not involved only in the ratio. Indeed,
each of the  $e^+e^-\ra K^+ K^-$   and $e^+e^-\ra K^0 \overline{K}^0$  cross sections
should keep  valid absolute scales separately. 
Moreover, as clear from Appendices \ref{EE}, \ref{FF},  \ref{GG}
and \ref{HH}, and from Eqs. (\ref{eq58}) and (\ref{eq60}) given above, this change of scale 
should also fit with all anomalous processes, including the $\pi^0 \ra \gamma \gamma$
partial width, now within the partial width data sample submitted to the global fit.

Before ending up this Section and this Part, let us remark that the upgraded breaking of the
HLS model allows to address the question of the $\pi^0-\eta-\eta^\prime$ mixing in an
unusually large context. Moreover, as seen in Subsection \ref{omgpipi}, the exact
structure of the $\omg \pi \pi$ coupling discussed several times in the literature
\cite{Maltman1996,Maltman1997,Maltman2009} can also be examined within the largest 
possible data set.

A last remark is worth being emphasized. The scale treatment and the partial width ratio
quoted in {\bf i} and  {\bf ii}, within the upgraded breaking of the HLS model show up
as two $different$ aspects of the $same$ mechanism. Indeed, the former proceeds from applying
the extended BKY--BOC breaking scheme to ${\cal L}_{V}$, while the latter follows from
applying the same mechanism to   ${\cal L}_{A}$.

\section{The Data Sets and Their Handling}
\label{DataSets}
\indent \indent In this Section, we outline the data sets submitted to the global fit
and the way  correlated and uncorrelated uncertainties  are dealt with. Nothing
really new is involved here compared to what is already stated in 
\cite{taupaper,ExtMod1,ExtMod2}, except for the data sets associated with
the  $e^+e^- \ra K^+ K^-$ and $ e^+e^-  \ra K^0\overline{K}^0$ cross sections.
One may, thus, consider that this Section is, to a large extent, a simple
reminder provided in order to ease the reading of the present paper.

\subsection{The $e^+e^- \ra \pi^+ \pi^+$ Data}
\indent \indent 
Four data sets have been collected recently in Novosibirsk at the VEPP2M ring. 
The first one  \cite{CMD2-1995,CMD2-1995corr}, covering the region from about 600 to  960  MeV,
is claimed to carry a remarquably small  systematic error ($0.6\%$).
Later, CMD--2 has  published two additional data sets, one \cite{CMD2-1998-1} --
covering the energy region from 600 to 970 MeV -- is supposed to reach a systematic error of 
$0.8\%$, and a second set \cite{CMD2-1998-2} closer to the threshold region (from 370 to 520 MeV)
has an estimated systematic error of $0.7\%$. 
On the other hand, the SND collaboration has published 
\cite{SND-1998} a data set covering the invariant mass region from 370 to 970 MeV. Except
for the two data points closest to threshold which carry a sizable systematic error
($3.2\%$), a reported systematic uncertainty of $1.3\%$ affects this spectrum. These four data sets 
may be referred to in the following as "new timelike data" \cite{taupaper}.

When dealing with these data sets, statistical and uncorrelated systematic uncertainties
have been added in quadrature as usual. However, these four data sets also carry a common
correlated systematic uncertainty estimated to $0.4\%$ which affects all of them in the same way 
\cite{simonPriv}. This is accounted for by modifying appropriately the  covariance matrix
as outlined  in \cite{taupaper,ExtMod1} -- see also Subsection \ref{fitproc} below -- and by accounting 
for the data set to data set 
correlations. This is performed by treating these four data sets altogether, as if they
were subsets of a single (merged) data set.  

In order to be complete, we have also included in our
fit all data on the pion form factor collected formerly by the OLYA
and CMD Collaborations as tabulated in \cite{Barkov} and the DM1 data \cite{DM1} collected
at ACO (Orsay). These data will be referred to globally as "old timelike data". 
The systematic uncertainties carried by OLYA data ($4\%$) and CMD ($2\%$) contain
an uncorrelated part which has been added in quadrature to the reported statistical
errors. A common correlated part of the systematics, conservatively estimated \cite{simonPriv} 
to 1\%,  has been dealt with appropriately. Instead,  the accuracy of the DM1 data 
set  being poor and its weight marginal,  we did not find any need to go beyond the published 
uncorrelated errors.

\subsection{The $e^+e^- \ra (\pi^0/\eta) \gamma$ Data}
\label{pi0eta_g}
\indent \indent 
Since 1999, several data sets on the anomalous annihilation channels
$e^+e^- \ra \pi^0 \gamma$ and $e^+e^- \ra \eta \gamma$ have been made available
by the CMD--2 and SND Collaborations.  In our analysis, we only use
the provided data points up to $\sqrt{s} = 1.05$ GeV.

The first one used is the data set from CMD--2  \cite{CMD2Pg1999} on  the $\eta \gamma$ final state
($\eta \ra \pi^+\pi^-\pi^0$) which carries a systematic error of 4.8\%. 
CMD--2 has also provided  \cite{CMD2Pg2001} a second data set  on the $\eta \gamma$ final state, 
tagged with the decay mode  $\eta \ra 3 \pi^0 $. 
The systematic uncertainty carried by this sample is estimated to 6.1\%  and 4.1\%
for, respectively, the energy regions below and above 950 MeV. 
More recently, CMD-2 has 
also published two more data sets \cite{CMD2Pg2005} covering both the $(\pi^0/\eta) \gamma$ 
final states, tagged with the 2--photon decay modes, in the energy region  from 600 to 1380 MeV.
These are reported to carry a 6 \% systematic error.

The SND Collaboration has recently published \cite{sndPg2007}
two different data sets for the  $\eta \gamma$ final state with an estimated 
systematic uncertainty of $\simeq$ 4.8 \%. The first one covers the energy region
from 600 to 1360 MeV and the second from 755 to 1055 MeV.  A sample covering the energy 
range from 600 to 970 MeV for the $\pi^0 \ra \gamma \gamma$ decay mode was also published 
\cite{sndPg2003}. Finally, two data sets for both $(\pi^0/\eta) \gamma$ final states
with 14 data points (from 985 MeV to  1039 MeV) from SND \cite{sndPg2000}  
are also available; these exhibit the much lower systematic error  of $2.5$\%.

Altogether, these two Collaborations have provided 86 measurement points for
the $e^+e^- \ra \pi^0 \gamma$  cross section and 182 for $e^+e^- \ra \eta \gamma$
for $\sqrt{s}\leq 1.05$ GeV. Preliminary analyses  \cite{ExtMod1} did not reveal any need to 
split up correlated and uncorrelated parts of the systematic errors for the 
$(\eta/\pi^0) \gamma$ data samples. Nevertheless, we have made a few checks by
comparing fit results derived by adding in quadrature statistical and systematic
uncertainties with fit results derived assuming the reported systematic
error to be 100\% bin--to--bin correlated. We did not observe any significant 
difference. Therefore, when analyzing the $e^+e^- \ra (\pi^0/\eta) \gamma$ data,
the reported statistical and systematic uncertainties have been simply added 
in quadrature as in  \cite{ExtMod1}.

\subsection{The $e^+e^- \ra \pi^0 \pi^+ \pi^-$ Data}
\indent \indent This channel is important as it provides a single
place where the box anomaly sector \cite{WZ,Witten} is present. Other
physics channels involving the box anomaly in the $\eta/\eta^\prime$ sectors exist
($\eta/\eta^\prime \ra \pi^+ \pi^- \gamma$) and may be relevant. However, the
overall experimental situation is unclear \cite{box,ExtMod1}, even if
the Crystal Barrel data sample \cite{Abele} may look secure. Therefore,
we find preferable to wait for confirmation with new data samples which 
could  come from BES and KLOE. 

There are several published data sets for the $e^+e^- \ra \pi^0 \pi^+ \pi^-$ annihilation
channel with various statistical and systematic uncertainties. We first included
in our data sample the data sets collected by CMD--2 which consist of a measured
sample covering the $\omg$ region \cite{CMD2-1995corr} affected with a global scale
uncertainty of  1.3\% and   two others which cover the $\phi $ region 
with a reported scale error of, respectively, 4.6\% \cite{CMD2KKb-1} and
1.9\%  \cite{CMD2-1998}. The most recent CMD--2  data sample  \cite{CMD2-2006} also covers the 
$\phi $ region  with a scale uncertainty of 2.5\%.
 
 SND has published two spectra covering altogether the region from 0.44 to 1.38 GeV,
 the former below 980 MeV  \cite{SND3pionLow}, the latter above  \cite{SND3pionHigh}.
  For both data samples, the correlated part of the systematic uncertainty has been
  extracted in order  to be treated as a scale uncertainty (3.4 \% for \cite{SND3pionLow}
  and 5\% for  \cite{SND3pionHigh}, respectively); the uncorrelated parts have been added in quadrature with
  the reported statistical errors.
 
Former data sets are also considered which cover the region in between the 
$\omg$ and $\phi $ peaks where physics constraints are valuable. The most useful  
has been collected by the ND Collaboration with 10\% systematics
and can be found in  \cite{ND3pion-1991},
the latter is a small data sample from CMD \cite{CMD3pion-1989} providing 5 measurement
points with 15\% systematics in the intermediate region. Concerning these two complementary
data samples, we perform as in \cite{ExtMod1} and do not extract the correlated
part of the systematics as the accuracy is poor enough that this could not lead
to visible effects in global fits. Finally, there also exists a small data
sample from DM1  \cite{DM13pion-1979} which has been used for illustrative purposes
only   \cite{ExtMod1}.

The analysis of these data samples has been performed in   \cite{ExtMod1}; however,
as the $N_5$ term which contributes to the cross section (see Eq. (\ref{eq70})) was missing, 
the analysis is redone and the conclusions revisited.

\subsection{The $\tau^\pm \ra \pi^\pm \pi^0 \nu_\tau$ Data}
\indent \indent In the collection  of data samples submitted to global fitting, 
 we also use the ALEPH \cite{Aleph}, CLEO \cite{Cleo} and BELLE \cite{Belle}
data sets. When dealing with $\tau$ data, it is important to note that
the relevant quantity, sensitive to the spectrum lineshape $and$ to its absolute normalization
is given by~:
\be
\displaystyle \frac{1}{\Gamma_{\tau}} \frac{d\Gamma_{\pi \pi}(s)}{ds} = 
{\cal B}_{\pi \pi} \frac{1}{N}\frac{dN(s)}{ds}
\label{eq72}
\ee
where $\Gamma_{\tau}$ is the full $\tau$ width, ${\cal B}_{\pi \pi}$ the branching
ratio to $\pi \pi \nu$, and $1/N dN(s)/ds$ is the normalized spectrum of yields as
measured by the various experiments. 

The data published by the ALEPH Collaboration correspond directly to the 
quantity shown in the left--hand side of Eq. (\ref{eq72}).
Instead, each of CLEO and BELLE has published separately 
the  normalized spectrum of yields and the  measured
branching ratio ${\cal B}_{\pi \pi}$. In the $\tau$ data handling,
we have considered the reported uncertainties on these measured ${\cal B}_{\pi \pi}$'s 
as bin--to--bin correlated scale uncertainties; these come into the 
various $\chi^2$ associated with each data set  in the way reminded in Subsection 
\ref{fitproc}. Stated otherwise, they are no longer fitted as previously done
\cite{ExtMod2}.

Following closely the experimental information provided 
by \cite{Aleph}, \cite{Belle},  \cite{Cleo}, the scale uncertainties have been estimated to
0.51\% (ALEPH), 1.53\% (Belle) and 1.74\% (CLEO). On the other hand, a possible
absolute energy scale uncertainty of 0.9\% r.m.s. affecting the CLEO data sample \cite{Cleo}
has not been found significant \cite{taupaper,ExtMod2} and is not considered in the present study.
All these experiments have provided their statistical and systematic error covariance matrices; 
these are the main ingredient of the $\chi^2$ functions used in the fits.

As the HLS model relies on the lowest mass vector meson nonet only, it cannot access 
$\Gamma_{\tau}$ which is therefore taken from the Review of Particle Properties \cite{RPP2008}. 
Finally, our model provides \cite{taupaper}~:
\be
\displaystyle \frac{d\Gamma_{\pi \pi}(s)}{ds} = \displaystyle
\frac{|V_{ud}|^2 G_F^2}{64 \pi^3 m_\tau^3} |F_\pi^\tau(s)|^2 
G_0(s) 
\label{eq73}
\ee
\noindent with~:
 \be
 \left \{
\begin{array}{lll}
G_0(s) &= \displaystyle \frac{4}{3} \frac{(m_\tau^2-s)^2(m_\tau^2+2 s)}{s^{3/2}} Q_\pi^3\\[0.5cm]
 Q_\pi &= \displaystyle \frac{\sqrt{[s-(m_{\pi^0}+m_{\pi^+})^2][s-(m_{\pi^0}-m_{\pi^+})^2]}}{2\sqrt{s}}
\end{array}
\right .
\label{eq74}
\ee
and $F_\pi^\tau(s)$ is given in Eq. (\ref{eq37}). Isospin symmetry breaking specific of the
$\tau$ decay will be considered 
and taken into account as emphasized in Section \ref{fit_tau}.

Of course, the published $\tau$ spectra extend much beyond the validity range of the HLS model,
as this presently stands. Therefore, when using it, we have to truncate at some $s$ value.
Consistency with the treatment of scan data would imply a truncation at 1.05 GeV. However, various
studies \cite{Belle,ExtMod2} showing the behavior of fit residual clearly observe that ALEPH data 
on the one hand and Belle and CLEO data, on the other hand, exhibit inconsistent behavior
 starting
in the $0.9 \div 1.$ GeV region. Therefore, we have preferred truncating the spectrum at $1.$ GeV,
where the three spectra are in reasonable agreement with each other.

\subsection{The $e^+e^- \ra K \overline{K}$ Data}
\label{KKbData}
\indent \indent 
 Several data sets have been collected by the CMD-2 and SND
Collaborations on both
annihilation cross sections $e^+ e^- \ra K^+ K^-$ and
$e^+ e^- \ra K^0 \overline{K}^0$. Here also, we have discarded the 
data points above 1.05 GeV.

The oldest data sets, published by CMD-2  \cite{CMD2KKb-1}, provide 
the spectra for both the neutral and charged decay channels with a systematic 
uncertainty of 4\%.  Recently CMD-2 has reanalyzed four data 
sets
for the neutral decay mode \cite{CMD2KKb-2} getting small systematic
errors (1.7\%). More recently, CMD-2 has also published two scans of the charged mode
spectrum \cite{CMD2KKb-3} with a systematic uncertainty of 2.2\%.

On the other hand, SND has published in 2001 several data sets \cite{SNDKKb}~:
2 for the charged decay channel with a systematic error of 7.1\%,
2 data sets in the neutral mode with $K_S \ra \pi^0 \pi^0$ and 2 more
with $K_S \ra \pi^+ \pi^-$, with respectively 4.2\% and 4.0\% systematics.

The quoted systematics are treated as correlated scale uncertainty as outlined
in Subsection \ref{fitproc} below.

\subsection{The Partial Width Data Set}
\label{decays}
\indent \indent In order to work out the fit procedure and get enough
constraints on the physics parameters of the model, an important input is 
the set of  decay partial widths \cite{taupaper}. All decay modes of the form 
$VP\gamma$ and $P \gamma \gamma$ not related with the cross sections listed 
above should be considered. This covers  the radiative partial widths
$\rho^\pm \ra \pi^\pm \gamma$,  $\eta^\prime \ra \omg \gamma$ and  
$\phi  \ra \eta^\prime \gamma$ on the one hand and
$ (\eta^\prime/\eta/\pi^0) \ra  \gamma \gamma $ on the other hand.
They have been extracted  from the
Review of Particle Properties \cite{RPP2008}. The accepted values for
radiative partial widths
for $K^{*\pm} \ra K^\pm \gamma$ and  $K^{*0} \ra K^0 \gamma$ have
also to be  used \cite{RPP2008}.

As  the currently available data on $e^+ e^- \ra \pi^+ \pi^-$ stop
slightly below 1 GeV, the  phase of the $\phi \ra \pi^+ \pi^-$ amplitude
and its branching ratio as measured by SND \cite{SNDPhi} are relevant pieces
of information, not included in the above listed annihilation data\footnote{
However, one might have to be cautious with these data. Indeed, as emphasized
in \cite{ExtMod1} -- see Section 13 therein -- the single piece of information
truely model independent is the product ${\cal B}_{ee} {\cal B}_{\pi \pi}$.
Therefore  $separate$ values for ${\cal B}_{ee}$ and ${\cal B}_{\pi \pi}$, given as
"experimental" values in the various releases of the Review of Particle Properties, 
 are actually model dependent to an unknown extent.   }. 
In contrast, the corresponding information for the
$ \omg$  meson is irrelevant as it is fully contained in the
amplitude for $e^+ e^- \ra \pi^+ \pi^-$ (see Eq. (\ref{eq41}))
and is already part of the data sample.

With respect to former studies within the same framework, the only new
piece of information included in the fit data set is the partial width
$ \pi^0 \ra  \gamma \gamma $. Indeed, as can be seen from Eqs. (\ref{eq58}),
the corresponding amplitude may constrain $\Delta_A$  as well as the 
$e^+e^- \ra K \overline{K}$ annihilation amplitudes.

In fits involving all the above quoted annihilation channels, one has no longer to
consider the leptonic widths $(\rho^0 /\omg/\phi) \ra e^+ e^-$ 
and the decay widths $(\rho^0 /\omg/\phi) \ra (\eta/\pi^0) \gamma$
as they are essentially extracted from some of the cross sections
listed above which permanently enter our fit procedure. 

Therefore, the additional decay information to be used as input to final fits
represents in total 10 more pieces of information.

\subsection{Outline of the Fit Procedure (The Method) }
\label{fitproc}
\indent \indent For all data sets listed above, one always has at one's disposal
the statistical error covariance matrix. For scan data, this may include the 
uncorrelated part of the systematic errors; if not done at start,
enough information is generally provided to allow one to perform this (quadratic) sum.
In the case of $\tau$ data, the systematic error covariance matrix may be provided 
by the experimental groups (as ALEPH \cite{Aleph}, for instance).

In this case, for each group of data sets ($\pi^+\pi^-$, $\pi^0\gamma$,
$\eta \gamma$, $\pi^+\pi^-\pi^0$, $K^+ K^-$,  $K^0 \overline{K}^0$, $\pi^\pm \pi^0 \nu$)
one computes the partial $\chi^2$~:
 \be
\chi^2_i= \displaystyle (m -M)^T V^{-1} (m -M) ~~~ {\rm (Experiment~\#~} {\it i}) 
\label{eq75}
\ee
using matrix notations, and denoting by $m$ and $M$ the  measurement  vector
and the corresponding model function vector. $V$ is the error covariance matrix 
already referred to. The function to minimize is simply the sum of the $\chi^2_i$.

Actually, this is the procedure to estimate $\chi^2_i$  when the corresponding
data sample is not subject to an overall scale uncertainty. If such a
scale uncertainty takes place for some data set, one should perform a modification.

Let us assume that the data set $i$ is subject to a  scale uncertainty; this 
is supposed\footnote{In practical use, a data set \# $i$, subject to a scale uncertainty
$\lambda_{i,0}$ is supposed to have been corrected in order to absorb a possible bias; this is the
reason why the corresponding random variable is supposed unbiased, e.g. carrying
zero mean. If not, Eq. (\ref{eq76}) should be modified by performing
$\lambda_i \ra \lambda_i -\lambda_{i,0}$.}
 to be a random variable $\varepsilon(0,\sigma)$ of zero mean (unbiased) and with 
r.m.s. $\sigma$, independent of $s$. Then any fit corresponds to getting $one$ 
sampling of $\varepsilon(0,\sigma)$, named $\lambda_i$. In this case, Eq. (\ref{eq75})
should be  modified to~:
\be
\chi^2_i= \displaystyle 
\left [ m- M -A \lambda_i\right ]^T V^{-1} \left [ m-  M -A \lambda_i\right ]
+\frac{\lambda^2_i}{\sigma^2}
\label{eq76}
\ee
where  \cite{D'Agostini} $A$ is traditionally the vector  of the model values $M$ and 
the other notations
are obvious. One can solve for $\lambda$, which turns out to perform the change~:
\be
  \displaystyle 
V^{-1} \Longrightarrow W^{-1}(\sigma^2) = \left [ V + \sigma^2 A A^T \right ]^{-1}=
V^{-1} -\frac{\sigma^2} {1+ \sigma^2 (A^T V^{-1} A)} (V^{-1}A) (V^{-1}A)^T
\label{eq77}
\ee
in Eq. (\ref{eq75}).
The modified covariance matrix $W$  depends on the vector $A$.
As just stated, the best motivated choice for  the vector $A$ is the model function $A=M$.
However, this implies a recursive determination of the modified
covariance matrix, and, therefore, recalculating (or inverting) large matrices
at each step of the minimization procedure (several hundreds of times for each 
fit attempt). It happens, however, probably because the experimental  data we deal with
are already accurate enough, that choosing $A=m$ ({\it i.e.} the measurement vector
of the corresponding experiment) does not sensitively affect the results and strongly
improves the convergence speed of the minimization procedure  \cite{ExtMod1}.
Therefore, unless otherwise stated, we always perform this approximation.

\subsection{The  Discarded Data Sets  }
\indent \indent There exists data sets which have been discarded for the present
study. The most important are the three data sets collected using the Initial State Radiation
(ISR) method by the KLOE \cite{KLOE08,KLOE10} and BaBar \cite{BaBar} Collaborations.
These suppose a specific statistical treatment as the structure of the reported systematic errors
is much more complex than for any set of scan data. The method used in \cite{ExtMod1}
for KLOE 2008 data \cite{KLOE08} allows to deal with, but should be studied carefully with 
each ISR data set separately. 

In order to keep clear the message of
the present study, we prefer avoiding using now data sets invoking delicate statistical methods.
Therefore, the ISR data sets \cite{KLOE08,KLOE10,BaBar}  will be treated in a forthcoming
publication. Because of their high statistics, if well understood, these data samples
may  improve  the physics results derived by using the model and the fit procedure presented
in this study.

Other data sets could have been useful~:
\begin{itemize}
\item
Those providing the pion form factor in the spacelike
region close to $s=0$ \cite{NA7,fermilab2}. Indeed such data could severely constrain the pion
form factor in the threshold region. This was illustrated in
\cite{taupaper} where an archaic form of our model has been used. 
However, we gave up using them -- especially \cite{NA7} -- because there is some suspicion 
concerning their estimated overall scale. Such a kind of data would nevertheless help in getting
more precise information on $g-2$.
\item
More data involving  the box anomaly, especially in the $\eta/\eta^\prime$ sectors may also help
in constraining the model parameters. For instance, the dipion spectra in 
$\eta/\eta^\prime \ra \pi^+ \pi^- \gamma $ provide such information. Some available data collected
in \cite{box}, especially those for $\eta^\prime \ra \pi^+ \pi^- \gamma $ provided by the Crystal
Barrel Collaboration \cite{Abele}, might be considered sometime. However, new data sets on this subject,
with larger statistics and better systematics should come from the KLOE and BES Collaborations, especially 
concerning the decay process $\eta \ra \pi^+ \pi^- \gamma $. These are certainly more easy to handle than
the  $e^+ e^- \ra \eta \pi^+ \pi^-$ annihilation data which {\it in fine} carry the same physics
information.
\end{itemize}

\subsection{The Physics Parameter Set}
\label{param}
\indent \indent 
It looks appropriate to give  the list of the free model parameters to be fitted from data.
The model parameters are of various kinds~:
 \begin{itemize}
\item The basic HLS (4)  parameters~:  the universal vector coupling $g$; the relative
weight $a$ of the Lagrangian pieces ${\cal L}_A$ and ${\cal L}_V$,
expected $a \simeq 2$ from most VMD models;  finally the weights  $c_3$, $c_4$ and $c_1-c_2$ of the
anomalous FKTUY Lagrangian pieces to be added to the HLS Lagrangian in order to
address the full set of data outlined in the above Subsections.
\item SU(3) breaking parameters which modifies the physics content of the HLS Lagrangian
($z_A$, $z_V$ and $z_T$), together with the parameter named $\lambda$ which accounts
for nonet symmetry breaking in the pseudoscalar sector. This amounts to a total of 4.
\item The isospin breaking parameters $\Delta_A$, $\Sigma_V$, $\Delta_V$ 
and $h_V$ which affect the non--anomalous HLS Lagrangian. These represent the Direct 
Isospin Breaking mechanism introduced in this paper through the BKY mechanism.

\item Some parameters \cite{leutw96} allowing the $\pi^0-\eta-\eta^\prime$ mixing.
The $\eta-\eta^\prime$ mixing angle $\theta_P$ and the parameters named above 
$\epsilon$ and $\epsilon^\prime$, which may account for, respectively, the $\pi^0-\eta$ and 
$\pi^0-\eta^\prime$ mixings. The last couple of parameters is not important for 
$g-2$ estimates but may provide interesting  physics information. One may anticipate
on fit results by saying that the condition $\theta_0=0$ is well accepted
by the data as in previous analyses \cite{WZWChPT}; as a matter of consequence
$\theta_P$ can be (and will be) chosen as entirely fixed by the nonet symmetry breaking
parameter $\lambda$ (see Eqs. (\ref{eq26})). One will also see that the pair
$\epsilon$ and $\epsilon^\prime$ can be safely replaced by a single free parameter
\cite{leutw96}. Therefore, the number of really free parameters accounting for
the $\pi^0-\eta-\eta^\prime$ will be reduced to one.

\item Some subtraction parameters (8) involved in the mixing functions of vector mesons,
in the $\rho$ meson self--energy and in the $\gamma-V$ transition amplitudes.

\item Some more parameters (4) describing the mass and width of the narrow $\omg$ and
$\phi$ mesons. As a detailed description of the loop corrections to their inverse propagators
is of little importance for the present purpose   ,  there is no need  to go beyond.
\end{itemize}

Stated otherwise, only the parameters  $\Delta_A$, $\Sigma_V$, $\Delta_V$ and $h_V$ are new
and all others have been already
dealt with in previous releases of the present model \cite{taupaper,ExtMod1,ExtMod2}.

 One may
be surprised to face a so large number ($\simeq 25$) of parameters to be fitted from data. 
This only reflects that the number of physics pieces of information and of processes to account
for is also exceptionnally large~: more than 900 data points, six annihilation channels
($\pi^+ \pi^-$, $\pi^0 \gamma$, $\eta \gamma$, $K^+ K^-$,$K^0 \overline{K}^0$, 
$\pi^+ \pi^- \pi^0 $), 10 radiative decay modes ($VP\gamma$, $P \gamma \gamma$
including now the $\pi^0 \ra \gamma \gamma$ partial width), the $\phi\ra \pi^+ \pi^-$
decay
and finally the dipion decay mode of the $\tau$ lepton. All these pieces of information
should get simultaneously a satisfactory description. 

It should be stressed that the parameter
space is sharply constrained, as will be confirmed and illustrated by the present study.
One should also stress that the $\pi^+ \pi^-$, $\pi^0 \gamma$ and $\eta \gamma$  cross sections,
together with the decay modes referred to above, allow already a good determination
of all fit parameters except for two~: $c_1-c_2$ and $\Delta_A$. The former
is derived from fitting the 3 pion cross section, the second from fitting
both $K\overline{K}$ annihilation channels. Actually, in order to accurately
determine $\Sigma_V$, the dipion spectrum in the decay of the $\tau$ lepton
 also  plays a crucial role.

This peculiarity leads us to a motivated critical analysis of the available $\pi^+ \pi^- \pi^0 $,
$K\overline{K}$  and $\tau$ data sets. As one plans to motivate a value for the hadronic contribution
to $g-2$, our dealing with the corresponding data should also be motivated.

As far as cross sections are concerned, it is already known from our previous studies
that the $\pi^+ \pi^-$, $\pi^0 \gamma$ and $\eta \gamma$  annihilation cross sections are very well
described within a simultaneous fit including also the decay data already listed. 
This can be seen in \cite{taupaper,ExtMod1}; indeed Figure 2 in 
\cite{taupaper} and Figures 1 and 2 in \cite{ExtMod1} are indistinguishable from
what is derived in the present study.

\section{Reanalysis of the $\pi^+ \pi^- \pi^0$ Annihilation Channel}
\label{3pions}
\indent \indent Taking into account the error described in Footnote \ref{cor3pion}, 
the analysis
of the model description of the $\pi^+ \pi^-\pi^0$  data is worth being redone. We take
profit of this case in order to exemplify how the dealing with data sets is done.

The available 3--pion data sets can be gathered into 3 different groups~:

{\bf i/} The former  data set collected by the Neutral Detector (ND)
at Novosibirsk and published in \cite{ND3pion-1991}~: we include in this group
the few data points from \cite{CMD3pion-1989}. These mostly cover the energy region 
in between the $\omg$ and $\phi$ peaks. 

{\bf ii/} A CMD--2 data set covering the $\omg$ region \cite{CMD2-1995corr}
together with a corresponding SND data sample  \cite{SND3pionLow} which actually
extends up to  980 MeV.

{\bf iii/} Several CMD--2 data sets covering the $\phi$ region and extracted from
\cite{CMD2KKb-1,CMD2-1998,CMD2-2006}, accompanied by a data set from SND 
\cite{SND3pionHigh} starting at 970 MeV.
 
 The small data
sample from DM1  \cite{DM13pion-1979} is used for illustrative purposes and is not included in
the fit procedures. It would not influence the fit results.

In fit procedures, it is very hard to run {\sc minuit} normally
because integrating the parameter dependent 3--pion cross section (see Eqs. (\ref{eq70}) and 
(\ref{eq71})) renders  prohibitive the execution time. Therefore, we still use here
the iterative method described and motivated in Section 10.3 of \cite{ExtMod1}.

The choice of the 3--pion data sets considered in the global fit was performed
in   \cite{ExtMod1} relying on the data sets listed in {\bf i}. Indeed, the 
$\pi^+ \pi^-$ data used in the global fit serve to fix all parameters, except 
for the $\omg$ and $\phi$
mass and width parameters which are derived from having included the $\pi^0 \gamma$
and $\eta \gamma$ cross sections; therefore, the ND data having a large lever arm
(see downmost Figure \ref{Fig:final3p}), they are alone able
to determine accurately the value for $c_1-c_2$ (see third line in Figure  \ref{Fig:c1Mc2}).

Here one proceeds otherwise in order to learn more as  each of the just above mentioned
data set carries intrinsically a value for  $c_1-c_2$. Nevertheless, the group of data sets
 needed in order to fix all parameters except for $c_1-c_2$ has been enlarged~: Beside
 the $\pi^+ \pi^-$, $\pi^0 \gamma$ and $\eta \gamma$ cross sections, we have included
 the $\tau$ decay information from ALEPH, Belle and CLEO. This will be justified 
 later on. On the other hand,  one assumes $c_3=c_4$ which is justified in 
 Section \ref{Kubis}.

Fits are performed by including
either the CMD--2 data sets or  SND data sets, each in isolation. On the other hand, 
separate (and independent) fits are performed in
either of the $\omg$  and $\phi$ regions. Therefore, in these fits,
the  $\omg$ region fits are not influenced by  the  $\phi$ region  information and conversely. 
Moreover, CMD--2 and SND data are not influencing each other. 
The data sets associated with the so--called $\omg$  and $\phi$ regions is not
ours; it has been performed by the experimental groups who published the 
corresponding data sets separately.

It should be stressed, especially in the present case, that the notion of data set covers, 
as importantly, the data points, the full
 error covariance matrices ({\it i.e.} including the correlations reflected by the
 non--diagonal entries),  and all the additional pieces of
information provided by the experimental groups. Among this last kind of information,
the   global scale uncertainty included in the systematics should be suitably
accounted for. As far as scan data are concerned, the statistical methods we use are
the standard (text--book) methods briefly reminded in Subsection \ref{fitproc}.
 
 \vspace{0.5cm}

\begin{figure}[!ht]
\begin{minipage}{\textwidth}
\begin{center}
\resizebox{\textwidth}{!}
{\includegraphics*{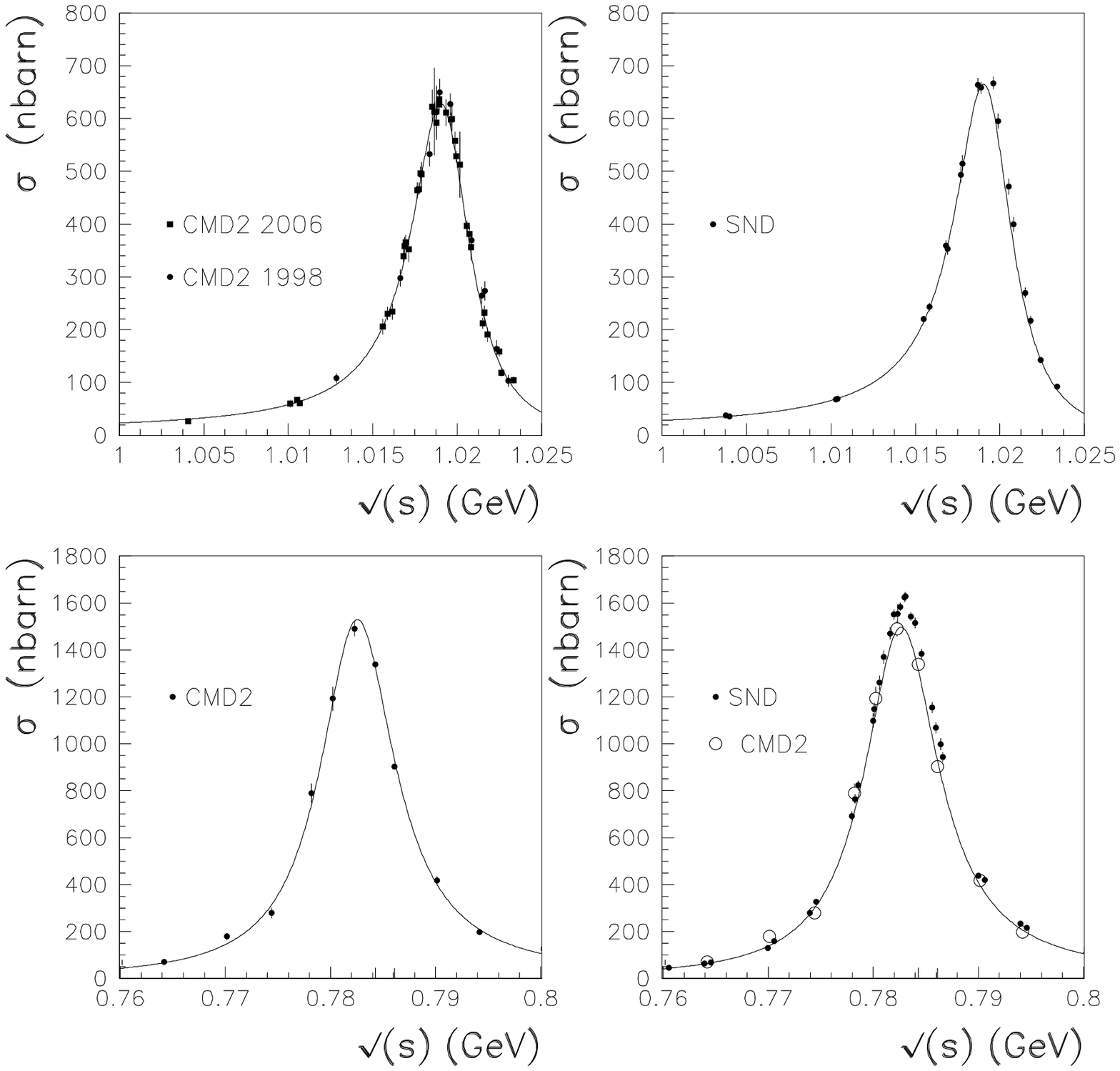}}
\end{center}
\end{minipage}
\begin{center}
\vspace{-0.3cm}
\caption{\label{Fig:bestfits3p}
Best fits to $e^+e^- \ra \pi^+ \pi^- \pi^0$ cross sections
for data sets in isolation.
Left column displays fits of the CMD--2 data, right column
displays fits of the SND data. Top shows the $\phi$ region,
bottom the $\omg$ region. The plotted data are extracted 
from  \cite{CMD2-1998,CMD2-2006} (CMD--2) and \cite{SND3pionHigh}
(SND) for the $\phi$ region and from  \cite{CMD2-1995corr} (CMD--2)
and  \cite{SND3pionLow} (SND) for the $\omg$ region. The empty circles
(bottom right plot) are superimposed on the SND fit results and are $not$
used in the fit displayed in $this$ Figure.}
\end{center}
\end{figure}

The results of these fits are 
summarized in Figure \ref{Fig:bestfits3p} and are commented on now.
As a word of caution, it should be noted that the experimental errors shown
in these plots are the quadratic sum of the reported statistical and systematic
errors, neglecting all correlations. As the error bars
 do not (and cannot) take into account the correlations, they should only be
considered as a visual indication of what is going on. The real distance of
data points to its best fit curve is instead accurately reflected by the $\chi^2$ 
values which, indeed, take appropriately into account all the reported pieces of information
about the error covariance matrix.

Top left Figure \ref{Fig:bestfits3p} shows the fit of only the CMD--2 data in the $\phi$ 
region; this
provides a good fit\footnote{The  numbers for $\chi^2/npoints$ are the 3--pion sample 
contributions to the global $\chi^2$ and its number of data points. One cannot
provide the number of degrees of freedom as several hundreds of (other) data points are involved
in each fit.}
 ($\chi^2/npoints=110/80$) returning  $c_1-c_2=1.21\pm 0.10$.
Top right Figure \ref{Fig:bestfits3p} shows the case for the SND data in the $\phi$ region in isolation; the
fit is much better ($\chi^2/npoints=26/33$) but returns $c_1-c_2=2.18\pm 0.13$.
These two fit values for  $c_1-c_2$ differ by $\simeq 10 \sigma$, clearly
tagging an inconsistency between the CMD--2 and SND data sets  in the $\phi$ region.

On the other hand, one has performed likewise for the  $\omg$  region in isolation.
One then gets for CMD--2 data a large  $\chi^2/npoints=26/13$ with $c_1-c_2=1.29\pm 0.04$
(bottom left Figure  \ref{Fig:bestfits3p}). A closer examination of these data
shows that an important part of this relatively large $\chi^2$ is due
to only the first data point which falls right on the vertical axis in
this Figure.

Instead, the SND $\omg$  region data yield $\chi^2/npoints=48/49$ and
$c_1-c_2=1.12\pm 0.06$ (bottom right Figure \ref{Fig:bestfits3p}). These two fit values for  $c_1-c_2$
differ by $\simeq 3 \sigma$; then, one may consider that the CMD--2 and SND data sets  in the 
$\omg$ region are in reasonable agreement with each other.

One should note from fitting the SND  $\omg$ data set, the important 
effect of correlations~: 
In the bottom right Figure  \ref{Fig:bestfits3p}, the large distance of the (SND) data
points to their fitting curve is compensated by the correlations in such a way that 
 $\chi^2/npoints$  remains quite reasonable. The high level of compensation
 can  be checked by computing
 the "diagonal" part\footnote{Denoting by $V$ the full covariance matrix 
 constructed as explained in Subsection \ref{fitproc}, the (full) $\chi^2$
 can be split up into its diagonal part 
 $\chi^2_{diag}= \sum_i V^{-1}_{i,i} (\Delta_i)^2$
 and its non--diagonal part 
 $\chi^2_{non~diag}= \sum_{i\ne j} V^{-1}_{i,j} \Delta_i \Delta_j$, where
 $\Delta_i$ is the difference of the $i^{th}$ measurement and the corresponding
 value of the theoretical cross--section.} of the $\chi^2$ which reflects
 the visual impression provided by  the bottom right Figure  
 \ref{Fig:bestfits3p}; one gets $\chi^2_{diag}= 554$~!

  In addition, one has found instructive
to plot  the CMD--2 data together with the SND ones and the fit performed to the SND 
data in isolation. Thus,  the bottom right Figure  \ref{Fig:bestfits3p} illustrates 
that the correlations reported by SND allow  
a reasonable reconstruction of the cross section valid for both the SND and CMD--2 data sets.

For information, the fit performed using only the ND data\footnote{As reminded 
above, this data set covers the region in between the $\omg$ and $\phi$ peaks.}  
yields $\chi^2/npoints=25/37$ and  $c_1-c_2=1.30\pm 0.06$, in good
accord with the previous fit result  $c_1-c_2=1.17\pm 0.07$, derived under
comparable conditions (see second data column in Table 3 of  \cite{ExtMod1});
the difference between these two estimates for $c_1-c_2$ can be attributed to
the influence of the $\tau$ data samples.

The  various estimates for $c_1-c_2$ derived from our fits  are gathered in 
Figure \ref{Fig:c1Mc2} using obvious notations. Using the fit values for $c_1-c_2$,
as tag of consistency, this plot clearly shows that the $\phi$ region SND data set
behaves differently from the other three--pion data sets.

From these considerations, one can conclude that~:
\begin{itemize}
 \item In the $\omg$  region, there is a good agreement between CMD--2 and SND data
 from within the filter of our model. 
 
  \item In the $\phi$ region, at minimum $\chi^2$, one can get a reasonable description
  of both CMD--2 and SND data, but with much different values for the fit parameters
  as reflected by their $c_1-c_2$ values.
\end{itemize}
Therefore, one observes a qualitative difference between  all CMD--2 data and
the SND data in the $\omg$  region, on the one hand, and the SND data in the $\phi$  region,
on the other hand.

\begin{figure}[!ht]
\begin{minipage}{\textwidth}
\begin{center}
\resizebox{\textwidth}{!}
{\includegraphics*{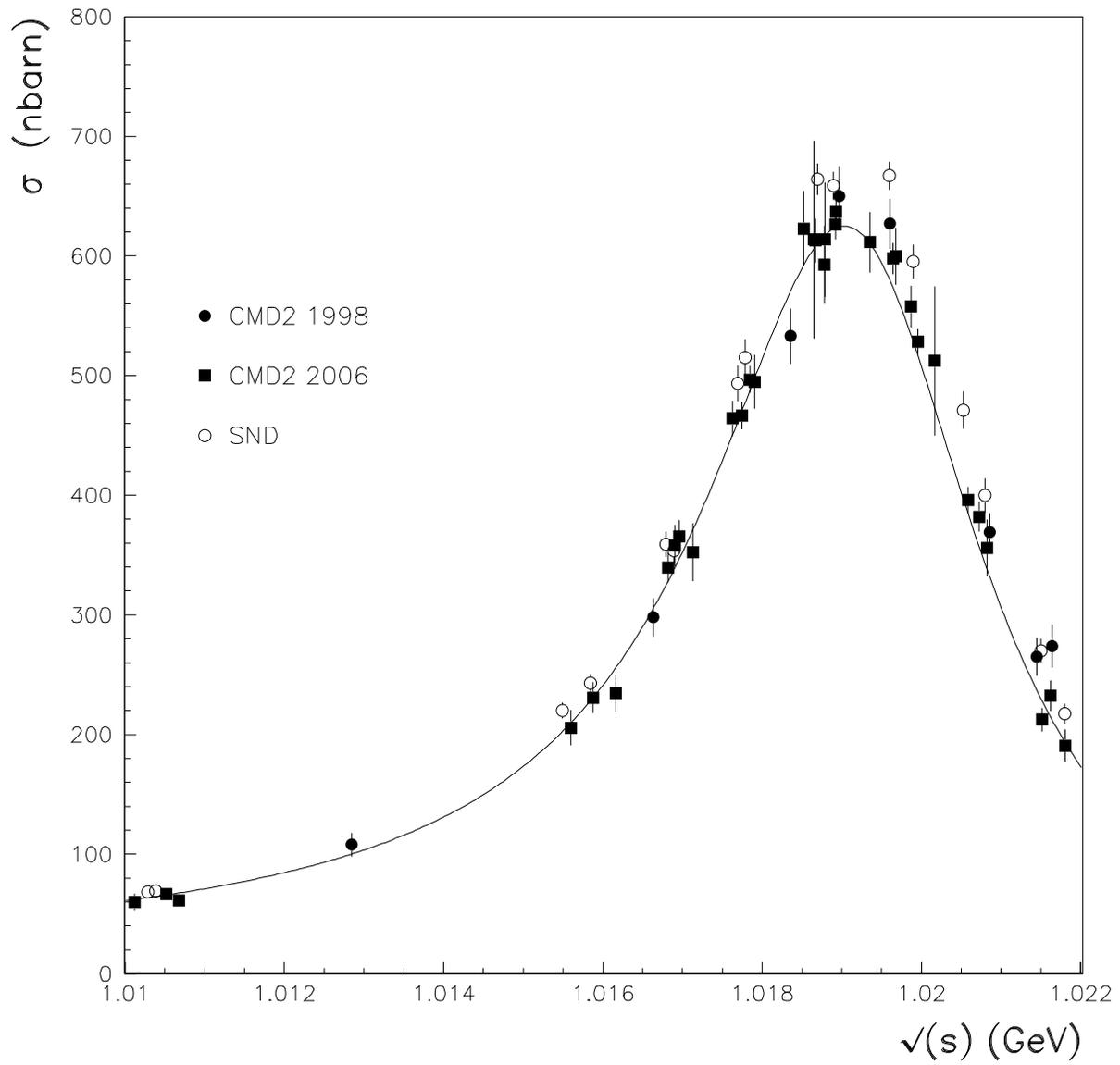}}
\end{center}
\end{minipage}
\begin{center}
\vspace{-0.3cm}
\caption{\label{Fig:commonfit3p_phi}
Simultaneous fit of $e^+e^- \ra \pi^+ \pi^- \pi^0$ cross section 
on the $\phi$ region data from \cite{CMD2-1998,CMD2-2006} (CMD--2) and \cite{SND3pionHigh}
(SND).
}
\end{center}
\end{figure}

\begin{figure}[!ht]
\begin{minipage}{\textwidth}
\begin{center}
\resizebox{\textwidth}{!}
{\includegraphics*{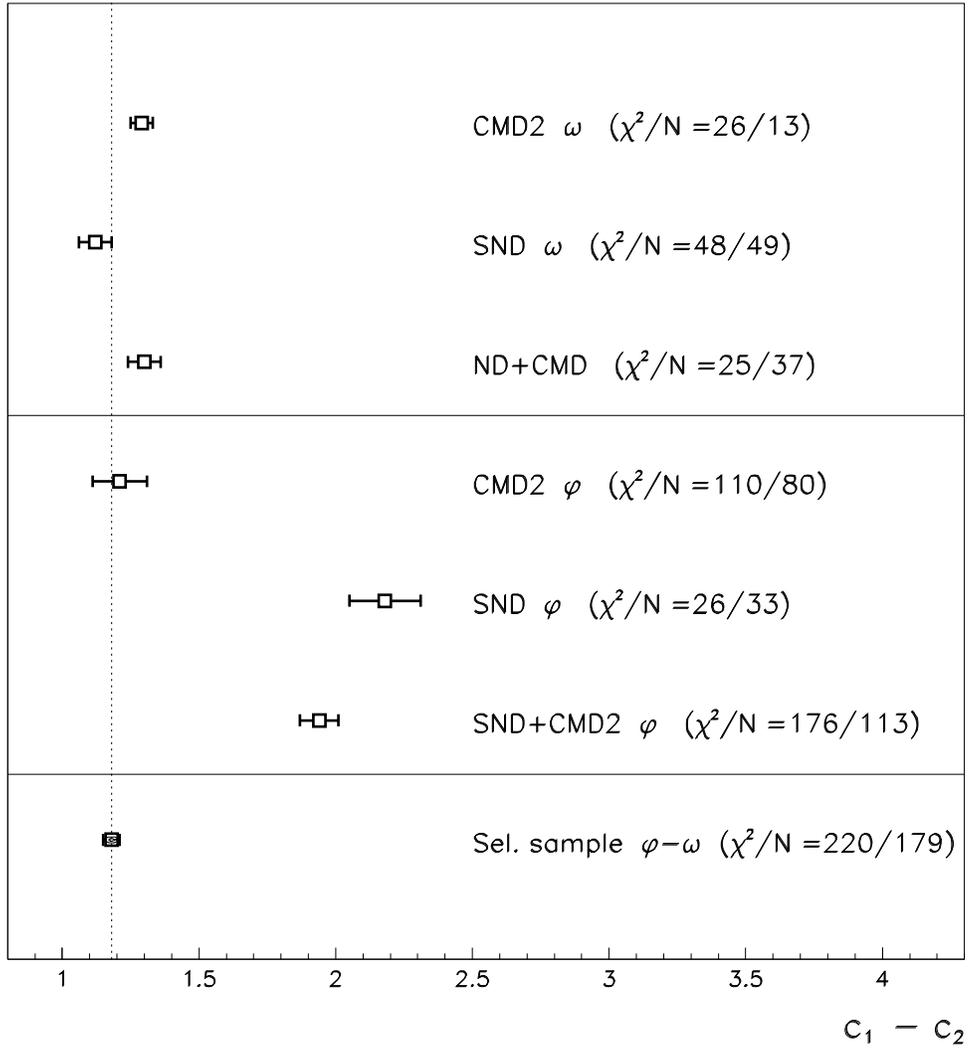}}
\end{center}
\end{minipage}
\begin{center}
\vspace{-0.3cm}
\caption{\label{Fig:c1Mc2}
$c_1-c_2$ values returned by fits. CMD2 $\omg$ denotes the fit result of the data from
\cite{CMD2-1995corr},
SND $\omg$ those from \cite{SND3pionLow}, ND+CMD the fit result to the
 merged data from  \cite{ND3pion-1991} and
\cite{CMD3pion-1989},  CMD2 $\phi$ indicates that only the merged data from  
\cite{CMD2KKb-1,CMD2-1998,CMD2-2006} have been used in the fit,  SND $\phi$ 
corresponds to the fit of the data from \cite{SND3pionHigh}
 and  SND+CMD2 $\phi$ provides the (simultaneous) fit result of 
\cite{CMD2KKb-1,CMD2-1998,CMD2-2006,SND3pionHigh}. Finally, the last line
shows the result for the selected data consisting of the sample reported in
\cite{ND3pion-1991,CMD3pion-1989,SND3pionLow,CMD2-1995corr,CMD2KKb-1,CMD2-1998,CMD2-2006}.
The vertical dotted line serves to show how the fits perform the averaging.
}
\end{center}
\end{figure}

\begin{figure}[!ht]
\begin{minipage}{\textwidth}
\begin{center}
\resizebox{\textwidth}{!}
{\includegraphics*{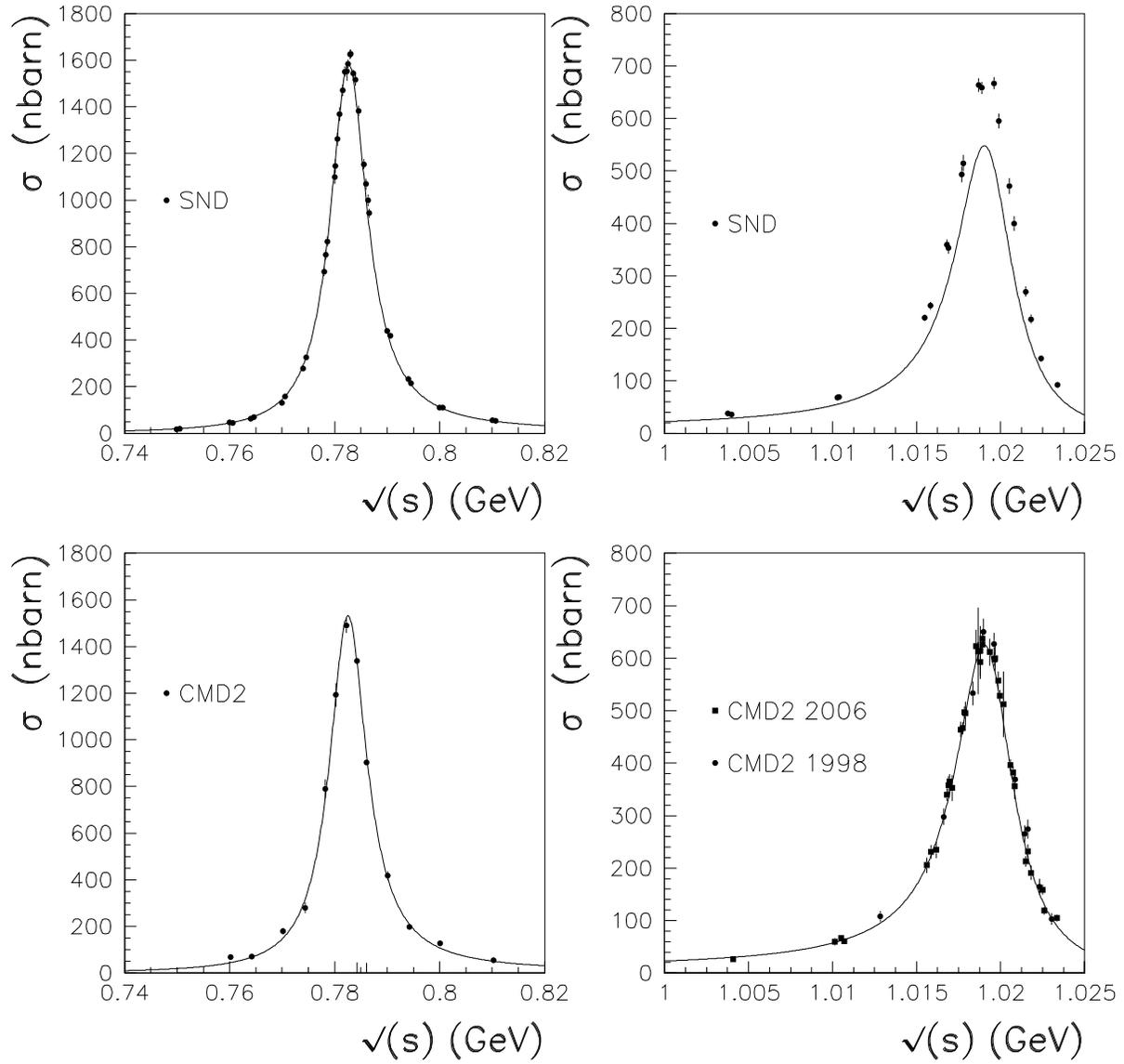}}
\end{center}
\end{minipage}
\begin{center}
\vspace{-0.3cm}
\caption{\label{Fig:omgphifit}
Simultaneous fit of the $e^+e^- \ra \pi^+ \pi^- \pi^0$ data in the
$\omg$ and $\phi$ regions. Top figures show the case
for the merged data from \cite{SND3pionLow,SND3pionHigh}.
Bottom  figures 
display the fit results for CMD--2 data
from \cite{CMD2-1995corr,CMD2-1998,CMD2-2006}.
}
\end{center}
\end{figure}

\begin{figure}[!ht]
\begin{minipage}{\textwidth}
\begin{center}
\resizebox{\textwidth}{!}
{\includegraphics*{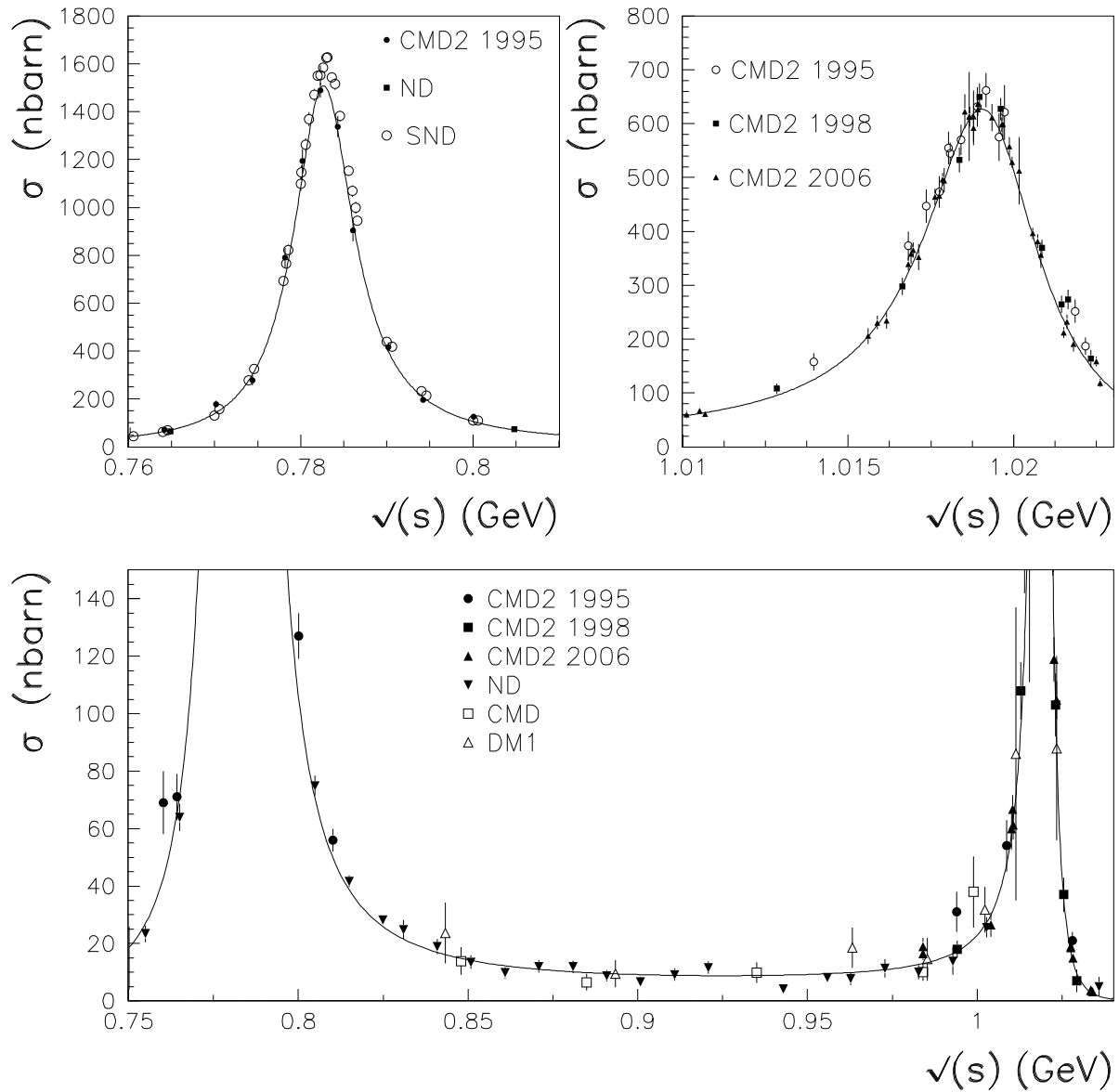}}
\end{center}
\end{minipage}
\begin{center}
\vspace{-0.3cm}
\caption{\label{Fig:final3p}
Global fit of the $e^+e^- \ra \pi^+ \pi^- \pi^0$ data . 
Top left enhances the $\omg$ region, top right the $\phi$ region.
The data superimposed are all fitted. Bottom plot shows
the intermediate region; all plotted data are included in the fit procedure,
except for the DM1 data set. The particular data sets used are described in the main 
text and in the captions to previous Figures.
}
\end{center}
\end{figure}

One has pushed a little further the analysis by two more series of fits~:
\begin{itemize}
 \item One has simultaneously submitted to fit the CMD--2 and  SND data but
 only in the $\phi$  region. One gets the result shown in Figure \ref{Fig:commonfit3p_phi}.
 The fit might look reasonable ($\chi^2/npoints=176/113$) and returns 
 $c_1-c_2= 1.94\pm0.07$, close to the SND value, as can be seen from Figure \ref{Fig:c1Mc2}.
 \item One has submitted $separately$ to fit the CMD--2 data
 and the SND ones but  $simultaneously$ in the  $\omg$  and $\phi$  regions. 
 The CMD--2 data return   $\chi^2/npoints=136/93$ with $c_1-c_2=1.31\pm 0.04$,
 while the SND data return $\chi^2/npoints=102/82$ with $c_1-c_2=1.23\pm 0.06$.
 Figure \ref{Fig:omgphifit} displays the corresponding best fit curves with data 
 superimposed\footnote{Here also, one may wonder that the top right Figure 
 corresponds to a quite reasonable fit quality. We thus remind once more that, for
 all figures shown, the effects of correlated uncertainties is not -- cannot be -- shown.
 In the case of SND, this is larger than 5\%. Along the same lines, one should
 mention that the errors plotted are always the quadratic sum of  statistical
 and uncorrelated systematic uncertainties. }. Even if the $\chi^2/npoints$
 and the fit value for $c_1-c_2$ are reasonable, top right Figure \ref{Fig:omgphifit}
leads us to avoid using the SND   $\phi$  region data\footnote{In this case, the 
so--called "diagonal" part of the $\chi^2$ at minimum is larger than 1100.
}.
\end{itemize}

From this series of fit, one can conclude that it is possible to fit simultaneously
the CMD--2 and SND data in the $\phi$ region and get a reasonable solution. However,
mixing the $\omg$  and $\phi$  regions returns, in the case of SND, an unacceptable solution, 
even if the $\chi^2/npoints$ may look reasonable.

Therefore, one is led to include in the set of data samples submitted 
to the global fit all 3--pion data referred to above,  except for the SND $\phi$ region data set.
The corresponding fit has been performed and is shown in Figure \ref{Fig:final3p} with 
$c_1-c_2=1.18\pm 0.03$; the 3--pion
data contribute to the global fit with $\chi^2/npoints=220/179$. The result shown at the last line
in Figure \ref{Fig:c1Mc2} shows that the global fit performs, as expected, 
a good (fitted) average of $c_1-c_2$. This also indicates that the data sets considered
are statistically consistent with each other.

\section{Analysis of the $K \overline{K}$ Annihilation Data}
\label{KKbar}
\indent \indent
As reminded in subsection \ref{KKbData} above, several data samples 
are available collected by the CMD--2 and SND Collaborations on VEPP--2M at Novosibirsk.
The CMD--2 data are extracted from \cite{CMD2KKb-1,CMD2KKb-2,CMD2KKb-3}
and the corresponding SND data from  \cite{SNDKKb}. The quoted systematics
are treated as a scale uncertainty and dealt with as explained in Subsection \ref{fitproc}.

The published data being cross sections, the fitting function is~:
\be
\sigma(s) =\displaystyle
\frac{8 \pi \alpha_{em}^2}{3 s^{5/2}} q_K^3  |F_K^e(s)|^2~~~{\rm with}~~~
q_K=\frac{1}{2}\sqrt{s-4 m_K^2}~~~, ~~(K=K^\pm,~K^0/\overline{K}^0)
\label{eq78}
\ee
for each of the 2--kaon annihilation channels; the kaon form factors $F_K^e(s)$
are given by Eqs. (\ref{eq49}). Both  cross sections are corrected for
the intermediate photon dressing. Moreover, for the charged kaon channel, the additional
Coulomb factor \cite{Gourdin,BGPter}  $Z(s)$, reminded in Eq. (\ref{eq53}), is 
understood
and is not "renormalized" as in \cite{SNDKKb,CMD2KKb} with  $Z(m_\phi^2)$. 

\subsection{Fitting the $K \overline{K}$ Data}
\indent \indent
In order to perform this analysis, we have done a first series of fits
using separately the CMD--2 neutral and charged $K\overline{K}$ channels
and the corresponding data from SND. In order to avoid $\phi$ peak information
not following from the $K\overline{K}$ data, we have decided to remove the 
data from the $\pi^0 \gamma$ and $\eta \gamma$ annihilation channels from the fit procedure.
However, anticipating on our final results, we have included the three data
sets from ALEPH, Belle and CLEO referred to above.


\begin{table}[ph]
\begin{center}
\begin{tabular}{|| c  || c  | c || c || c ||}
\hline
\hline
\hhhc
\hhhd ~~~~ & \hhhv $\chi^2_0/N$   &  \hhhv $\chi^2_c/N$ 
&  \hhhv $\Delta_A$ (\%)&  \hhhv Fit Prob (\%)\\
\hline
\hline
 \hhhv $K^0\overline{K}^0$ +
 $K^+ K^-$  (SND ~ stand--alone)    & 60.10/60 & 56.54/26 &  $8.54 \pm 1.93$&  33.7\\
\hline
 \hhhv  $K^0\overline{K}^0$ +
 $K^+ K^-$ (CMD--2 stand--alone) & 59.30/59 & 29.00/36&  $5.98 \pm 0.86$&  85.8\\
 \hline
 \hhhv $K^0\overline{K}^0$ (SND \& CMD--2) & 115.68/119&  -- & $5.51 \pm 3.21$  & 81.8\\
 \hline
 \hhhv $K^0\overline{K}^0$ +
 $K^+ K^-$ (SND \& CMD--2) & 119.83/119&  88.09/62 & $6.29 \pm 0.80$& 40.4\\
\hline
\hline
  \hhht $K^0\overline{K}^0$ (SND \& CMD--2) & ~~& ~~& ~~& ~~\\
  \hhhv  + $K^+ K^-$ (CMD--2)&118.54/119& 29.27/36&  $6.09 \pm 0.79$& 80.8\\
\hline
\end{tabular}
\end{center}
\caption {
\label{T2}
Fit quality of  the $K^+K^-$ and $K^0 \overline{K}^0$ data. Beside
the additional data sample (see text), each line in the first column
tells which $K\overline{K}$ data samples have been included in the fit procedure.
$\chi^2_0$ is the $\chi^2$ value for $K^0 \overline{K}^0$ data,
 $\chi^2_c$ is the corresponding information for $K^+ K^-$ data. The
 $N$'s are the respective numbers of data points. The last data column provides
 the global fit probability for each case.}
\end{table}

Therefore, the additional data sample is composed of all $e^+e^- \ra \pi^+ \pi^+$
data, all $\tau^\pm \ra \pi^\pm \pi^0 \nu$ data and 18 partial width decays (all
$VP\gamma$ and $ P \gamma \gamma$ modes and also the three leptonic decays
$(\rho/\omg/\phi) \ra e^+e^-$ modes). None among these pieces of information has
any direct influence on the description of $e^+e^- \ra K\overline{K}$, even
through the $ \phi$ mass and width parameters which are, thus, solely determined
by the $K\overline{K}$  data.
 
\begin{figure}[!ht]
\begin{minipage}{\textwidth}
\begin{center}
\resizebox{\textwidth}{!}
{\includegraphics*{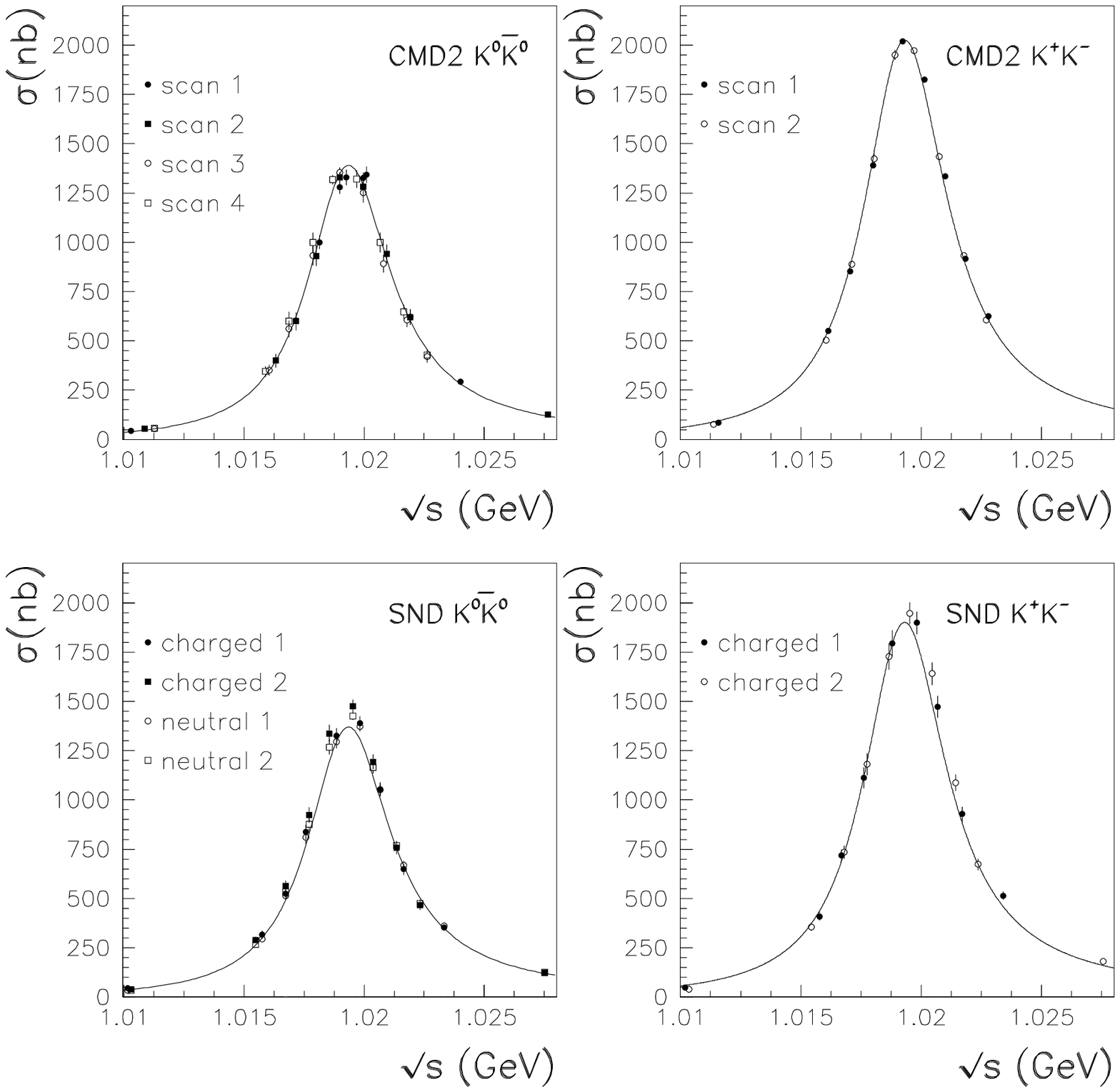}}
\end{center}
\end{minipage}
\begin{center}
\vspace{-0.3cm}
\caption{\label{Fig:check1_kk}
Fit of the $e^+e^- \ra K \overline{K}$ data.
Left side are  $K^0 \overline{K}^0$, right side $K^+ K^-$
}
\end{center}
\end{figure}

The results are shown in Figure \ref{Fig:check1_kk}, left side for the 
$K^0\overline{K}^0$ data and right side for $K^+ K^-$. One observes a good description of 
the $K^0\overline{K}^0$ data for each of the samples provided by the 
SND or CMD--2 Collaborations. The picture is quite different for the  $K^+K^-$ data; the CMD--2
data sample is well fitted, while the SND sample is poorly fitted.
Additional information for these peculiar fits is displayed in the first 
two lines of Table \ref{T2}. One can see there, that  the value for $\chi^2/N$
associated with the $K^0\overline{K}^0$ data are the same for both data samples,
while they differ significantly for the corresponding $K^+ K^-$ data samples.

Fitting simultaneously both CMD--2 and SND $K^0\overline{K}^0$ data samples
only, returns the same  $\chi^2$ information, illustrating that the corresponding
data samples are perfectly consistent with each other. Simultaneous fits
of all $K\overline{K}$ data confirm this property (see third line in 
Table \ref{T2}). Interestingly, the  $\chi^2$'s at best fit in the
third and fourth lines practically coincide with the sum of the corresponding
information in  the first two lines of the same Table. 
This illustrates that the so--called additional data set
sharply constrain the $K\overline{K}$ cross sections. Moreover, 
in view of the fit results for CMD--2 data, one can consider that 
the constraints are well fulfilled by data, giving a strong support
to our modelling.

\begin{figure}[!ht]
\begin{minipage}{\textwidth}
\begin{center}
\resizebox{\textwidth}{!}
{\includegraphics*{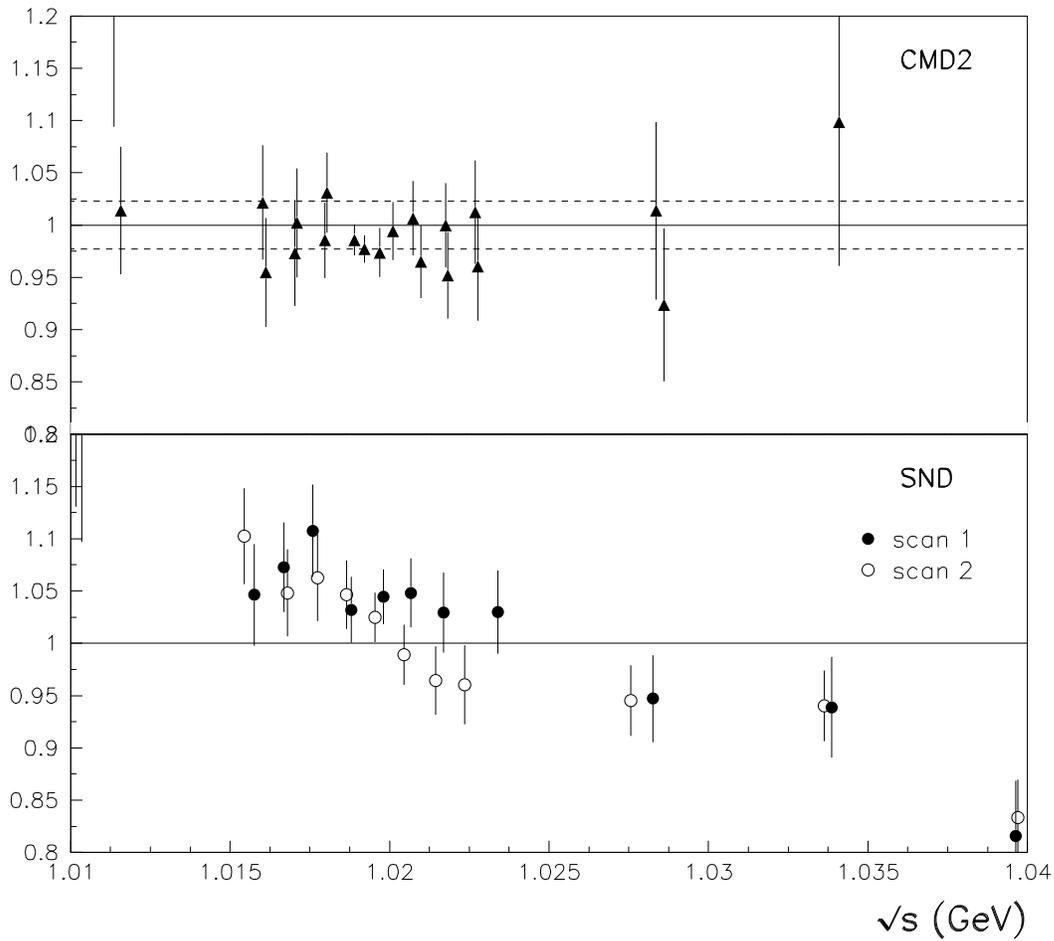}}
\end{center}
\end{minipage}
\begin{center}
\vspace{-0.3cm}
\caption{\label{Fig:check2_kk}
Ratio of the $e^+e^- \ra K^0 \overline{K}^0$ and 
$e^+e^- \ra K^+ K^-$ cross sections normalized to the
model ratio. Top panel displays the case for CMD--2 data, Bottom
panel 
those for SND data. In the top panel, the residual experimental
systematics band (2.3\%) is figured by dashed lines. Correlated
systematics between charged and neutral modes are expected to cancel
out in the experimental ratios.
}
\end{center}
\end{figure}

The ratio of cross sections 
$\sigma(e^+e^- \ra K^0\overline{K}^0)/\sigma(e^+e^- \ra K^+K^-)$ is observed to
provide a valuable piece of information, as it allows to magnify
the effects mentioned just above. This is shown in Figure \ref{Fig:check2_kk}, where
the data for this ratio are plotted normalized to the ratio of cross
sections as coming out from our fits. The data ratio plotted in
the top Figure \ref{Fig:check2_kk} is derived from the information
given in \cite{CMD2KKb} and one can estimate its uncertainty to $\simeq 2.3 \div 2.4$ \%.

The CMD--2 data points normalized to the fit expectations bin per bin
is perfectly consistent with 1 over the whole $s$ region covered by the $\phi$ 
resonance.  The dotted lines in top Figure \ref{Fig:check2_kk} represent the 
experimental scale uncertainty
and do not take into account the uncertainties on the fitting functions.
This also illustrates that our modified Breit--Wigner lineshape 
is very well accepted by the data. 

In contrast,
the SND data exhibit a behavior reasonably well averaged by the fit function ratio;
however,  it does not look consistent with flatness -- at least as well as for CMD--2 data.

It follows from these considerations that the largest 
self--consistent data set for the $K\overline{K}$ channel is made
by merging all CMD--2 data and the $K^0\overline{K}^0$ data provided by SND (see last line in Table \ref{T2}).

As a matter of information, beside  getting an 
appropriate description of both   $e^+e^- \ra K^0\overline{K}^0$ 
and $e^+e^- \ra K^+K^-$ cross sections, it is worth remarking that
the radiative partial widths included in the fitted data set are
also well accounted for. For instance, including also 
the $e^+e^- \ra (\pi^0/\eta) \gamma $ cross sections in the fitted data set, 
the remaining set of 10 radiative decays yields a quite remarkable $\chi^2/n=6.5/10$,
with estimated $\Gamma(\pi^0\ra\gamma\gamma)$, $\Gamma(\eta \ra\gamma\gamma)$ 
and $\Gamma(\eta^\prime \ra\gamma\gamma)$ at respectively $0.27 \sigma$, $1.77 \sigma$
and $0.23 \sigma$ from their accepted values \cite{RPP2010}. As the corresponding
couplings are strongly affected -- especially $g_{\pi^0\gamma\gamma}$ -- by 
$\Delta_A$ (see Eqs. (\ref{eq58})), we consider that physics validates our model.

\subsection{The HLS Solution of  $\phi \ra K \overline{K} $ Puzzle} 
\label{phiKKbar}                      
\indent \indent The partial width decays $\phi \ra K \overline{K}$ are 
defined by~:
\be
\displaystyle
\Gamma(\phi \ra K \overline{K})=\frac{q_K^3}{6 \pi} |g_{\phi K \overline{K}}|^2~~~,~~
(q_K=\frac{1}{2}\sqrt{m_\phi^2-4m_K^2})
\label{eq79}
\ee

Therefore, one has~:
\be
\displaystyle
\frac{\Gamma(\phi \ra K^+K^-)}{\Gamma(\phi \ra K^0\overline{K}^0 )}=
\frac{{\rm Br}(\phi \ra K^+K^-)}{{\rm Br}(\phi \ra K^0\overline{K}^0 )}
= R \left|\frac{g_{\phi K^+K^- }}{g_{\phi K^0 \overline{K}^0}} 
\right|^2 Z(m_\phi^2) \simeq R  Z(m_\phi^2) (1-2\Delta_A)
\label{eq80}
\ee
where $R=1.528$ originates from  the ratio of momenta and the Coulomb factor
computed at the $\phi$ peak is $Z(m_\phi^2)=1.049$. The ratio of couplings
has been given in Eq. (\ref{eq54}). Therefore, using $\Delta_A$ from the last line 
in Table \ref{T2}, one gets ~:

\be
\displaystyle
\frac{{\rm Br}(\phi \ra K^+K^-)}{{\rm Br}(\phi \ra K^0\overline{K}^0 )}=1.41 \pm 0.03
\label{eq81}
\ee

The same ratio can be computed from information
given by CMD--2 in a recent paper \cite{CMD2KKb} and amounts\footnote{The uncertainty might be
somewhat overestimated, as one has assumed independent the  errors
for ${\rm Br}(\phi \ra K^+K^-)$ and ${\rm Br}(\phi \ra K^0\overline{K}^0)$.
} to $1.47 \pm 0.04$. The difference between the CMD2 estimate and ours
amounts to about $2\sigma$. Our final result, obtained by using the largest possible 
ensemble of data sets, provides $\Delta_A=(6.34 \pm0.70)~10^{-2}$ and then the ratio 
of branching ratios becomes $1.40 \pm 0.02$.

Therefore, the HLS model,  equipped with the (BKY) direct isospin symmetry breaking 
mechanism, provides a solution to the long--standing puzzle concerning the 
$\phi \ra K \overline{K}$ decays as thoroughly analyzed in \cite{BGPter} and
more recently discussed in \cite{SNDKKb}. In our approach, the mechanism  responsible for 
this is, {\it in fine}, the kaon field renormalization which should be performed 
within the HLS model once isospin symmetry breaking is performed \`a la BKY--BOC.
Indeed, as the neutral and charged kaon field renormalization factors play in
opposite directions (see Eq. (\ref{eq18})), they pile up in the ratio.

The relatively large value found for $\Delta_A$ indicates that several sources
contributes to the BKY breaking of isospin symmetry. The contribution to $\Delta_A$
due to the light quark mass mass difference \cite{GassLeutw}  ($\simeq  1 \%$) is
certainly not the single source and others -- like electromagnetic corrections --
are certainly absorbed within the numerical value for $\Delta_A$. Moreover, it is also likely that 
different corrections at the $V K^+ K^-$ and $V K^0 \overline{K}^0$ vertices may influence the 
fit value for $\Delta_A$. Being global, the BKY breaking mechanism
cannot allow to disentangle the various contributions to $\Delta_A$ which
share a common order of magnitude (each at the percent level). The situation
is quite different from the breaking of SU(3) symmetry which is widely dominant
numerically and can motivatedly be compared to ChPT expectations \cite{WZWChPT}. 

\section{Analysis of the $\tau$ Decay Data}
\label{fit_tau}
\indent \indent Using $F_\pi^\tau(s) $, the pion form factor in  the decay of the $\tau$ lepton
(see Subsection \ref{pionFF}), the partial width of the two--pion decay is 
given by Eq. (\ref{eq73}).
On the other hand, the quantity which encompasses the full experimental information
in this field is Eq.(\ref{eq72})~:
 $$ H(s)=
\displaystyle \frac{1}{\Gamma_{\tau}} \frac{d\Gamma_{\pi \pi}(s)}{ds} S_{EW} G_{EM}(s)= 
{\cal B}_{\pi \pi} \frac{1}{N}\frac{dN(s)}{ds} 
$$
as, indeed, the lineshape and the absolute magnitude of each experimental spectrum
are merged together. The full width $\Gamma_{\tau}$ is taken from the RPP \cite{RPP2010}.
The last two factors in the middle expression above account for isospin symmetry
breaking effects specific of the $\tau$ decay~: $S_{EW}=1.0235$ for short range corrections
\cite{Marciano}, $G_{EM}(s)$ for long range corrections \cite{Cirigliano1,Cirigliano2,Cirigliano3}.

In former studies, it was shown that the lineshape alone was
perfectly consistent with annihilation data \cite{taupaper,ExtMod2}. However, if one
also takes into account the absolute magnitude -- represented by the branching ratio
${\cal B}_{\pi \pi}$ in
the formula reminded just above -- the agreement is poor. In order to reach a satisfactory
description of the data, Ref. \cite{ExtMod2} introduced a mass difference $\delta m^2$
and a coupling difference $\delta g$ between the neutral and charged $\rho$ mesons, which
underlays all reported stand--alone fits to $\tau$ spectra \cite{DavierHoecker}. 
However, additional scale factors were needed and their fitted  values were found
consistent with the reported scale uncertainties \cite{Cleo,Belle,Aleph}.

However, the present study, as reflected by Table \ref{T1} above, has clearly demonstrated
that isospin breaking of the HLS model does not necessarily result in non--vanishing
  $\delta m^2$ and $\delta g$ at leading order\footnote{Our present results 
as well as formers \cite{ExtMod2} tend to indicate that an electromagnetic correction
to the $\rho$ mass does not give a significant effect (see Footnote \ref{cottingham}).}. 
As emphasized above, the BKY--BOC breaking scheme instead leads to a difference
between the universal vector coupling ($g$) as it comes in the anomalous sector and in the
non--anomalous sector of the HLS Lagrangian ($g(1+\Sigma_V)$). We prove, here, that this
provides a much better account of all data than only assuming some mass and width differences 
supplemented with some residual rescaling. Stated otherwise, it is because Direct
Isospin Breaking acts differently in the anomalous and non--anomalous sectors that
the model yields an almost perfect description for all data, without any need for
some additional rescaling.  In this mechanism,
the single sensible difference between the pion form factor in $e^+e^-$ annihilations
and in $\tau$ decays resides in the difference between the transition 
amplitudes $\gamma-V$ and $W-V$.

\begin{table}[ph] 
\hspace{-1.cm}
\begin{tabular}{|| c  || c  | c  || c |c |c ||}
\hline
\hline
\hhhc
\hhhd  $\chi^2/N$ & \hhhv ($\delta m^2$, $\delta g$,~$c_3=c_4$)   &\multicolumn{4}{|c|}{  \hhhv  Statistical Information}\\
\hhhd ~~~~ & \hhhv  \cite{ExtMod2}   &  \hhhv excl. ${K \overline{K}}$ &  \hhhv  excl. $\pi^+ \pi^- \pi^0$&  \hhhv  $A$&  \hhhv  $B$\\
\hline
\hline
 \hhhv Decays    			& 16.20/9    	& 5.53/10	& 6.13/10 	& 11.36/10 	& 5.94/10 	\\
\hline
 \hhhv New Timelike $\pi^+ \pi^- $ 	& 126.47/127 	& 119.73/127 	& 130.33/127	& 127.50/127 	& 129.65/127	\\
\hline
 \hhhv Old Timelike $\pi^+ \pi^- $ 	& 60.45/82   	&51.64/82	& 56.36/82	& 56.09/82 	& 56.60/82	\\
\hline
 \hhhv $\pi^0 \gamma$ 			& 66.07/86   	&66.84/86 	& 61.19/86 	& 67.21/86 	& 66.93/86	\\
\hline
 \hhhv $\eta \gamma$ 			& 135.78/182   	&128.89/182	& 122.64/182	& 122.62/182	&121.37/182	\\
\hline
 \hhhv $\pi^+ \pi^- \pi^0$ 		& {\bf 139.44/126}   	& {\bf 200.92/179}   	& -- 	
& {\bf 230.98/179}&{\bf 105.91/99}\\
\hline
 \hhhv $K^+ K^- $ 			& --   		& --   	  	& 29.93/36	& 35.16/36	&29.85/36	\\
\hline
 \hhhv $K^0 \overline{K}^0 $ 		& --   		& --   		& 120.07/119	& 117.94/119	&119.99/119	\\
\hline
 \hhhv ALEPH 				& 36.51/(37+1)   & 21.25/37 	& 15.92/37	& 16.80/37	&16.16/37	\\
\hline
 \hhhv Belle 				& 28.29/(19+1)   & 27.02/19 	& 34.19/19 	& 32.22/19	&33.62/19	\\
\hline
 \hhhv CLEO				& 39.46/(29+1)   & 35.12/29 	& 35.86/29 	& 36.09/29	&36.03/29	\\
\hline
\hline
  \hhht $\chi^2/dof$  & 648.68/680&656.93/726  &612.63/703&853.98/881&722.05/801\\
  \hhhv Global Fit Probability & 80.1\%&  96.8\% &  99.4\%&  73.7\%& 97.9\%\\
\hline
\end{tabular}
\caption {
\label{T3}
Comparison of the fit qualities between the fit results of the model as it was in
\cite{ExtMod2} (second data column) and as it is now (third data column).  $K  \overline{K}$
data were not submitted to fit in \cite{ExtMod2}. The '+1' added to the 
number of data points for $\tau$ data stands for the experimentally given r.m.s. affecting
the (fitted) global scale. The 3--pion data set information is displayed boldface in order
to show the difference in the fit data set~: In the second data column, 
 the 3--pion data set from SND \cite{SND3pionLow} has been (newly) introduced and in the last data column
 only the 3--pion data sets collected below the $\phi$ region are considered.
}
\end{table}

\begin{figure}[!ht]
\begin{minipage}{\textwidth}
\begin{center}
\resizebox{\textwidth}{!}
{\includegraphics*{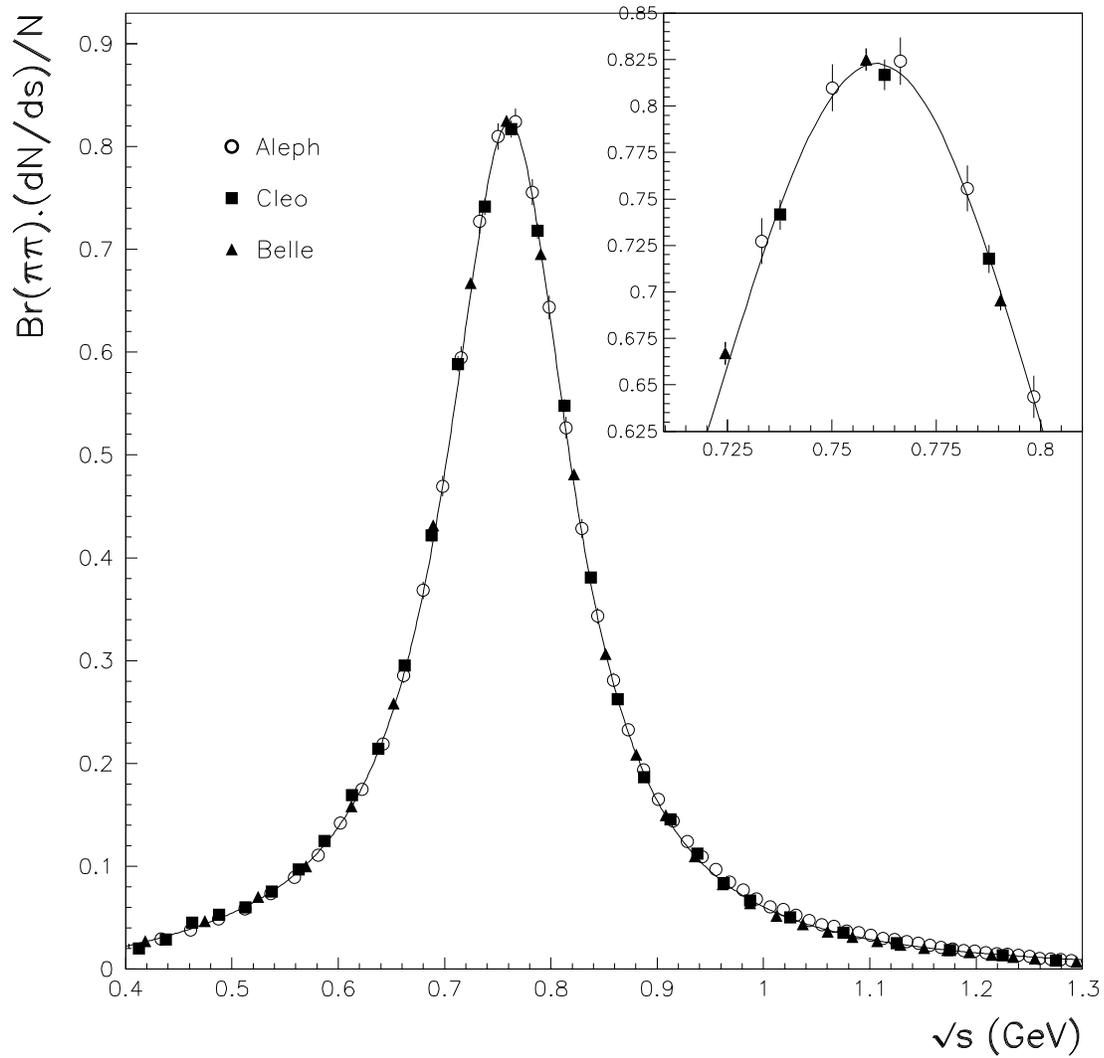}}
\end{center}
\end{minipage}
\begin{center}
\vspace{-0.3cm}
\caption{\label{Fig:FFtau}
Global Fit of the dipion spectrum in the decay of the $\tau$ lepton.
The data points are those from ALEPH \cite{Aleph}, Belle \cite{Belle}
and CLEO \cite{Cleo}. The inset magnifies the $\rho$ peak region. }
\end{center}
\end{figure}

\begin{figure}[!ht]
\begin{minipage}{\textwidth}
\begin{center}
\resizebox{\textwidth}{!}
{\includegraphics*{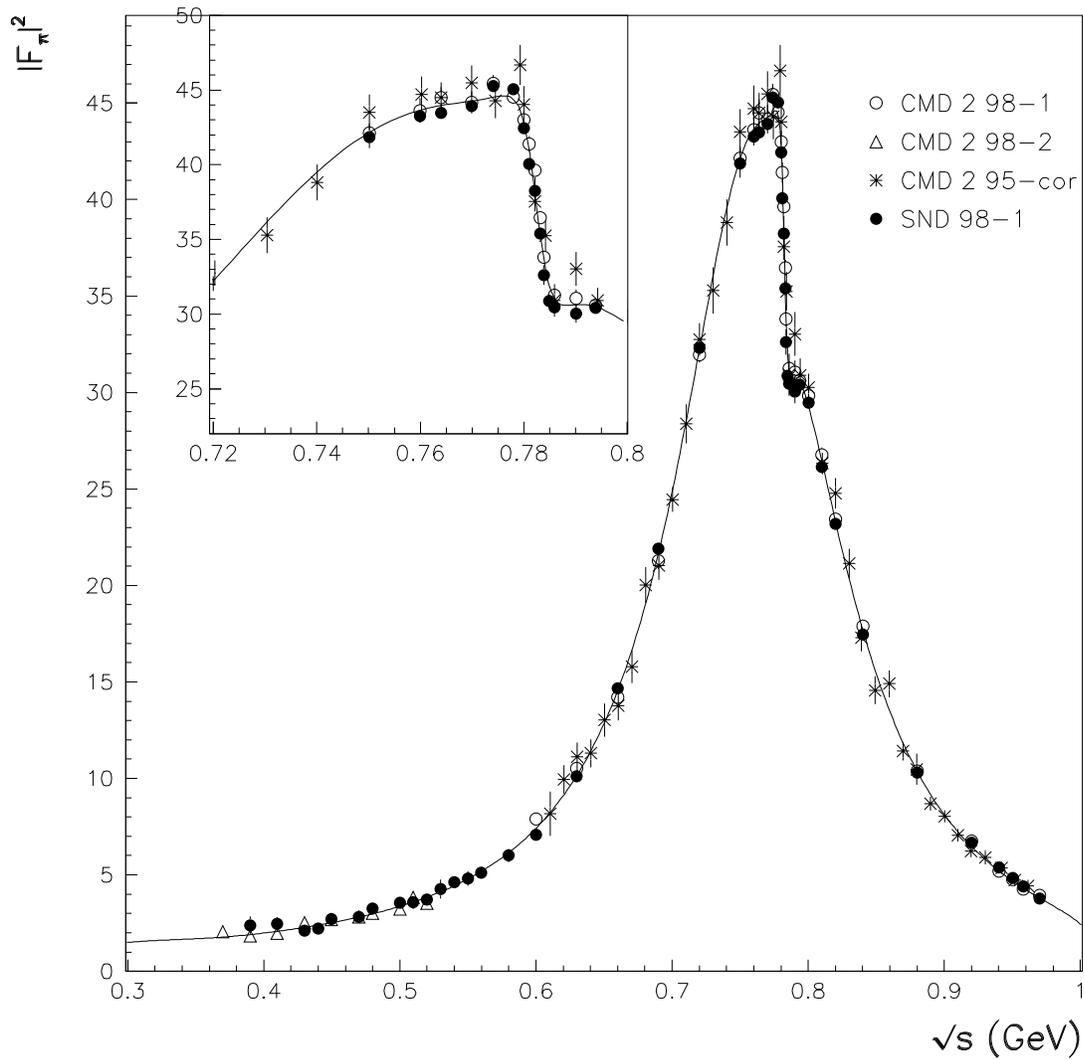}}
\end{center}
\end{minipage}
\begin{center}
\vspace{-0.3cm}
\caption{\label{Fig:FFee}
Global Fit of the pion form factor squared in $e^+e^-$ annihilations.
The data points are those from CMD--2 \cite{CMD2-1995corr,CMD2-1998-1,CMD2-1998-2}
and SND \cite{SND-1998}. One has not plotted the so--called "old timelike" data
also (mostly) collected at Novosibirsk. The inset magnifies the $\rho$ peak region
and the behavior at the $\rho-\omg$ interference region. }
\end{center}
\end{figure}

Figure \ref{Fig:FFtau} shows the global fit result  for the
function $H(s)$ defined just above together with the data points 
from ALEPH \cite{Aleph}, Belle \cite{Belle} and CLEO \cite{Cleo} Collaborations
\footnote{When dealing with $\tau$ plots, the error bars represent 
the diagonal errors, {\it i.e.} no account of bin--to--bin correlations
is attempted.}.
The inset magnifies the $\rho$ peak region. One can clearly conclude to a nice
agreement between model and data, all along the fitted region -- from threshold
to 1 GeV. The corresponding 
pion form factor in  $e^+e^-$ annihilations coming out of the global fit
is represented in Figure \ref{Fig:FFee}.
These two Figures illustrate that the simultaneous description of $e^+e^-$ and $\tau$ 
data allowed by the model is, indeed, as successfull in both sectors.

\begin{figure}[!ht]
\begin{minipage}{\textwidth}
\begin{center}
\resizebox{\textwidth}{!}
{\includegraphics*{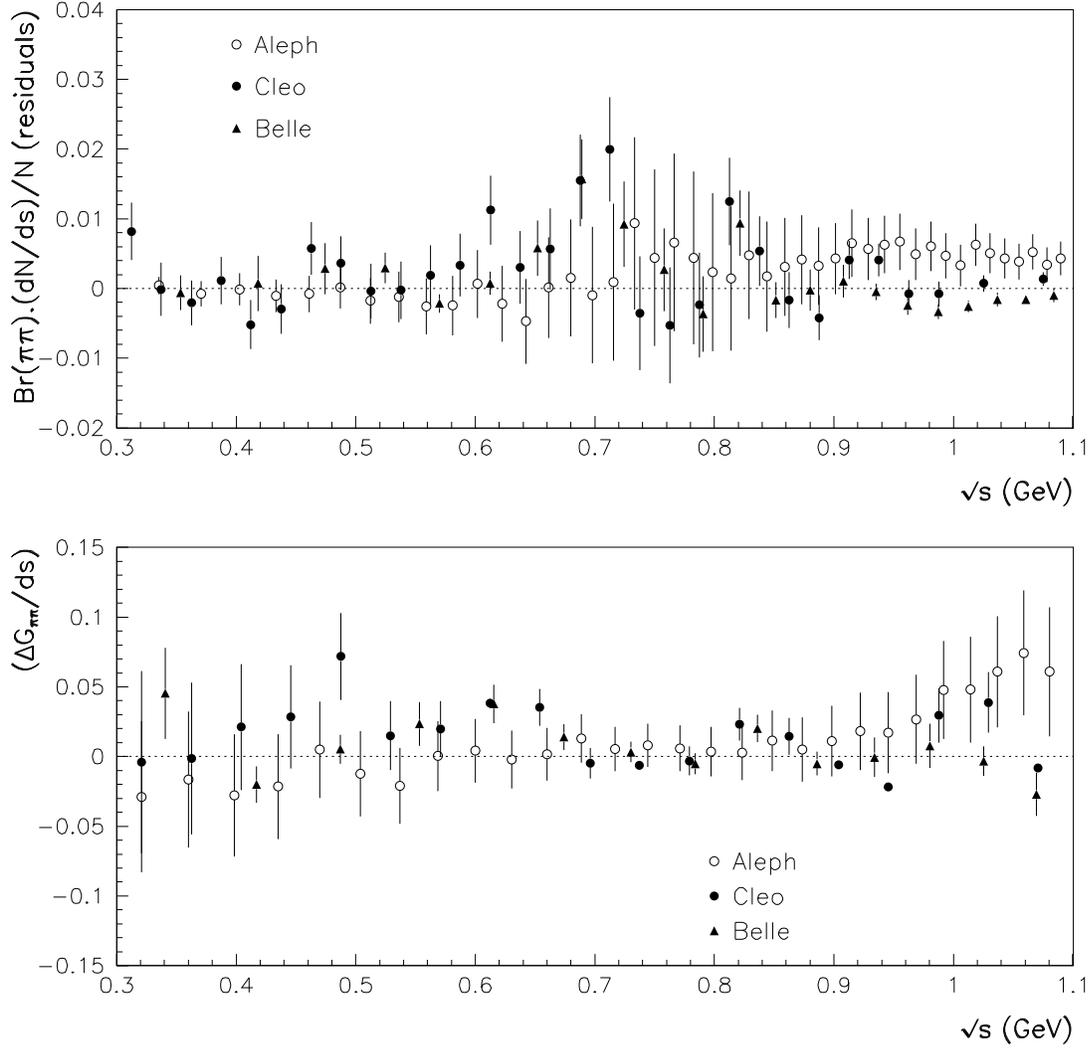}}
\end{center}
\end{minipage}
\begin{center}
\vspace{-0.3cm}
\caption{\label{Fig:ResTau}
Global Fit of the function $H(s)={\cal B}_{\pi \pi}/N dN/ds$ in $\tau$ decays.
Top Figure shows the residuals as a function of $s$; downmost Figure shows the
function $(H_{fit}(s) -H_{data}(s))/H_{fit}(s)$. The fitted region extends from
threshold to 1.0 GeV/c, {\it i.e.} over the region where the behavior of the data 
sets from ALEPH \cite{Aleph}, Belle \cite{Belle}and CLEO \cite{Cleo} reach some
 agreement.}
\end{center}
\end{figure}

Figure \ref{Fig:ResTau} shows in two different manners the $\tau$ residual
behavior. Top Figure \ref{Fig:ResTau} displays the usual residuals for the function $H(s)$,
while downmost Figure  \ref{Fig:ResTau}
represents $(H_{model}(s)-H_{data}(s))/H_{model}(s)$.  These can be compared 
with respectively
Figure 3 and Figure 4 from \cite{ExtMod2} where the ($\delta m^2$, $\delta g$) 
parametrization of isospin breaking was used. The comparison clearly indicates
that the present model better performs for all $\tau$ data sets and, especially,  
for the ALEPH \cite{Aleph} data.

In order to allow for a deeper comparison  with the previous release \cite{ExtMod2} 
of the present model, we reproduce  in Table \ref{T3} (first data column) the fit results reported 
in \cite{ExtMod2} together with our new fit results under various conditions.

The second data column in Table \ref{T3} is derived excluding the $K\overline{K} $ data sets 
in order to be as close as possible to \cite{ExtMod2}.
One observes, for almost all data sets, better fit results than in the former release
of our model \cite{ExtMod2} . There is no effect in introducing the 3--pion data set
from SND \cite{SND3pionLow} (covering the $\omg$ region) 
as the  $\chi^2_{3\pi}/dof= 1.11$ is unchanged.
It is also worth noting that the partial width for $\eta \ra \gamma \gamma$ 
is found at $0.43 \sigma$ from its accepted value \cite{RPP2010}; the distance 
is $0.11 \sigma$ for $\eta^\prime \ra \gamma \gamma$  and  $0.47 \sigma$ only
for the newly introduced  $\pi^0 \ra \gamma \gamma$ decay mode.

One may conclude therefrom that the HLS model, equipped with the mixing schemes 
provided by loops and by the direct isospin breaking procedure, provides
a fully satisfactory solution  to the $e^+e^--\tau$ puzzle, 
both in magnitude and in shape. The relatively poorer fit quality for the BELLE data
might be related with the absolute scale issue revealed by the stand--alone 
fit\footnote{The fit published by BELLE reveals a very significant improvement
if the absolute normalization of their spectrum is left free; instead of returning
an absolute scale of 1, the best fit exhibits a significant $\simeq$ 2\% shift. } 
provided by BELLE \cite{Belle}. Therefore, one can confirm  that~:
\begin{itemize}
\item The main drawback of the breaking model in  \cite{ExtMod2} was a too
tight correlation between the universal coupling in anomalous and in non--anomalous processes.
This has been cured by defining the Direct Isospin Breaking mechanism substantiated by a
highly significant value for    $\Sigma_V =(3.74 \pm 0.42)\%$. 

\item The breaking model in  \cite{DavierHoecker} may account
insufficiently for the difference between the $\rho^0-\gamma$ and $\rho^\pm-W^\pm$
transition amplitudes. 
\end{itemize}

\begin{figure}[!ht]
\begin{minipage}{\textwidth}
\begin{center}
\resizebox{\textwidth}{!}
{\includegraphics*{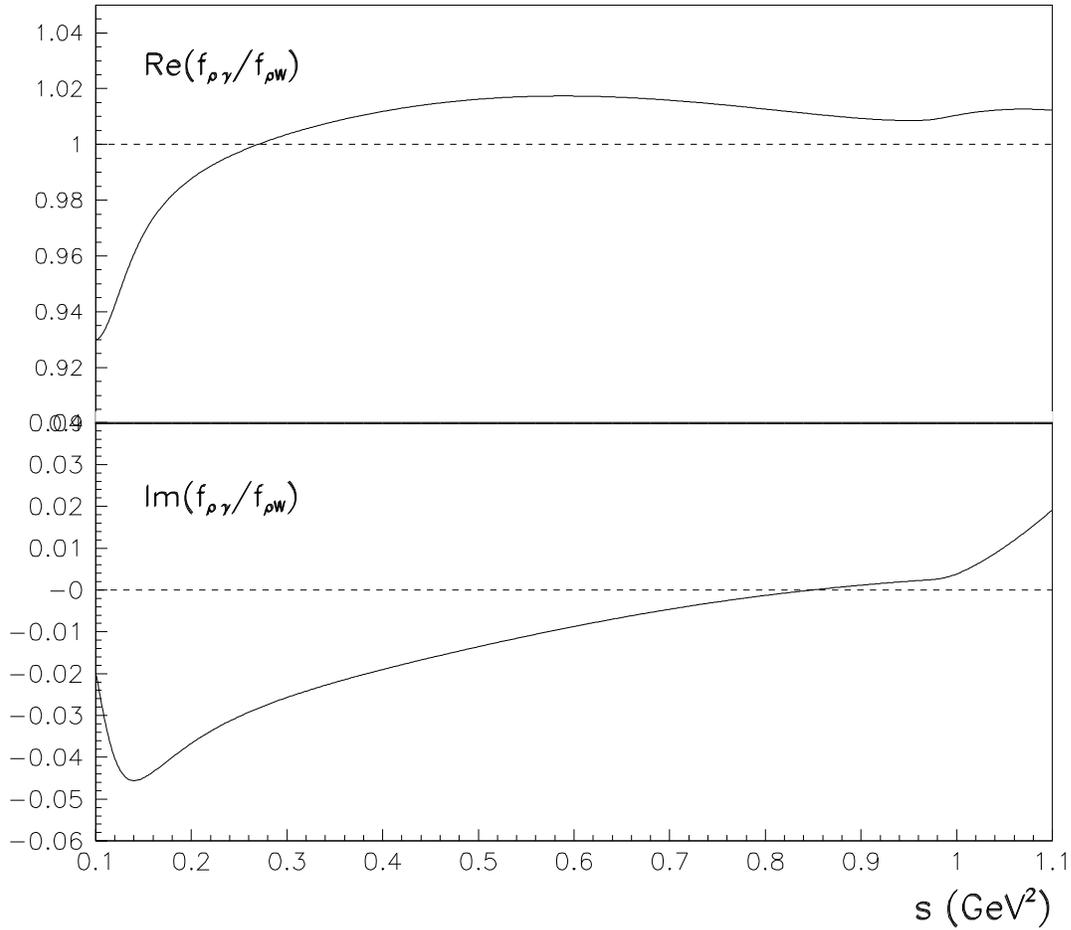}}
\end{center}
\end{minipage}
\begin{center}
\vspace{-0.3cm}
\caption{\label{Fig:Brkeetau}
Ratio of the transition amplitudes $\rho^0-\gamma$ and $\rho^\pm -W^\pm$, 
$f_{\rho \gamma}/f_{\rho W}$ following from the global fit and
neglecting loop corrections. This corresponds to the ratio shown in Table \ref{T1}
and reproduced in Section \ref{fit_tau}. 
Top Figure shows the real part as a function of $s$, bottom Figure the imaginary part.
Uncertainties due to fit parameter errors are not given; the uncertainty band for
$f_{\rho \gamma}/f_{\rho W}-1$ can estimated to a few percent.}
\end{center}
\end{figure}

Therefore,  the reported discrepancies between the pion form factor
in $e^+e^-$ annihilations and in  $\tau$ decays 
can always be attributed to an incomplete treatment  of isospin symmetry breaking.
For information, Figure \ref{Fig:Brkeetau} displays the ratio of the transition
amplitudes $f_{\rho \gamma}$ and $f_{\rho W}$ as coming from the global fit
and already given in Table \ref{T1}~:

 $$    \displaystyle \frac{f_{\rho}^\gamma}{f_{\rho}^\tau}=
 1 +\frac{h_V \Delta_V}{3}+\frac{\alpha(s)}{3}+\frac{\sqrt{2}\beta(s)}{3} z_V $$

 \vspace{0.5cm}
 We have found appropriate to provide in the third data column of Table \ref{T3}
 the results of the fit obtained keeping the $K \overline{K}$ data sets, while
 excluding all the $\pi^+ \pi^- \pi^0$ data sets. 
The  fourth data column reports on the fit quality reached using the full data set
we considered safe. This means all data sets discussed above, except for 
two SND data sets~: The $e^+e^- \ra 3 \pi$
data set collected above 970 MeV \cite{SND3pionHigh} and the 
$e^+e^- \ra K^+ K^-$ data set. These have been shown to provide either an unacceptable behavior
for the fit solution \cite{SND3pionHigh}, or a  poor $\chi^2$ \cite{SNDKKb}.
In this configuration, one fits 906 data points (including the 10 
individual decay modes) corresponding to 881 degrees of freedom. The global
fit probability is highly favorable (71\%). This configuration will be referred to in the following
as "Solution A" or "Configuration  A".

In this Solution A, one observes some tension between the $K \overline{K}$ and $\pi^+ \pi^- \pi^0$
data groups.  Indeed, comparing its content with the second data column, one observes
that the $\pi^+ \pi^- \pi^0$ data group yields a $\chi^2$ increased by 30 units.
Instead, comparing  Solution A with the third data column in Table \ref{T3}, 
one does not observe any significant degradation the fit quality of
the $K \overline{K}$ data group~:   
The $\chi^2$ for the $K^0 \overline{K}^0$ data group is improved by 2 units, while
the  $\chi^2$ for $K^+ K^-$ data group is worsened by 6 units.

As this 30 unit increase of the  $(\pi^+ \pi^- \pi^0)$ $\chi^2$ may look abnormal, we have tried
tracking its origin. This issue is clearly related with having introduced the $K \overline{K}$ 
data which influence the model description of the $\phi$ region. Therefore, we have redone fits
excluding all the $\pi^+ \pi^- \pi^0$ data data sets covering the  $\phi$ region. One obviously remarks a 
significant effect; this configuration will be named hereafter "Solution B" or "Configuration  B".

In the following, any differential effect between what has been named Solutions A and B is
examined carefully and commented. 

\section{Structure of the $\omg \ra \pi \pi$ Coupling}
\label{omgdecay}
\begin{figure}[!ht]
\begin{minipage}{\textwidth}
\begin{center}
\resizebox{\textwidth}{!}
{\includegraphics*{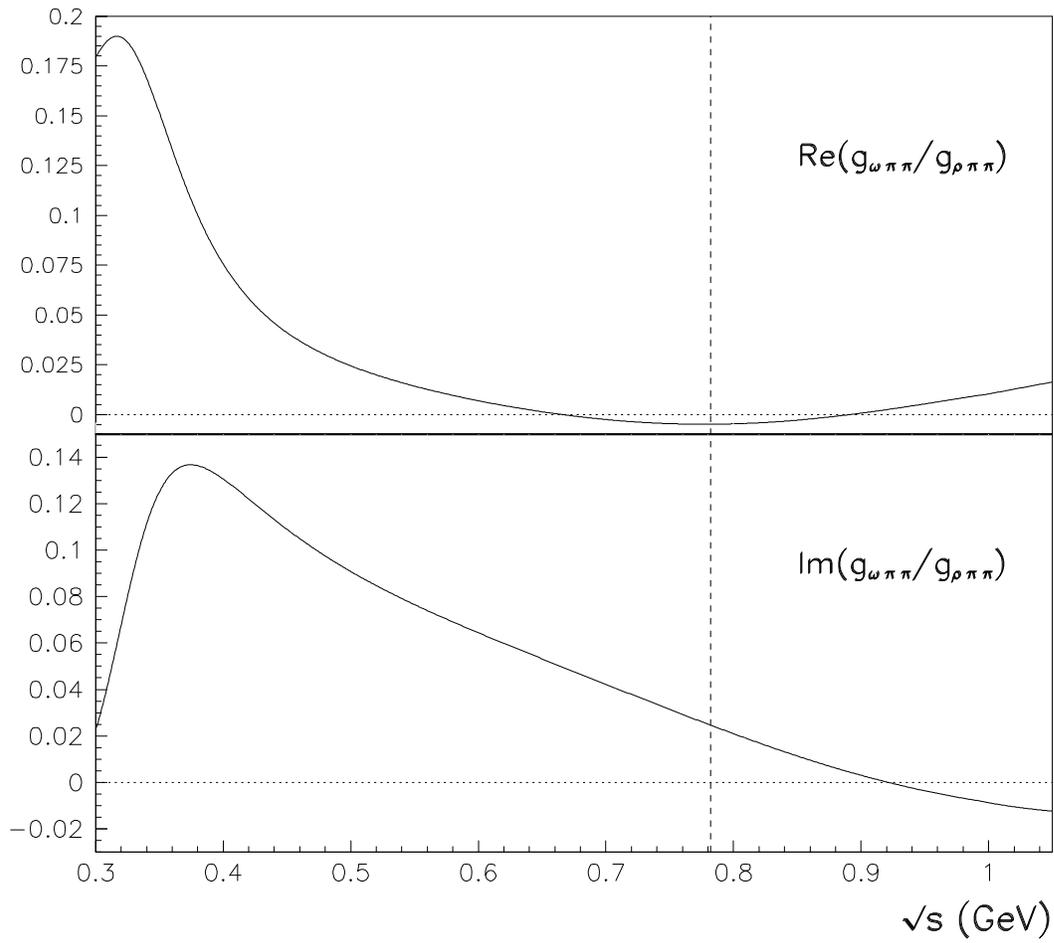}}
\end{center}
\end{minipage}
\begin{center}
\vspace{-0.3cm}
\caption{\label{Fig:g_omgSg_rho}
Ratio of the couplings  $g_{\omg \pi \pi}/g_{\rho \pi \pi}$ as a function of 
$\sqrt{s}$, as coming from the global fit (this ratio is explicitly 
given in Section \ref{omgdecay}).   
The vertical line  locates the PDG mass of the $\omg$
meson. The uncertainty band due to fit parameter errors is not shown.}
\end{center}
\end{figure}

\indent \indent As noted in Subsection \ref{omgpipi}, the coupling  $\omg \ra \pi \pi$
in the upgraded broken HLS model is given by~:
\be
\displaystyle g_{\omega \pi \pi} = \frac{a g}{2} \left [(1-h_V)\Delta_V-\alpha(s) \right ]~.
\label{eq82}
\ee
This expression exhibits two contributions of different origin. The first part 
is a constant term generated by the Direct 
Isospin Breaking procedure defined at the beginning of this paper, the second is generated
by the kaon loop mixing procedure already defined in \cite{taupaper,ExtMod1} and 
reminded above. This structure resembles that given in \cite{Maltman1996,Maltman2009}.
It is interesting to examine the behavior  of the ratio~:
 
$$\displaystyle \frac{g_{\omega \pi \pi}}{g_{\rho \pi \pi}}
=\left [(1-h_V)\Delta_V-\alpha(s) \right ]$$
\noindent
as a function of $\sqrt{s}$. It is given in Figure \ref{Fig:g_omgSg_rho}, where
the vertical line figures the $\omg$ mass location. Of course, the effective part
of this function is determined by the  $\omg$  Breit--Wigner distribution and is concentrated 
within a few tens of MeV's apart from the  $\omg$ peak position.

From the best fit discussed in the above Section (see the second data column in Table \ref{T3}), 
one gets the central values for the fit 
parameters and their error covariance matrix. These have been used to generate  $g_{\omega \pi \pi}$ 
by Monte Carlo methods. Computed with using the RPP \cite{RPP2010} mass for the $\omg$ meson, this 
gives\footnote{The quoted uncertainties for $\Delta_V$, $h_V$,  $a$  and $\Sigma_V$
are the improved uncertainties returned by the routine {\sc minos} of
the {\sc minuit} package \cite{minuit}.}~:
\be
\left \{
\begin{array}{ll}
\displaystyle g_{\omega \pi \pi} =(-0.071 \pm 0.003)+i(0.150 \pm 0.002)
\\[0.5cm]
\displaystyle \Delta_V=(-5.22 \pm 0.75) ~~10^{-2}~~~,~~h_V=1.690\pm 0.107
\\[0.5cm]
\displaystyle a=2.288 \pm 0.006~~~,~~g=5.556  \pm 0.014~~~,~~\Sigma_V =(3.74 \pm 0.50) ~~10^{-2}
\end{array}
\right.
\label{eq83}
\ee
The observed useful correlations are $<\delta \Sigma_V \delta \Delta_V>=-0.056$,
 $<\delta \Sigma_V \delta h_V>=0.028$ and  $<\delta h_V \delta \Delta_V>=0.232$.
 
In order to stay consistent with \cite{Maltman1996,Maltman2009} definitions, one can
consider that  $g_{\rho \pi \pi}^I=ag(1+\Sigma_V)/2$ and
 $ g_{\omega \pi \pi}^I= a g(1-h_V)\Delta_V/2$ are the couplings of the
 ideal fields, defined as such before applying the loop mixing. Therefore, the quantity $G$~:
 \be
 \displaystyle
 G=\frac{g_{\omega \pi \pi}^I}{g_{\rho \pi \pi}^I}=(1-h_V)\Delta_V(1-\Sigma_V)
\label{eq84}
\ee
should be close to the parameter carrying the same name in \cite{Maltman2009}.
One finds $G=(3.47  \pm  0.64) ~~10^{-2}$ to be compared with the two estimates
of the same parameter given in  \cite{Maltman2009}~: $G=(7.3  \pm  3.2) ~~10^{-2}$
when relying on the data from \cite{CMD2-1995corr}
and $G=(4.4  \pm  0.4) ~~10^{-2}$ when using, instead, the 
\cite{CMD2-1998-1,CMD2-1998-2} data.

Referring to Eqs. (\ref{eq28}), one can conclude\footnote{The isospin 0 component inside the
physical $\rho$ meson is given by $h_V\Delta_V$, while the isospin 1 part inside the $\omg$
is given by $(1-h_V)\Delta_V$}
 that there is much more
isospin 0 inside the physical $\rho$ than isospin 1 inside the physical $\omg$.
In this case, one also gets for the direct term $a g(1-h_V)\Delta_V/2=-0.332 \pm 0.024$.
Comparing this number with ${\cal R}e(g_{\omega \pi \pi})$, it is clear that
$a g(1-h_V)\Delta_V/2$ and ${\cal R}e(\alpha(m_\omg^2))$ compensate to a large extent,
in such a way that $g_{\omega \pi \pi}$ is highly dominated by its imaginary part\footnote{
In traditional fits with  the Orsay phase parametrization of the $\omg$ contribution
to the pion form factor,  this property is reflected by a value
for this phase close to $\pi/2$}.

\section{The $\pi^0-\eta-\eta^\prime$ Mixing Properties}
\label{PSproperties}
\indent \indent The  mixing of pseudoscalar neutral mesons has been addressed in Section
\ref{brkLa}, especially in Subsections \ref{PSmixing} and \ref{mixingAngl}. The present
Section is devoted to examining how the upgraded breaking scheme developed in this
paper performs compared to the results previously derived in this field \cite{WZWChPT}.
In order to perform this study, we let free the pseudoscalar mixing angle $\theta_P$,
which mostly determines the relationship between the physical $\eta$ and $\eta^\prime$
fields and their underlying octet and singlet components $\eta^0_R$ and $\eta^8_R$.
The parameters $\epsilon$ and $\epsilon^\prime$ which account for, respectively, the
$\pi^0-\eta$  and $\pi^0-\eta^\prime$  mixing are also let free. 

\begin{table}[ph]
\begin{center}
\begin{tabular}{|| c  || c  | c ||}
\hline
\hline
\hhhd ~~~~ & \hhhv General Fit   &  \hhhv Constrained Fit  				  \\
\hline
\hline
 \hhhv $\theta_0$  &    $-1.11^\circ \pm 0.39^\circ$    &  {\bf 0}			  \\
\hline
 \hhhv  $\theta_8$ &    $-23.88^\circ \pm 0.34^\circ$   &  $-23.82^\circ \pm 0.34^\circ$  \\
\hline
 \hhhv	$\theta_P$ &    $-12.66^\circ \pm 0.35^\circ$   &  $-12.91^\circ \pm 0.18^\circ$  \\
\hline
 \hhhv $\lambda$   &    $(8.52 \pm 3.55) ~10^{-2}$   	&  $(8.52 \pm 3.55) ~10^{-2}$	  \\
\hline
\hline
\end{tabular}
\end{center}
\caption {
\label{T4}
Some parameter values derived when leaving free $\theta_P$ and $\lambda$ (first data column)
or when relating them by imposing $\theta_0 =0$ to the fit (second data column).
}
\end{table}

 As shown in \cite{WZWChPT} and revisited in Subsection \ref{PSmixing} above, the 
 ChPT mixing angles \cite{leutwb,leutw}  $\theta_0$ and $\theta_8$ can be expressed
 in terms of the nonet symmetry breaking parameter $\lambda$ (or, better, using
 instead $v$ 
 defined in  Eq.(\ref{eq22})), $z_A$ the SU(3) breaking parameter of the Lagrangian
 ${\cal L}_A$ and the singlet--octet mixing angle $\theta_P$. Therefore, they can be
 estimated from fitting the data already defined.
 
 \subsection{The Mixing Angles $\theta_0$, $\theta_8$ And $\theta_P$}
 \indent \indent The mixing angles $\theta_0$ and $\theta_8$ have been recently
 introduced with the 2--angle description of the $\eta/\eta^\prime$ mixing
 \cite{leutwb,leutw}. The broken HLS model provides expressions for these
 in terms of the singlet--octet mixing angle  $\theta_P$ and of the breaking parameters 
 $z_A$ and $\lambda$ (see Eq. (\ref{eq26}) and also \cite{WZWChPT}).

 Therefore, using the fit results (parameter central values and their error covariance matrix)
 one can reconstruct the values for  $\theta_0$ and $\theta_8$. Having left free 
 $\theta_P$, one obtains the results shown in the first data column of Table \ref{T4}.
 Therefore, as in former studies, one observes that $\theta_0$ is small and its
distance to zero is only $2.8\sigma$; this should be compared with the estimate 
$\theta_0=-4^\circ$   given with no quoted uncertainty in \cite{leutw}. 
The value for $\theta_8$ is numerically
as expected from other kinds of data  \cite{leutw}. The 'tHooft parameter
\cite{tHooft} $\lambda$ is found of the order 10 \%, twice smaller than in \cite{WZWChPT}
where an approximate treatment of nonet symmetry breaking was used.
Finally,  the singlet--octet mixing angle
 $\theta_P$ is still found twice smaller than $\theta_8$, as in the former study 
 \cite{WZWChPT}.

As the distance to zero of $\theta_0$ is  $2.8\sigma$, the non--identically vanishing
of $\theta_0$ is on the border of statistical significance. Therefore, imposing the
condition $\theta_0=0$ is worth being considered; this turns out to algebraically relate 
$\theta_P$ to $z_A$ and $\lambda$ by $\tan{\theta_P}=\tan{B}$ (see Eqs. (\ref{eq26})).
Performing such a fit returns the results shown in the second data column of Table
 \ref{T4} with a quite comparable probability.

 It is interesting to observe that the value for $\theta_8$ is nearly
 unchanged and that the value for $\lambda$ is affected below the $10^{-4}$ level only.
 One also observes that the value for $\theta_P$ generated by the appropriate
 Eq. (\ref{eq26}) is found in agreement with its fitted value (when this parameter is left free).
 We conclude therefrom that assuming  $\theta_0=0$ does not degrade the fit quality 
 and is consistent with data.
 
 One should also note that the nonet symmetry breaking parameter   $\lambda=8.5\%$ 
has a statistical significance of $2.4\sigma$. 
 Performing an approximate nonet symmetry breaking
 \cite{WZWChPT}, the value for  $\lambda$ was overestimated by a factor of 2.

\subsection{The $\pi^0-\eta$ and $\pi^0 -\eta^\prime$ Mixing Properties}

\indent \indent These mixing properties are reflected by the parameters named 
respectively
$\epsilon$ and $\epsilon^\prime$ as displayed in Eqs. (\ref{eq23}). 
Comparing analogous fits performed by letting free and unconstrained
$\theta_P$, $ \epsilon$ and $\epsilon^\prime$, we did not find sensitively different 
results than those obtained by imposing the constraint on $\theta_P$ resulting
from the condition $\theta_0 =0$.  Therefore, from now on, all presented fit
results will refer to this configuration. One should note that the numerical results 
given in the above Sections have also been derived under these conditions.

The global fit returns $\epsilon=(4.89 \pm 0.44) ~10^{-2}$
and $\epsilon^\prime = (1.68 \pm 0.44) ~10^{-2}$, reflecting that 
the $\pi^0-\eta$ mixing is certainly much more important than
the $\pi^0 -\eta^\prime$ mixing phenomenon. With the concern
of reducing the number of free parameters, we have also assumed  \cite{leutw96}~:
\be
\left \{
\begin{array}{ll}
\displaystyle
\epsilon =& \displaystyle \epsilon_0 \cos{\theta_P}
\frac{\sqrt{2}\cos{\theta_P}-\sin{\theta_P}}{\sqrt{2}\cos{\theta_P}+\sin{\theta_P}} \\[0.5cm]
\displaystyle
\epsilon^\prime =& \displaystyle -2 \epsilon_0 \sin{\theta_P}
\frac{\sqrt{2}\cos{\theta_P}+\sin{\theta_P}}{\sqrt{2}\cos{\theta_P}-\sin{\theta_P}} 
\end{array}
\right.
\label{eq85}
\ee
with $\theta_P$ still determined by the constraint $\theta_0=0$. This reduces the
number of free parameters by one more unit. The fit returns 
$\epsilon_0=(3.16 \pm 0.23) ~10^{-2}$ with an unchanged probability; this corresponds
to values for  $\epsilon$ and $\epsilon^\prime$ very close ($2 \sigma$ each) from
the corresponding fitted values, while the global fit probability is unchanged.
The partial widths for the three decays $P \to \gamma  \gamma$ are all
well accounted for~: $1.64\sigma$ ($\eta$), $0.11\sigma$ ($\eta^\prime$) and
$0.06\sigma$ ($\pi^0$). Additional fit detail can be found  in Table \ref{T3}. 

The question of whether the present  $\epsilon_0$ can be identified
with the variable carrying the same name in \cite{leutw96} is 
unclear\footnote{The quantity named $\epsilon_0$ in \cite{leutw96}
is related with $R=(m_s-\tilde{m})/(m_d-m_u)$ by $\epsilon_0=\sqrt{3}/(4 R)$.
For instance, \cite{Colangelo} gives $R=37.2\pm 4.1$, while \cite{Dominguez} relying
on QCD sum rules proposes  $R=33\pm 6$. These provide respectively
$\epsilon_0=(1.16\pm 0.13)\%$ and $\epsilon_0=(1.31\pm 0.24)\%$, which have little 
to do with our fit result. }. Indeed, an important part of isospin symmetry breaking 
effects are already included in the definition of the renormalized PS fields
 (see Eqs. (\ref{eq19})  and  (\ref{eq21})) which undergo the
rotation defined by Eqs. (\ref{eq23}). Therefore, our  $\epsilon$, $\epsilon^\prime$
and $\epsilon_0$ carry only a part of the isospin breaking effects, while another part (governed by
$\Delta_A$) has been propagated to all sectors of the effective Lagrangian.
 
\section{The Values of the FKTUY Parameters }
\label{Kubis}
\indent \indent Our global fit modelling is in position to provide the most accurate information
concerning the parameters $c_3$, $c_4$ and $c_1-c_2$ defining the scales of the
various FKTUY anomalous pieces \cite{FKTUY} of the HLS Lagrangian.

In order to get the most accurate
results, we have explored the parameter behavior and found that the least correlated
combinations are $c_4+c_3$, $c_4-c_3$ and $c_1-c_2$. Running under the configuration A defined
above, one gets~:
 \be
\begin{array}{llll}
\displaystyle
c_+ \equiv \frac{c_4+c_3}{2}= 0.962 \pm 0.016~,
&\displaystyle
c_- \equiv \frac{c_4-c_3}{2}=(-3.98^{+1.88}_{-1.96})~10^{-2}~,
&\displaystyle
c_1-c_2=1.208^{+0.058}_{-0.054} 
\end{array}
\label{eq86}
\ee
with $g=5.541\pm0.016$, while configuration B leads to~:
 \be
\begin{array}{lll}
\displaystyle
c_+ \equiv \frac{c_4+c_3}{2}= 0.978 \pm 0.020~,
&\displaystyle
c_- \equiv \frac{c_4-c_3}{2}=(-6.75^{+2.74}_{-2.81})~10^{-2}~,
&\displaystyle
c_1-c_2=1.123^{+0.063}_{-0.060}
\end{array}
\label{eq87}
\ee
with $g=5.530\pm0.015$. The  correlation coefficients
are similar in both cases~: 
$<[\delta c_+][\delta (c_1-c_2)]>\simeq -0.20$,
$<[\delta c_+][\delta c_-]> \simeq -0.10$ 
and $<[\delta  c_-][\delta(c_1-c_2)]>\simeq 0.80$. Therefore, our global
fit yields quite consistent numerical values wathever the configuration\footnote{
Running our code excluding the $K \overline{K}$ data (see second data column in Table
\ref{T3})  yields $c_+ =0.967 \pm 0.021$, $c_- =(-5.18^{+2.80}_{-3.23})~10^{-2}$ and
$c_1-c_2=1.074^{+0.064}_{-0.068})$ with $g=5.530\pm0.015$. This configuration pushes
the significance for a non--zero $c_-$ at the $\simeq 1.7 \sigma$ level.
} for the FKTUY parameters.

These values can be compared with existing estimates. Using the
$\pi^0 \gamma \gamma^*$ form factor,  \cite{HLSRef} yields 
$c_+ =1.06 \pm 0.13$, while the partial width
$\omg  \to \pi^0 \gamma$ provides $c_+ =0.99 \pm 0.1-$
when using $g=5.80\pm0.91$. Our own estimates are consistent with these with, however,
({\sc minos}) uncertainties five times more precise.

 A rather unprecise value for the ratio $\tilde{c}=c_-/c_+$
has also been derived \cite{HLSRef} relying on the decay $\omg \to \pi^0 \mu^+ \mu^-$,
$\tilde{c}=0.42 \pm 0.56$, consistent
with our results but still much less precise. 

From our results, which happen to be the most precise in this field, one may conclude 
that data only favor a partial fulfilling of the VMD assumptions \cite{HLSRef}, in the
sense that $c_3-c_4 =0$ is in agreement with data  at the $2 \sigma$ level,
while $c_1-c_2+c_4=4/3$ is badly violated. This can be rephrased as follows~: the 
 VMD assumptions \cite{HLSRef} are experimentally fulfilled in the triangle anomaly sector
 and strongly violated in the box anomaly sector. This confirms the previous parent analysis
 \cite{ExtMod1} and  former studies on the box anomaly in the 
 $\eta/\eta^\prime \to \pi^+ \pi^- \gamma$ decays \cite{box}. 

In order to go beyond, better data on the annihilation channels involving anomalous couplings
($[\pi^0/\eta] \gamma$,  $\pi^+ \pi^- \pi^0$) are needed; including new processes like the
$\eta/\eta^\prime \to \pi^+ \pi^- \gamma$ decay spectra or information on the
$l^+ l^- \pi^0$ annihilation channels may also help as their dependence upon $c_3-c_4$ or
$c_1-c_2$ is more important than in the previous channels.

It thus follows from the present analysis that assuming  $c_3=c_4$ is justified.
In this case, one obtains the following results~:
 \be
\begin{array}{llll}
c_4=c_3= 0.950 \pm 0.014~~,&~~c_1-c_2=1.194\pm 0.060~~,&~~g=5.556 \pm 0.014~~,
\end{array}
\label{eq88}
\ee
for Configuration A and~:
 \be
\begin{array}{llll}
c_4=c_3= 0.951 \pm 0.016~~,&~~c_1-c_2=1.169 \pm 0.060~~,&~~g=5.553 \pm 0.012~~ 
\end{array}
\label{eq89}
\ee
for Configuration B.

In both cases, the correlation coefficient is $<[\delta c_3][\delta (c_1-c_2)]>\simeq -0.20$.
Therefore, the condition $c_4=c_3$ drastically reduces the correlation among the surviving FKTUY
parameters. Moreover, the fit quality is not significantly changed while assuming $c_4=c_3$.
Indeed, configuration A yields $\chi^2/dof=858.08/882$ (71.2 \% probability) instead of
$\chi^2/dof=853.98/881$ (73.7 \% probability) and configuration B $\chi^2/dof=728.38/882$ 
(97.0 \% probability) instead of $\chi^2/dof=722.05/881$ (97.9 \% probability), where the
difference mostly affects the set of partial widths which is always well fitted. Therefore, the
improvement obtained with the upgraded breaking model is not due to releasing the
condition $c_4=c_3$. From these considerings, it is justified to impose $c_3=c_4$ 
for the rest of this study.

\section{Hadronic Contributions to $g-2$}
\label{gmoins2}
\indent \indent In \cite{ExtMod2}, one analyzed in full detail 
the hadronic contribution to $g-2$ of most of the data sets
used in the present study.  The framework was the
previous release of the present model studied in detail in 
\cite{ExtMod1,ExtMod2}. Within this framework, only the
simultaneous account of both  annihilation channels  to $K\overline{K}$ 
was missing. On the other hand, one might find unsatisfactory
that some  global rescaling of experimental $\tau$  dipion spectra
was still playing an important role, even if this rescaling was in
accord with expectations. These two issues motivated the present study.

As shown above, the upgraded model allows by itself a satisfactory account
of all considered spectra simultaneously. It is therefore worth reexamining within our
upgraded framework, how the hadronic contribution to $g-2$ is estimated and how
this estimate evolves depending on the various kinds of data groups considered. 
\begin{table}[!htb]
\begin{tabular}{|| c  || c | c  | c ||}
\hline
\hline
\hhhu Data Set  &    Fit Solution & \multicolumn{2}{|c|}{Statistical Information} \\
\hline
\hhhu  ~~& ~~&  $\chi^2/\rm{dof}$ & Probability\\
\hline
\hline
\hhhu $e^+e^- \to \pi^+ \pi^-$ &   $360.00\pm 1.64$ & $177.38/208$ & 93.3\%\\
\hline
\hhhu + $[\tau]$ data (ABC)  & $359.8 \pm 1.47$  & $262.94/293$  & 89.6\%\\
\hline 
\hline
\hhhu  ++ ($e^+e^- \to [\pi^0/\eta] \gamma$)  & $ 360.09\pm 1.60$ &  $436.94/549$ & 99.9\% \\
\hline
\hhhu  ++ ($e^+e^- \to \pi^+\pi^-\pi^0$) & $ 360.91 \pm 1.45 $ &  $661.22/727$ & 96.1\% \\
\hline
\hline
\hhhu  ++ ($e^+e^- \to K \overline{K} $) & $ 362.79\pm 1.43$ &  $858.08/882$ & 71.2\% \\
\hline
\end{tabular}
\caption{
\label{T5} The contribution  to $10^{10} a_\mu(\pi \pi)$ from the invariant mass region $0.630-0.958$ GeV/c. 
The first line provides the fit results using all the $e^+e^- \to \pi^+ \pi^-$ annihilation
data set group. The next line uses the previous data group and the three $\tau$ spectra.
By "++" at any given line, we always mean all data sets belonging to the groups referred to in 
the preceding lines, plus the data set group indicated at this line. FSR corrections 
are taken into account. An appropriate set of radiative decays is always understood. 
The last line refer to what has been named Solution/Configuration A.}
\end{table}

\subsection{The $\pi^+\pi^-$ Contribution to  $g-2$~: VMD Estimates}
\indent \indent The most important hadronic contribution to $g-2$ is the 
$\pi^+\pi^-$ channel. Several experiments \cite{CMD2-1995corr,CMD2-1998-1,SND-1998} 
and some analyses \cite{DavierHoecker,ExtMod2,DavierHoecker2} give the $\pi^+\pi^-$
contribution to $g-2$ provided by the energy region $[0.630-0.958]$ GeV.  Therefore, it is
worth considering  the information provided by this reference region; this allows to
 substantiate the  improvement which can be expected from VMD--like models.
 Indeed, several kinds of information are  worth considering~:
 \begin{itemize}
 \item While unifying the description of $e^+e^-$ annihilation and $\tau$ decays,
 one expects an increased precision on the anomalous magnetic moment of the muon $a_\mu(\pi \pi)$.
 \item While having a framework which encompasses most of the physics up to
 the $\phi$ region, the stability and the robustness of the $a_\mu(\pi \pi)$
 estimates can be examined. The relative statistical consistency of the various data groups
 is also an issue which can be addressed, relying on their behavior under global fits. 
\end{itemize}

Table \ref{T5} displays our estimate for the $\pi \pi$ contribution to $a_\mu=(g-2)/2$ 
provided by the reference energy range
under various fit configurations. In each case, the fitted (central) parameter values
and their error covariance matrix are used in order to sample several thousand  
 parameter  vectors, assuming a $n$--dimensional gaussian error distribution. Each 
vector of sampled parameter values is, then, used to compute  $a_\mu(\pi \pi)$. The corresponding distribution
of the  $a_\mu(\pi \pi)$'s is then fitted to a Gaussian function. The results displayed in Table \ref{T5}
are the central values and the standard deviations of this distribution which intrinsically
takes into account the correlations among the fitted parameters. 
Unless   otherwise stated, the FSR correction is included in all reported contributions
of the $\pi^+\pi^-$ channel to $a_\mu$.

Beside the experimental spectra, there is always a set of partial width decays 
submitted to fit. These have been defined in Subsection \ref{decays}. In the results
reported below, one should keep in mind that the accepted values \cite{RPP2010} for the
$(\rho/\omg/\phi)\to (\pi^0/\eta) \gamma$ and $(\omg/\phi)\to e^+e^-$
partial widths are included in the set of partial widths submitted to the fit as long as the 
experimental spectra for the 
$e^+e^- \to (\pi^0/\eta) \gamma$ annihilation channels are not used. As emphasized in
\cite{ExtMod1}, this hides some model dependence which might be somewhat conflicting
with our own model. This explains why one should prefer any configuration where the
$e^+e^- \to (\pi^0/\eta) \gamma$ data are submitted to the global fit.

Also, when the data for the two annihilation channels $e^+e^- \to K \overline{K} $ are
not considered in the fit, one chooses to fix $\epsilon_0=\Delta_A=0$, as we have no 
real sensitivity to them. Likewise, $c_1-c_2$ is absent from fits as long as
the $e^+e^- \to \pi^+\pi^-\pi^0$ data are not considered. Finally, the parameters fixing
the mass and width of the $\phi$ meson are left free  only when the fitted data allow to
constrain them.

In the first line of Table \ref{T5}, one finds the value for $a_\mu(\pi \pi)$ derived 
by submitting to fit  the
scanned data for the annihilation process  $e^+e^- \to \pi^+ \pi^-$ -- together with
the full set of partial width decays.
This result compares well with the value derived using the previous release of our broken HLS 
model\footnote{Even if expected, this proves that the effects produced by having introduced $\Sigma_V$
do not modify the fit description of the $e^+e^- \to \pi^+ \pi^-$ data.},
as can be seen by comparing with the relevant piece of information reported in Table 4 of 
\cite{ExtMod2}. 

As there is no longer any mismatch between $e^+e^-$ and $\tau$ data, both in magnitude 
and in lineshape (see Section \ref{fit_tau}), it is legitimate to merge them.
This merging provides the new and important result given in  the second line of 
Table \ref{T5}. One clearly observes that the merged $\pi  \pi $ data
give  a result perfectly consistent with the $e^+e^- \to \pi^+ \pi^-$
data alone with a quite nice probability. The central value for $a_\mu(\pi \pi)$
is nearly unchanged and the uncertainty slighty improved.

This is, of course, the main effect of having upgraded our symmetry breaking procedure 
of the HLS  Lagrangian. In this new framework, there is no
 need for an auxiliary rescaling \cite{ExtMod2}
of the  $\tau$ spectra and the net result is a perfect 
consistency of the $e^+e^- \to \pi^+ \pi^-$ data with/without the $ \tau$ data considered as constraints.
This statement can be substantiated by comparing this result with those reported 
in the entry "NSK+ A B C" of \cite{ExtMod2} ($a_\mu(\pi \pi)=(364.48 \pm 1.34)~10^{-10}$)
which exhibited a shift of about $5~10^{-10}$ produced by the three $ \tau$ data
sets, a $\simeq 3.6 \sigma$ effect.

\vspace{0.5cm}

The third line in Table \ref{T5}, displays the effect of replacing the
$(\rho/\omg/\phi)\to (\pi^0/\eta) \gamma$ and  the $(\omg/\phi)\to e^+e^-$
partial widths by the cross sections for $e^+e^- \to (\pi^0/\eta) \gamma$.
The central value for $a_\mu(\pi \pi)$ is practically unchanged, while its
standard deviation is increased by $\simeq 9\%$.
The following line  in Table \ref{T5} displays the effect of including
the full  $e^+e^- \to \pi^+ \pi^- \pi^0$ data group already defined. As in \cite{ExtMod2}, one
observes a perfect consistency of the results for $a_\mu(\pi \pi)$. In total,
the standard deviation is slightly reduced ($\sigma(a_\mu(\pi \pi))\simeq 1.5 ~10^{-10} $). 
At this point one may conclude that the central value is marginally modified
by fully including the  $(\rho/\omg/\phi)\to (\pi^0/\eta) \gamma$ and
$e^+e^- \to \pi^+ \pi^- \pi^0$ data groups within the fit procedure. The variations of
the uncertainty returned by the fits might rather reveal
statistical fluctuations.

The last line  in Table \ref{T5} displays the effect of including
the two $e^+e^- \to K \overline{K} $ cross sections into the fitted data set.
One observes some effect, as  $a_\mu(\pi \pi)$ undergoes a $1.9~10^{-10}$
shift upwards while the fit probability remains quite good. This fit configuration 
 -- referred to  as Solution/Configuration A --
encompasses the largest set of data samples considered safe. This turns out
to consider  that the 30 unit increase of the $\chi^2$ associated with the  
$\pi^+ \pi^- \pi^0$ data group, even if large, is not abnormal (see the fourth data column
in Table \ref{T3}).
 
 \vspace{0.5cm}
 
The result shown in the last line of Table  \ref{T5}, may reveal some tension
among the data set groups. In order to explore this issue, one has redone
fits excluding the $\pi^+ \pi^- \pi^0$ data group, and examined the effects of
using the selected $e^+e^- \to K^0 \overline{K}^0 $ and $e^+e^- \to K^+ K^- $ data, either 
separately or together. The corresponding results are displayed in Table \ref{T6}. Comparing
the statistical information here with those in the  last line in  Table \ref{T5} renders somewhat suspicious the 
quoted   30 unit increase of $\chi^2_{\pi^+ \pi^- \pi^0}$.

\begin{table}[!htb]
\begin{tabular}{|| c  || c | c  | c ||}
\hline
\hline
\hhhu  ~~&Fit Solution  	&  $\chi^2/\rm{dof}$ & Probability\\
\hline
\hline
\hhhu only $e^+e^- \to K^+ K^-$ 	&   $360.79\pm 1.49$ & $474.69/585$ & 99.97\%\\
\hline
\hhhu  only    $e^+e^- \to K^0 \overline{K}^0 $ & $ 362.83\pm 1.47$ &  $580.78/668$ & 99.34\% \\
\hline
\hline
\hhhu  both $e^+e^- \to K \overline{K} $ & $ 362.81\pm 1.47$ &  $613.29/704$ & 99.40\% \\
\hline
\end{tabular}
\caption{
\label{T6} The contribution  to $10^{10} a_\mu(\pi \pi)$ from the invariant mass 
region $0.630-0.958$ GeV/c using    $K \overline{K} $ data sets under various
conditions. All $\pi^+ \pi^- \pi^0$ data have been excluded from fit. FSR corrections
have been performed.
}
\end{table}

A final piece of information is provided by performing the fit using
the $\pi^+ \pi^- \pi^0$ data group data amputated from the data points 
collected in the region
above 1 GeV (therefore, excluding the $\phi$ region). This fit configuration has already
been referred to  as  Solution/Configuration B. 
The reason which motivates this removal is that the $\pi^+ \pi^- \pi^0$ data
before introducing the $K \overline{K}$ data is only constrained 
in the $\phi$ region by the relatively unprecise data on the
$\pi^0 \gamma $ and $\eta \gamma$ channels. One then obtains~:
\be
\left \{
\begin{array}{lllll}
{\rm Solution~} B~~:~~a_\mu(\pi \pi)=(362.44 \pm 1.49)~10^{-10} 
& \chi^2/dof=722.05/801 & {\rm  Prob.}= 97.9\% \\[0.5cm]
{\rm Solution~} A~~:~~a_\mu(\pi \pi)=(362.19 \pm 1.44)~10^{-10} 
& \chi^2/dof=854.00/881 & {\rm  Prob.}= 73.7\% 
\end{array}
\right .
\label{eq90}
\ee 
where the result for Solution A is reminded.

These differences indicate that all physics channels covering the $\phi$ region
are worth to be reconsidered, as already argued from discussing the fit results 
in Table \ref{T3}. Indeed, the difference in fit quality between Configurations
A and  B reveals some tension between the $K \overline{K}$ data  and the
$\pi^+ \pi^- \pi^0$ data collected in the $\phi$ region. Fortunately, the physics in 
the $\phi$ region is
still accessible at VEPP--2M. It seems also in the  realm of the KLOE detector,
as this turns out to run DA$\Phi$NE within a $\pm$ 20 MeV interval apart from the $\phi$ mass peak 
value. 

\subsection{The $\pi^+\pi^-$ Contribution to  $g-2$~: Comparison with Data}
\indent \indent An interesting piece of information comes from comparing
our (VMD) estimates derived from global fitting 
with the corresponding estimates provided by the various experimental groups.

Table \ref{T7} displays the published experimental results concerning the contribution
of the $0.630-0.958$ GeV/c region to $a_\mu(\pi \pi)$.  We first list the three 
important results from CMD--2 and SND; as we also use the data sets
from OLYA and CMD \cite{Barkov}, we also
 give at the line flagged by   "OLD"  our average using these data sets
together with those from NA7 \cite{Amendolia}, TOF \cite{vasserman81}, 
M2N \cite{M2N}, DM1 \cite{DM1}, 
all collected before those from \cite{CMD2-1995corr,CMD2-1998-1,CMD2-1998-2,SND-1998}.

\begin{table}[ph]
\hspace{-1.cm}
\begin{tabular}{|| c  || c  | c | c  ||}
\hline
\hline
\hhhv Data Set  & Experimental Result &   Average & Fit solution \\
\hline
\hhhv CMD--2 (1995)\cite{CMD2-1995corr}& $362.1 \pm (2.4)_{stat} \pm (2.2)_{syst}$   &
~~~~ & ~~~~  \\
\cline{1-2} 
\hhhv CMD--2 (1998)\cite{CMD2-1998-1,CMD2-1998-2} & $361.5 \pm (1.7)_{stat} \pm (2.9)_{syst}$    
& ~~~~  &~~~~   \\
\cline{1-2} 
\hhhv  SND (1998) \cite{SND-1998} & $361.0 \pm (1.2)_{stat} \pm (4.7)_{syst}$ 
&~~~~   & ~~~~ \\
\hline 
\hline 
\hhhv Average  &~~~~  & $361.26 \pm (2.66)_{tot}$   & ~~~~    \\
\cline{1-3} 
\hhhv  OLD & $354.1 \pm (3.3)_{stat} \pm (8.1)_{syst}$
&~~~~   & ~~~~ \\
\cline{1-3} 
\hhhv  
Average (excl. ISR) &~~~~  & $360.65 \pm (2.55)_{tot}$   & ~~~~    \\
\cline{1-3} 
\hhhv Fit Solution A &~~~~  & ~~~~ & $A~~:~~ 362.79 \pm 1.43_{tot}$ \\
\cline{4-4} 
\hhhv Fit Solution B &~~~~  & ~~~~ & $B~~:~~363.16 \pm 1.47_{tot}$ \\
\hline
\hline
\hhhv KLOE--2008 \cite{KLOE08}&$356.7 \pm (0.4)_{stat} \pm (3.1)_{syst}$ &~~~~  & ~~~~  \\
\cline{1-2}
\hhhv KLOE--2010 \cite{KLOE10}
 &$353.3 \pm (0.6 )_{stat} \pm (3.2)_{syst}$ &~~~~  & ~~~~  \\
\cline{1-2}
\hhhv BaBaR \cite{BaBar,DavierHoecker2}&$365.2 \pm (1.9)_{stat} \pm (1.9)_{syst}$ &~~~~  & ~~~~  \\
\cline{1-3}
\hhhv Total Average &~~~~  
& $360.53 \pm (1.44)_{tot}$   & ~~~~    \\
\hline
\hline
\end{tabular}
\caption{
\label{T7} The various published estimates of the contribution to $10^{10} a_\mu(\pi \pi)$ 
from the invariant mass region $0.630-0.958$ GeV/c. 
The quoted averages always refer to all $experimental$ results displayed in the preceding lines.
The line "OLD" information refers to our average performed using the data sets
collected before those of CMD--2 and SND (see text). 
Our fit solutions A and B are derived using the $\tau$ spectra from \cite{Aleph,Cleo,Belle}.
KLOE--2010 estimate for $a_\mu(\pi \pi)$ is ours, as the experimental spectrum
stops  slightly below $\sqrt{s}=0.958$ GeV \cite{KLOE10}.
}
\end{table}

The third data column provides, first, our average derived using the data sets from
\cite{CMD2-1995corr,CMD2-1998-1,CMD2-1998-2,SND-1998} and, next, also those including 
the older data sets referred to just above. Our results are directly  comparable with 
these as we do not yet use ISR data.
 
 Both solution $A$ and solution $B$ results favorably compare with the scan  
 $(\pi \pi)$ data averaging as the uncertainty is reduced by a factor close to 2.
 
The following lines of Table \ref{T7} display, for information, the experimental  
results derived from the data sets collected using the ISR method  and the global average 
of the ISR and scan data. 

One should stress that our results for $a_\mu(\pi \pi)$, derived excluding the ISR data,
provide information already comparable in precision to those obtained using them. 
This motivates to examine the ISR data in view of including them into the fit procedure.

One may also compare our estimates with the weighted average of the $\tau$ data \cite{Aleph,Belle,Cleo}
which gives  $10^{10} a_\mu(\pi \pi)=365.21 \pm 2.67_{exp}$ in the reference region, 
including FSR corrections; applying the $\rho-\gamma$ 
corrections proposed in \cite{Fred11},
this becomes  $10^{10} a_\mu(\pi \pi)=361.66 \pm 2.67_{exp}$ and provides 
$10^{10} a_\mu(\pi \pi)=361.15 \pm 1.76_{exp}$ when averaged with the 
$e^+ e^-$ data. This indicates that examining the idea proposed in
 \cite{Fred11} in a wider context is an interesting issue. Indeed, this could lead
 to another successful VMD--like model  and, therefore,  may contribute to
 a motivated evaluation of the model dependence of $a_\mu$ estimates.

As a summary, one may conclude that our global model provides a good determination of the
contribution to  $a_\mu(\pi \pi)$ from the invariant mass region $0.630-0.958$ GeV/c. The 
accuracy of our VMD estimates is found much improved compared to direct averaging of the 
experimental data and  their central values are found consistent within uncertainties. 
By including ISR data at a later stage, the precision of the result might be further
increased.

\subsection{Hadronic Contribution  to  $g-2$}
\indent \indent 
In Table \ref{T8}, one displays the  contribution of 
each of the examined channels to $a_\mu$ from their respective thresholds
up to 1.05 GeV/c, {\it i.e.} slightly above the $\phi$ peak.
    
\begin{table}[ph]
\hspace{-0.5cm}
\begin{tabular}{|| c  || c | c ||c |c ||}
\hline
\hline
Process  \hhhu  		&  Solution $B$  	& Solution $A$  & Data (excl. ISR) & Data (incl. ISR)\\
\hline
\hhhu   $\pi^+ \pi^-$ 		& $498.54 \pm 1.97 $  	& $ 497.98 \pm 1.76$  	& $498.53\pm 3.73$ & $497.72\pm 2.12$\\
\hline
\hhhu   $\pi^0 \gamma$  	& $4.64\pm 0.04 $  	& $4.28 \pm 0.04$  	& \multicolumn{2}{|c|}{$3.35 \pm 0.11_{tot}$ }\\
\hline
\hhhu  $\eta \gamma$  		& $0.65 \pm 0.01 $  	& $0.67 \pm 0.01$ 	& \multicolumn{2}{|c|}{$0.48 \pm 0.02_{tot}$} \\
\hline
\hhhu $\eta^\prime \gamma$  	& $0.01 \pm 0.00$  	& $0.01 \pm 0.00$ 	&  \multicolumn{2}{|c|}{ -- }  		 \\
\hline
\hhhu  $\pi^+ \pi^-\pi^0$ 	& $42.03 \pm 0.60 $  	& $40.88\pm 0.52$ 	& \multicolumn{2}{|c|}{$43.24 \pm 1.47_{tot}$ } \\
\hline
\hhhu  $K^+K^-$ 		&$16.87 \pm 0.20 $  	& $16.93\pm 0.18$ 	& \multicolumn{2}{|c|}{$17.88 \pm 0.54_{tot} $}\\
\hline
\hhhu  $K^0 \overline{K}^0$ 	&$12.02 \pm 0.09 $  	& $12.07 \pm 0.08 $     & \multicolumn{2}{|c|}{$12.31\pm 0.33_{tot}$} \\
\hline
\hline
\hhhu  Total Up to 1.05 GeV	& $574.76 \pm 2.10$ 	& $572.82 \pm 1.90 $ 	& $575.79 \pm 4.06_{tot}$ & $574.98 \pm 2.66_{tot}$\\
\hline
\hline
\end{tabular}
\caption{
\label{T8} Contributions to $10^{10} a_\mu$ from thresholds up to 1.05 GeV/c
 The  experimental errors merge  the reported statistical and systematic uncertainties in quadrature.
 FSR effects ($3.43 ~10^{-10}$) have been included into the $\pi^+ \pi^-$ contribution.
 The first two data columns display our fit results and
the last two data columns report the direct numerical integration of the relevant data.
}
\end{table}

The first two data columns show the results corresponding to the so--called 
configurations/solutions $A$ and $B$.
These have been derived by fitting the data sets referred to in the preceding Sections and the
motivation to consider both solutions valid can be emphasized from Table \ref{T3}.

The last two data columns exhibit the  averages  of experimental data for each of the measured channels
submitted to the global fit. These differ by excluding (third  data column) or including (fourth  data 
column) in the averaging the ISR data sets collected by KLOE \cite{KLOE08,KLOE10} and BaBar \cite{BaBar} 
for the $\pi^+ \pi^-$  final state. As we have excluded for now the ISR data from our
analysis, the gain  due to the global fit can be directly inferred by comparing with the third  data column;
nevertheless, it is interesting to compare the accuracy of solutions $A$ and $B$
to the averages  derived using the high statistics ISR data. 

As expected,  the improvement generated by the global fit affects all the channels considered
and is always  a factor of 2 or more (see the $\pi^+ \pi^-\pi^0$ channel) better than the average
of the same data. The first line even shows that our accuracy is comparable -- actually slightly
better -- than the average derived using  the ISR data. 
 
 It is interesting to  note that the sum of all contributions for solution $B$ is in accordance with 
 the result expected from the standard sum  as reported in the third (or fourth) data column. Solution $A$, 
 instead, gives a smaller sum than the experimental average of the same data; the distance is
 $2.97~10^{-10}$, {\it i.e.}   $\simeq 1.6 \sigma_{theor.}$ or  $\simeq 0.7 \sigma_{exp}$.
 
It is interesting to examine the individual channel contributions.
Those from the $\pi^0\gamma$ and $\eta\gamma$ channels, as calculated from data,
rely on pretty poor statistics and generally cover  restricted energy ranges 
\cite{CMD2Pg1999,CMD2Pg2001,CMD2Pg2005,sndPg2000,sndPg2003,sndPg2007} (see
Subsection \ref{pi0eta_g}); instead, our model results are estimated (significantly) larger
and cover precisely the full energy range from thresholds to 1.05 GeV.  This especially
concerns the region in between the $\omg$ and $\phi$ peaks. 

Our model
estimates  for the $\pi^+ \pi^-\pi^0$ and $K^+K^-$ channels are found smaller than
the experimental averages at the 1 or 2 $\sigma_{exp}$ levels, while the $K^0 \overline{K}^0$
contribution corresponds to the experimental expectation.
This confirms the need for a better experimental knowledge of all annihilation channels 
in the $\phi$ region.

 \begin{table}[ph]
\hspace{-1.cm}
\begin{tabular}{|| c  | c || c  | c || c  | c ||}
\hline
\hline
\hhhu Final State  &    Range (GeV) & \multicolumn{2}{|c|}{Contribution (incl. $\tau$)} 
& \multicolumn{2}{|c|}{Contribution (excl. $\tau$)}\\
\hline
\hhhv ~~~~~  &    ~~~~~ & Solution A & Solution B & Solution A & Solution B\\
\hline
\hhhv $e^+e^-\to$ hadrons &   threshold $\to$ 1.05 		&   572.82[1.90] & 574.76[2.10]	& 569.86[2.15]&  571.40[2.27]\\
\hline
\hhhv missing channels 	&  threshold $\to$ 1.05 		&  \multicolumn{4}{|c|}{$1.55 (0.40)(0.40)[0.57]$} \\
\hline
\hhhv $J/\psi$ 		& ~~~  					&  \multicolumn{4}{|c|}{$8.51(0.40)(0.38)[0.55]$} \\
\hline 
\hhhv $\Upsilon$ 	& ~~~  					&  \multicolumn{4}{|c|}{$0.10(0.00)(0.10)[0.10]$}  \\
\hline  
\hhhv hadronic		& (1.05, 2.00)				& \multicolumn{4}{|c|}{$60.76(0.22)(3.93)[3.94]$}  \\
\hline 
\hhhv hadronic		& (2.00, 3.10)				& \multicolumn{4}{|c|}{$21.63(0.12)(0.92)[0.93]$}  \\
\hline 
\hhhv hadronic		& (3.10, 3.60)				& \multicolumn{4}{|c|}{$3.77(0.03)(0.10)[0.10]$}   \\
\hline
\hhhv hadronic		& (3.60, 5.20)				& \multicolumn{4}{|c|}{$7.64(0.04)(0.05)[0.06]$}   \\
\hline
\hhhv  pQCD	        & (5.20, 9.46)				&\multicolumn{4}{|c|}{$6.19(0.00)(0.00)[0.00]$}   \\
\hline
\hhhv hadronic     	& (9.46, 13.00)				& \multicolumn{4}{|c|}{$1.28(0.01)(0.07)[0.07]$}   \\
\hline
\hhhv  pQCD		& (13.00,$\infty$)			& \multicolumn{4}{|c|}{$1.53(0.00)(0.00)[0.00]$}   \\
\hline
\hhhv Total		& 1.05 $\to \infty$			& \multicolumn{4}{|c|}{$112.96 \pm 4.13_{tot}$}   \\
\hhhe ~~~		& + missing channels			&  \multicolumn{4}{|c|}{~~~} \\
\hline
\hline
\hhhv Total Model	&threshold $\to \infty$ &  $685.78 \pm 4.55$&$687.72 \pm 4.63$& $682.82 \pm 4.66$ & $684.36 \pm 4.71$   \\
\hline
\end{tabular}
\caption{
\label{T9} Hadronic VP contributions to $10^{10} a_\mu$ with FSR corrections included. 
Numbers within brackets
 refer to respectively statistical and systematic errors. Numbers within square brackets are the total
 uncertainties.
 }
\end{table}

\vspace{0.5cm}

The first data line in Table \ref{T9}  reports the results derived from fits with our global model.
The second line ("missing channels") provides  the experimental averaged  contribution  to $a_\mu$
from the channels unaccounted for within our model (the $4 \pi$, $5 \pi$, $6 \pi$, $\eta \pi \pi$ and $\omg \pi$ 
final states). This has been computed using the trapezoidal integration rule. As the corresponding data are 
sparse below 1.05 GeV, this estimate might have to be improved.

The line "Total Model"  provides the estimate of the full hadronic vacuum polarization (HVP),
merging our model results with the additional listed contributions. 

The corresponding experimental average taking into account all available ISR data sets \cite{KLOE08,KLOE10,BaBar}
has been estimated \cite{Fred11} to $a_\mu(e^+e^-)=(690.75 \pm 4.72_{tot}) ~~10^{-10}$, including the contributions
above $5.2 $ GeV calculated using perturbative QCD. For comparison,  the corresponding total
average provided by \cite{DavierHoecker2} is  $a_\mu(e^+e^-)= (695.5  \pm 4.0_{exp} \pm 0.7_{QCD}) ~~10^{-10}$ 
(not accounting for the recent KLOE data set \cite{KLOE10});
 accounting for all the available ISR data sets, \cite{DavierHoecker3} yields as experimental average
$a_\mu(e^+e^-)= (692.3  \pm 4.2_{tot}) ~~10^{-10} $. 

\vspace{0.5cm}

In order to illustrate the impact of $\tau$ data,  we present separately the fit results derived when
including or when excluding the
$\tau$ data sets from the fitted data sets, keeping for the rest the configurations leading to solutions
$A$ and $B$ as previously defined.

Including $\tau$ data sets results in an increased value of the hadronic VP by $\simeq 3 ~~10^{-10}$. 
This will be commented on below.  
One also remarks that our uncertainties are comparable to the experimental one,
even if  our estimates are penalized by having -- provisionally -- discarded the ISR data. 
Our estimates also compare favorably with the revised estimate excluding all ISR data given by \cite{DavierHoecker}~:
$a_\mu(e^+e^-)= (690.9  \pm 5.2_{exp+rad}  \pm 0.7_{QCD}) ~~10^{-10}$. 

\begin{table}[ph]
\hspace{-2.cm}
\begin{tabular}{|| c | c  | c ||c  | c ||}
\hline
\hline
\hhhv $10^{10} a_\mu$  &    \multicolumn{2}{|c|}{Values (incl. $\tau$)}  &    \multicolumn{2}{|c|}{Values (excl. $\tau$)}\\
\hhhv ~~~  		&   Solution A 			&   Solution B 		&   Solution A 			&   Solution B 				\\
\hline
\hhhv LO hadronic  	&    $685.78 \pm 4.55$   &   $687.72 \pm 4.63$& $682.82 \pm 4.66$ & $684.36 \pm 4.71$ 			\\
\hline
\hhhv HO hadronic	&   \multicolumn{4}{|c|}{ $-9.98 \pm 0.04_{exp} \pm 0.09_{rad} $}			\\
\hline
\hhhv LBL 				&   \multicolumn{4}{|c|}{ $10.5 \pm 2.6$ }			  		\\
\hline
\hhhv QED			&   \multicolumn{4}{|c|}{$11~658~471.8096 \pm 0.016_{tot}$} 	   		\\
\hline 
\hhhv EW 			&   \multicolumn{4}{|c|}{$15.32\pm 0.10_{hadr} \pm 0.15_{Higgs}$ }  		\\
\hline  
\hline 
\hhhv Total	Theor.	& $11~659~173.43 \pm  5.25 $	&  $11~659~175.37 \pm  5.31 $	
					& $11~659~170.47 \pm 5.34 $&   $11~659~172.0\pm 5.39 $	 \\
\hline 
\hline 
\hhhv Exper. Aver. 	&  \multicolumn{4}{|c|}{$11~659~208.9 \pm 6.3_{tot} $ } 		 \\
\hline 
\hline 
\hhhv 	$\Delta a_\mu$		& $35.47\pm 8.20 $	& $33.53\pm 8.24 $ & $38.43 \pm 8.26 $	& $36.89\pm 8.29 $	  \\
\hline
\hhhv Significance ($n \sigma$)	& $4.33 \sigma $ 	& $4.07 \sigma $ & $4.65 \sigma $	& $4.45 \sigma $ 	  \\
\hline
\hline
\end{tabular}
\caption{
\label{T10} The various contributions to $10^{10} a_\mu$. 
$\Delta a_\mu= (a_\mu)_{exp}-(a_\mu)_{th}$ is given in units of $10^{-10}$
and the last line displays its significance.
}
\end{table}

\subsection{The Anomalous Magnetic Moment of the Muon $a_\mu$}
\indent \indent Table \ref{T10} displays our final results concerning $a_\mu$. We still report
on the results derived in the fit configurations $A$ and $B$, using or not the $\tau$ data
in the fit procedure. The leading--order (LO) hadronic VP discussed in the 
previous Subsection
is reminded in the first line. In order to yield our estimate of $a_\mu$ under the various
quoted configurations, one should add the effect of higher--order hadronic loops  taken from
\cite{Fred11}, the light--by--light contribution \cite{LBL}; we took the latest estimate
of the pure QED contribution\footnote{The recent \cite{Kinoshita} 
value $a_\mu[QED]=11658471.8096(0.0044)$ displayed in Table \ref{T10} 
should be updated to $a_\mu[QED]=11658471.8960$ (in units of $10^{-10}$).
In order to compare with already published results we prefer keeping the former
value for our estimates of the HVP and of $g-2$.} \cite{Passera06} and the electroweak (EW) contribution
is taken from \cite{Fred09}. Summing up all these, one obtains the values given as "Total Theor."
which should be compared with the average \cite{BNL} of the different measurements for $a_\mu$,
recently updated  \cite{BNL2}.	

The difference between our theoretical estimates and the experimental average  \cite{BNL2}
is finally given together with their respective statistical significance.
The significance of this difference varies between $4.07\sigma $ (solution $B$ including $\tau$'s)
to $4.65 \sigma $ (solution $A$ excluding $\tau$'s).  The difference between including
$\tau$'s and excluding them is a $\simeq 0.4 \sigma$ effect. \cite{DavierHoecker} provides
an estimate excluding the KLOE data\cite{KLOE08} --  and the more recent ISR data sets not
available at that time -- reaching a difference with the BNL average  \cite{BNL2}
of $(30.1 \pm 8.6)~~10^{-10}$, a $3.5 \sigma$ significance. Our least significant estimate (solution $B$
including $\tau$'s) is, instead, $4.07 \sigma$.

\begin{figure}[!ht]
\begin{minipage}{\textwidth}
\begin{center}
\resizebox{\textwidth}{!}
{\includegraphics*{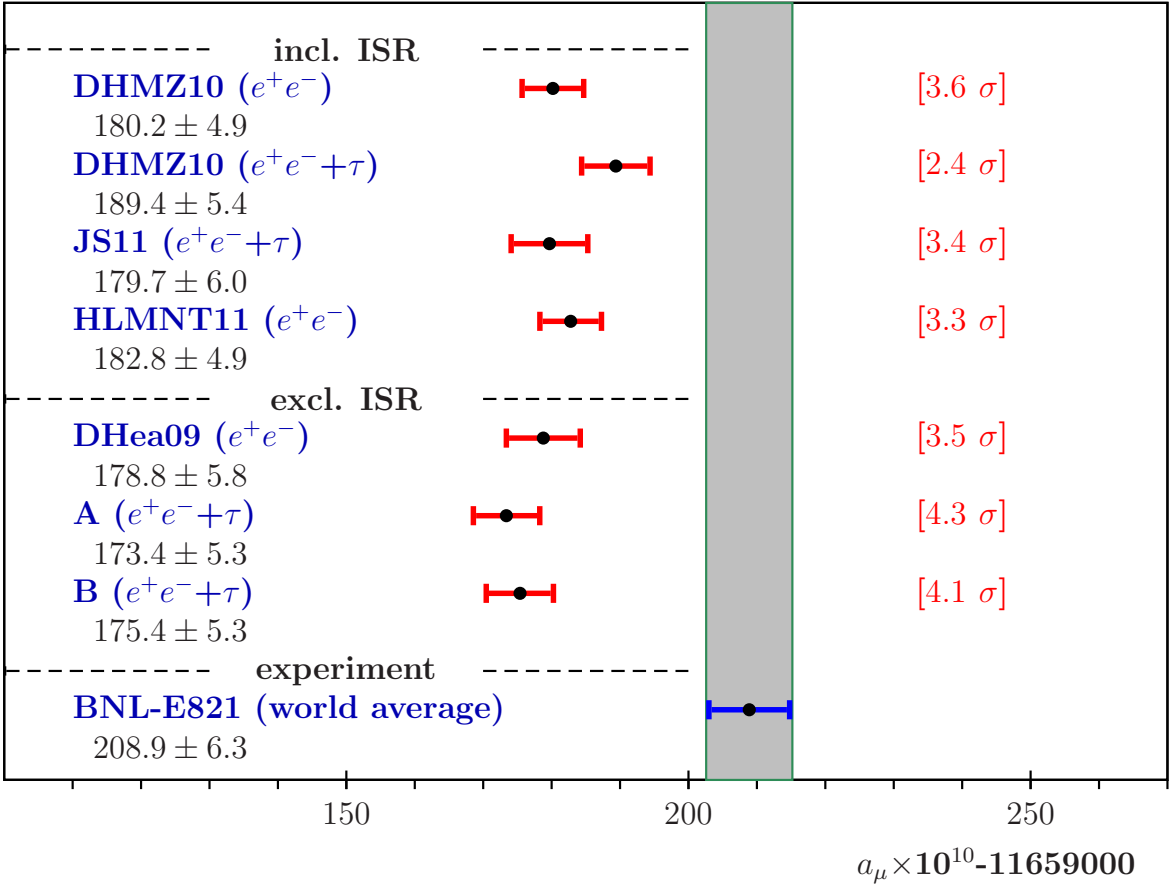}}
\end{center}
\end{minipage}
\begin{center}
\vspace{-0.3cm}
\caption{\label{Fig:fredFig}
A set of recent estimates of the muon anomalous magnetic moment
$a_\mu$ together with the BNL average value \cite{BNL,BNL2}. These are
extracted from  \cite{DavierHoecker3} (DHMZ10), \cite{Fred11} (JS11),
 \cite{Teubner} (HLMNT11) and \cite{DavierHoecker} (DHea09). Our own
 results are figured by A and B for respectively solutions A and B.
The statistical significance
of the difference between the estimated and measured values of $a_\mu$ 
is displayed on the right side of the Figure for each of the reported analyses.
}
\end{center}
\end{figure}

 Figure \ref{Fig:fredFig} displays our results together with the most recently published estimates.
On top of the Figure, one finds the estimates using or not the $\tau$ data provided in
\cite{DavierHoecker3}. The following entry is the estimate given in \cite{Fred11} which
combines $e^+e^-$ and $\tau$ data (after correcting for the $\rho^0-\gamma$ mixing).
The last entry \cite{Teubner} is derived including the ISR data (HLMNT11); this is the latest result
using the final KLOE \cite{KLOE10} and BaBar \cite{BaBar} data.

We have also displayed the latest result \cite{DavierHoecker} derived excluding ISR data
which directly compares to ours. This indicates that
the improvement provided by the global fit method corresponds to
increase the discrepancy of the BNL measurement \cite{BNL2}
with the Standard model prediction by $\simeq 0.6 \div 0.8 \sigma$. 
Therefore, the  discrepancy starts reaching an interesting significance.

\subsection{Influence of Data Set Choices on the Estimate for $a_\mu$}
\indent \indent In order to derive our estimates for $a_\mu$, we have defined
a paradigm, unusual in this field. Indeed, one usually performs the average using
all data sets contributing to a given final state in isolation; the prescription used
is the S--factor technics of the Particle Data Group. However, this supposes the 
simultaneous handling of statistical and systematic uncertainties. The most common way
of performing this handling is to use as weights the quadratic sum of statististical
and systematic uncertainties \cite{FredAndSimon}. 

In our approach, especially in this paper, the underlying paradigm is different and can be formulated in the following way~:
\begin{itemize}
\item All different channels are correlated by their underlying common physics and
an Effective Lagrangian approach is presently the best tool to deal with the non--perturbative 
QCD regime.
\item All data sets, covering or not  the same physics channel are considered by
taking into account the peculiarities of their uncertainties as reported by the experimental 
groups.  There is, in principle, no real difficulty in order to deal with statistical uncertainties. 
It is commonly assumed that uncorrelated systematics and statistical uncertainties could be added 
in quadrature and we followed this rule. Other systematics involving bin--to--bin or experiment---to--experiment 
correlations should be treated as such; the method is standard \footnote{In the scan experiments we deal with
in the present paper, all reported correlated systematics can be considered as global scale uncertainties for which
the standard method applies. For ISR experiments \cite{KLOE08,KLOE10,BaBar}, the situation is different as
several independent sources of systematics are defined which, additionally, vary all along the spectra. The standard method can be
extended to this case \cite{ExtMod1}; however, it should better be  reformulated in a way which avoids introducing
as many scale factors to be fitted as sources of different systematics. Indeed, this may produce fit instabilities
and, on the other hand, one has to deal with correlations between physics parameters and these scale factors
which may be uneasy to handle. } and has been sketched in Subsection 
\ref{fitproc}.
\item The Lagrangian model should allow for a good description of a large number of data sets
in as many different physics channels as possible. The goodness of the global fit should be accompanied
by a good description of each group of data sets -- ideally  each data set. As tag for
this property, we choosed the $\chi^2/n_{points}$ value for each data set group; this tag should 
not too much exceed 1.
Referring to our case,  the $\pi^+ \pi^-$, $\pi^0 \gamma$, $\eta \gamma$ physics channel data 
and the reported partial width decays already represent  an acceptably good starting point,
allowing a critical examination of the data associated with further additional channels.

\item Including a new data set,  or a new group of data sets, should not result in a significant
degradation of the already accounted for data sets. This should be observed at the global level
$and$ at the local levels ({\it i.e.} for each group). Following from the analyses
 in Sections \ref{3pions} and \ref{KKbar}, peculiarities of their fit behavior
led us to discard from our global fit the $K^+ K^-$ data set and one of the $\pi^+ \pi^-\pi^0$ 
data sets provided by SND. This turns out to require that the (large) set of data samples 
considered be statistically self--consistent~: Only 2 data sets out of 45 did 
not pass this consistency criterium. 
\end{itemize}

At this point, given the (broken) Lagrangian one uses, the selection criteria
are only the global fit quality and the "local" (data set specific) fit properties
reflected by the various $\chi^2/n_{points}$ values, discarding any possible consequence 
for the value for $a_\mu$. With Solutions A and B, one has also avoided any kind of data set 
reweighting by discarding the two data sets exhibiting some faulty behavior compared to the rest.

\vspace{0.5cm}

Nevertheless, it is a simple exercise to switch on the two discarded SND data sets within our fitting code.
For information, this leads to $\Delta a_\mu= (a_\mu)_{exp}- (a_\mu)_{th}= (34.00 \pm 8.21)~~10^{-10} $,
a $4.14\sigma$ effect. However, this is associated with an exceptionally poor global fit
probability (1.75\%)  and
 to  $\chi^2_{\pi^+ \pi^-\pi^0}/n_{points}=331/212=1.56$ and
$\chi^2_{K^+ K^-}/n_{points}=93/62=1.50$. Interestingly, and somewhat unexpectedly, 
the $\chi^2/n_{points}$
for the other data sets are practically unchanged compared to Table \ref{T3}, except for the decay
data set account which is sharply degraded~:  $\chi^2_{decays}/n_{points}=20.5/10\simeq 2 $. 
This may reflect that our broken
HLS model is so sharply constrained that poor data sets are mostly reflected by poor 
global fit probabilities.

\vspace{0.5cm}

A tag value of $\chi^2/n_{points}=1.3$, as yielded for the chosen $\pi^+ \pi^-\pi^0$ final
state data, is on the border of what could look reasonable to us (see third data column in Table \ref{T3}).
Nevertheless, compared with  $\chi^2/n_{points}=1.1$ (see second data column in Table \ref{T3}), it looks acceptable;
however, this corresponds to an  increase  by 30 units of the absolute magnitude of  $\chi^2_{\pi^+ \pi^-\pi^0}$, 
when introducing the selected kaon data. One may, indeed, consider that this indicates some tension
within the $\phi$ region data calling for a closer experimental examination which can 
be performed at the existing facilities covering the $\phi$ region.
  
Awaiting for better data in the $\phi$ region, we have been left with two challenging solutions~: Solution
A which uses all the data sets we have considered as secure, and solution B
obtained by removing all $\pi^+ \pi^-\pi^0$ data sets above the $K \overline{K}$  threshold.

\subsection{The Differential Effect of the Various $\tau$ Data Samples}
\label{vartau}
\indent \indent
In view of the discussions above, we have chosen to display all our final results for $a_\mu$, 
 in the fit configurations corresponding to solutions A and B. On the other hand, as can
be read off Table \ref{T3}, at the fit properties level, one can consider that the so--called 
$e^+e^- - \tau$ puzzle is over. 

However, one still observes a $(2\div 2.5)~~10^{-10}$ increase  of the
returned values for $a_\mu$ produced by the $\tau$ data.  
As stated already above, the $\tau$ data are essential in order to return a reasonably
precise value for our fit parameter\footnote{ The numerical accuracy of the scan $e^+e^-$ data
$alone$ does not permit a precise determination of  $\Sigma_V$ which is returned by {\sc minuit}
with large errors.} $\Sigma_V$. Therefore, the shift attributable to the $\tau$ data can be considered as
a normal consequence when fitting a model with a more constraining set of data samples.

Nevertheless, Table \ref{T3} indicates that the $\chi^2/n_{points}$ are sensitively different for ALEPH
($\simeq 0.43$), CLEO ($\simeq 1.26$) and BELLE\footnote{
\label{bellefit} Leaving free the absolute normalization 
of their dipion spectrum improves the stand--alone fit of the  BELLE Collaboration \cite{Belle}
from 80/52 to 65/51. This corresponds
to a best normalization of $1.02 \pm 0.01$. Such a re-normalization of their absolute scale
has some influence on the value for $a_\mu$. One should remind that we do not have any longer
fitted rescaling factors in our fitting functions.} ($\simeq 1.77$). This difference of
fit quality leads us to examine
the effects of removing the CLEO data sample and/or the BELLE  data sample for our fitted data set.
  
When keeping only the ALEPH data sample, we get  $\Delta a_\mu = 38.47 \pm 8.22$ 
(a $4.68 \sigma$ significance) 
and $\Delta a_\mu = 36.81 \pm 8.90$  (a $4.13 \sigma$ significance) 
for respectively solutions A and B. As can be seen 
from Table  \ref{T10}, these strikingly resemble the corresponding values for 
$\Delta a_\mu$ derived when keeping only $e^+e^-$ data in our fit procedure 
({\it i.e.} excluding all  $\tau$ data). In these peculiar configurations, the ALEPH data fit 
quality which was already very good ($\chi^2/n_{points}\simeq 16/37$),  
becomes impressively  better ($\chi^2/n_{points}\simeq 4/37$).

Going a step further, we have examined the effect of considering only ALEPH and
CLEO data. In this case, our fit returns $\Delta a_\mu = 36.02 \pm 8.22$ ($4.38 \sigma$ significance) 
and $\Delta a_\mu = 34.74 \pm 8.26$  ($4.21 \sigma$ significance) 
for respectively solutions A and  B.
One can check with Table  \ref{T10} that these values become closer to their
partners when fitting excluding $\tau$ samples.

Therefore, using only the $\tau$ data samples from ALEPH \cite{Aleph} and/or CLEO \cite{Cleo}
returns values for $\Delta a_\mu$  consistent well within errors with those derived using 
only $e^+e^-$ data.  The slightly different behavior of BELLE data may be related
with the normalization issue sketched in footnote \ref{bellefit}.

\subsection{On the Significance of the HLS Value for  $\Delta a_\mu$}
\indent \indent
In view of the considerations developed in the two preceding Subsections, one can certainly
consider that the most conservative estimates for  $\Delta a_\mu$ are those derived
while including $\tau$ data as they are reported by ALEPH, BELLE and CLEO. This corresponds
to the information provided in the first two data columns of Table \ref{T10}. 

This means
that the disagreement between the BNL measurement \cite{BNL2} and the Standard
model prediction for $\Delta a_\mu$ lays in beween 4.07 and 4.33 $\sigma$.
Moreover, from our analysis of the differential effects of the various available $\tau$
data samples, one may consider these bounds as conservative and that the significances
in the right part of Table  \ref{T10} cannot be discarded.

In view of this, in the perspective of taking into account relatively poor data
set group, one has rerun our code in order to get the solution when
weighting the contributions of\footnote{The weights used in this Subsection
refer to partial $\chi^2$'s obtained by fitting under
Configuration A with assuming $c_3=c_4$; it is the reason why they slightly
differ from the corresponding numbers given in Table \ref{T3}.}~:
\begin{itemize}
\item  all $\pi^+ \pi^-\pi^0$ data in our global sample by 179/232.41,
\item  the BELLE data sample by 19/32.31,
\item  the CLEO data sample by 29/36.48,
\end{itemize}
 in the global $\chi^2$ while leaving the other weights (all equal 1) unchanged.
This turns out to rescale globally the uncertainties associated with the corresponding
data sets by the inverse of these weights, assuming that their relatively poor quality
is only due to an overall underestimate of the uncertainties by a factor 
of respectively
1.14 ($\pi^+ \pi^-\pi^0$), 1.30 (BELLE) and 1.12 (CLEO). This may look as
a way to infer some sort of S--factors inside the global fit procedure.

This reweighting procedure\footnote{We have also made a fit leaving free
scale factors affecting the covariance matrices of the 3--pion data as a whole,
of the BELLE and CLEO data. The hadronic VP we get is $(686.73\pm 4.49) ~10^{-10}$, quite 
similar to this value.} 
provides as total hadronic VP contribution to $ a_\mu$
$(686.32 \pm 4.60) ~10^{-10}$ and $\Delta a_\mu = (34.93 \pm 8.23) ~10^{-10}$,
a $4.25 \sigma$ significance.

Going a step further, another check may look appropriate. As 
the contributions of the $\pi^+ \pi^-\pi^0$, BELLE and CLEO data
to the total $\chi^2$ have been weighted in order to reduce their influence, one can do alike
with those groups of data which exhibit too favorable individual $\chi^2$'s.
Still referring to fitting with configuration A, this turns out to weight the 
"Old Timelike" data by 82/56.61, the $\pi^0 \gamma$ data group by 86/68.37,
the $\eta \gamma$ data group by 182/123.31, the ALEPH data by 37/15.92
while keeping unit weights for the "New Timelike" and both $K \overline{K}$
data groups. This leads to an hadronic VP of $(685.00\pm 4.58) ~10^{-10}$
and to $\Delta a_\mu = (36.25 \pm 8.21) ~10^{-10}$ corresponding to a
a $4.41 \sigma$ discrepancy. This is almost identical to the value
found with Solution B, excluding $\tau$'s, as can be seen from Table \ref{T10}.

Therefore, these exercises enforce  our conclusion that the most conservative value
for $\Delta a_\mu$ exhibits a discrepancy of $4.07 \sigma$ and values
as large as $\simeq (4.30 \div 4.50) \sigma$ are not unlikely.

\section{Conclusion and Perspectives}
\label{Conclusion} 
\indent \indent
Several aspects should be emphasized. They can be grouped into two
items~: Low energy hadronic physics description and $g-2$ related topics.

Concerning the first item, the present study indicates that
the HLS model suitably broken is able to encompass most 
low energy physics in an energy range extending up to
the $\phi$ meson mass. More precisely, among the non--baryonic
possible final states, one covers\footnote{Among these, only the process 
$e^+e^- \ra \eta \pi \pi$ has not been examined; however, the
good description of the $\eta/\eta^\prime \ra  \pi \pi \gamma$ decays 
reported in \cite{ExtMod1} indicates that it could be successfully
considered. On the other hand, the $e^+e^- \ra  \omg \pi^0$ annihilation
is too much influenced by high mass vector resonances  \cite{GLi,Arbuzov}
to be accounted for by the standard HLS model.
}  most channels with multiplicity $n < 4$. 

More precisely, equipped with the so--called upgraded direct
symmetry breaking -- in the $u$, $d$ and $s$ sectors -- and including the
mixing of neutral vector mesons produced at one--loop, the HLS model
accounts quite satisfactorily for all the examined 
physics pieces of information. This covers the 6 annihilation channels 
having significant cross sections up to the $\phi$ meson mass and
a few more spectra like the dipion spectrum in the $\tau$ decay and,
also, an additional list of partial width decays. Previous studies \cite{box,ExtMod1}
have also shown that the dipion spectra in the $\eta/\eta^\prime \ra \pi \pi\gamma $
decays fall inside the scope of the HLS model. 

It is an attractive feature of this framework to exhibit a parent character
between the long reported issues represented by the $e^+e^- - \tau$ and the
$\phi \ra K \overline{K}$ puzzles~: Indeed, it is the same breaking
mechanism implemented in the ${\cal L}_A$ and in the ${\cal L}_V$ 
pieces of the HLS Lagrangian which provides a solution to both.
 It permits -- together with the $s$--dependent
vector meson mixing -- to finalize the consistency
of the $e^+e^- $ and $\tau$ physics  and to reproduce the branching fraction ratio
$\phi \ra K^+  K^-/\phi \ra K^0 \overline{K}^0$. This is materialized
by a satisfactory simultaneous fit of both $e^+e^- \ra K \overline{K}$
cross sections and of the pion form factor in both $e^+e^- $ annihilation
and $\tau$ decay.

The upgraded model thus provides a tool allowing a simultaneous treatment
of a large number of experimental spectra. It also permits a critical analysis
of the fit behavior of any data set in consistency with the others. Then, one
is in position to discard motivatedly some data samples which do not behave satisfactorily
within a global fit procedure and could then put some shadow on derived numerical
results. We have shown that such data samples are only few~: 2 out of the 45
considered spectra. It should be stressed  that discarded data sets 
are always identified because of their full redundancy with some other data sets, which are
found to behave normally  within the global model; stated otherwise, this
removal is not expected to produce a bias and, {\it a contrario}, any effect
resulting of keeping them is suspicious.

\vspace{0.5cm}

The model provides a tool which has the virtue of exhibiting
 the physics relationship between the various
physics channels. Within the global fit procedure involving the data on each channel, 
the model parameters yield
a better accuracy which propagates to all the reconstructed pieces of information, 
especially   the photon hadronic vacuum polarization and, thus, improves
significantly $g-2$ estimates.

Indeed, we have shown that the various components of the HVP yield
central values in accordance with expectations and an uncertainty improved
by a factor of 2 quite uniformly within the fit range.  This has been shown for
the $\pi^+\pi^-$, $\pi^0 \gamma$, $\eta \gamma$, $\pi^+\pi^-\pi^0$, $K^+  K^-$ and
$K^0 \overline{K}^0$ channel contributions up to $1.05$ GeV. Up to this energy, these
channels represent altogether more than 80\% of the hadronic VP and one of the two
dominant sources of uncertainty\footnote{The other dominant error comes from
 the hadronic VP between 1.05 and 2 GeV.}.
 
 In order to figure out the gain in terms of statistics, one can make the following
 statement~:  considering $globally$ the existing data sets is equivalent to having
 $\times 4$ more statistics $simultaneously$ in each of the considered channels without any increase of 
 the systematics. Therefore, considering additionally the high statistics ISR data 
 leaves some room for improved estimates of the HVP, provided the dealing with systematics
 can be reasonably well performed. One should nevertheless stress that  the global
 method we advocate, used with only the standard scan data samples provides already
 as good results as all scan $and$ ISR data using the standard numerical integration
 of the experimental cross sections.
 
 One may also try to figure out the improvement expected from   including the high statistic ISR
 data samples \cite{KLOE08,KLOE10,BaBar} within the fit procedure.  Being optimistic,
 one may think that the uncertainty on the HVP contribution up to 1.05 GeV could be divided
 by 2, from $\simeq 2 \times 10^{-10}$ (see Table \ref{T8}) to $\simeq 1 \times 10^{-10}$.
 Let us also assume that the ISR data samples will not rise unsolvable bias problems.
 Taking into account the rest of the HVP, which carry an uncertainty of $\simeq 4 \times 10^{-10}$
 (see Table \ref{T9}), the uncertainty on the full HVP would decrease from  $\simeq 4.60 \times 10^{-10}$
 (see Table \ref{T9}) to $\simeq 4.25 \times 10^{-10}$. Using the information collected in
 Table \ref{T10}, the total uncertainty on $a_\mu$ would decrease from $\simeq 5.30 \times 10^{-10}$
 to $\simeq 5.00 \times 10^{-10}$ and the uncertainty on  $\Delta a_\mu$ would
 decrease from $\simeq 8.20 \times 10^{-10}$ to $\simeq 8.00 \times 10^{-10}$.
 This may look a marginal improvement; the reason for this is the large value
 for the systematics generated by hadronic HVP in the region $1.05 \div 3.10$ GeV
 (see Table \ref{T9}), which thus becomes a prominent issue for future 
 significant improvements\footnote{Actually, even if the uncertainty
 on the  HVP contribution coming from the energy region 
 up to 1.05 GeV vanishes, this would not entail
 a significant improvement of the global uncertainty for $a_\mu$~!
 Stated otherwise, reducing the HVP error in the region
 from threshold to  1.05 GeV from  $\simeq 4 \times 10^{-10}$ to  
 $\simeq 2 \times 10^{-10}$ has much more dramatic effects than reducing it
 from $\simeq 2 \times 10^{-10}$ to $\simeq 1 \times 10^{-10}$.
 This is a pure  algebraic effect following from having to perform 
 quadratic sums for final uncertainties.}.

 However, this is not the end of the story. In the course of the paper, 
 and this is well expressed by Tables \ref{T9} and \ref{T10}, we saw that
 below 1.05 GeV systematics may produce significant shifts of the central values
 for the HVP and thus for $a_\mu$. This was observed, for instance, in the A and
 B configurations, where the shift for the HVP -- and for $a_\mu$ --
 amounts to $\simeq 2.00 \times 10^{-10}$ (see also
 Subsection \ref{vartau}).
  Because of this, there is still valuable experimental work to do also in the sub-GeV
 domain to decrease and/or better understand systematic  errors.
   More precisely,  a better experimental knowledge of all channels in the  $\phi$ mass region
 -- $0.95 \div 1.05$ GeV -- may result in improving quite significantly our 
 estimate on $g-2$ and in resolving some of the ambiguities discussed in the main text.
 As stated above, the information in this mass region
  has an important influence down to the threshold regions. This is
  certainly within the scope of existing machines and detectors\footnote{
  One may remark that scan data for the $e^+e^- \ra \pi^+ \pi^-$ cross section in
  the $\phi$ region are still not available.}.

\vspace{0.5cm}
What are the prospects for the future?

A new muon $g-2$ experiment at Fermilab is expected to come into operation in 5 years from now.
The accuracy is expected to improve to 0.14 ppm from its current 0.54 ppm. This also requires 
a factor 4 improvement of the hadronic vacuum polarization. As demonstrated by our analysis, it is
possible to improve the low energy part up to and including the $\phi$ by a systematic application of
 effective field theory methods in form of a resonance Lagrangian approach. However, as mentioned above, the
main effort will be required in the range above the $\phi$ up to about 3 GeV. In this range, major progress is
expected from CMD3 and SND at VEPP 2000 at Novosibirsk, from BESIII at Beijing, as well as from exploiting additional yet
unanalyzed ISR data from BaBar and Belle.
Within the 5 years available until a new experimental result for $a_\mu$ will be realized, lattice QCD
is expected to be able to produce results which are competitive with standard evaluations based on data.
This also would  provide important cross checks for the present results and, more generally, for
the effective Lagrangian approach.

\vspace{0.3cm}
 For now, 
 one can conclude that the paradigm represented by a global model which encompasses
 the  largest possible set of data indeed results in a highly significant improvement of the
 photon HVP uncertainty and of the uncertainty on $g-2$.  As the global  model allows to detect
 problematic data sets susceptible of generating biases, it must be accompanied by
 the most accurate possible treatment of the reported experimental systematics.

    Taking into account the ambiguities generated by a limited number of data sets, the
    most conservative estimate for the hadronic vacuum polarization leads to a  
    significance   for  a non--zero  $\Delta a_\mu$  of $4.1 \sigma$. Solving these
    ambiguities discussed in the main text may result in a significant increase
    of this conservative bound.

\newpage
\section*{\Large{Appendices}}
\appendix
\section{The Full HLS Non--Anomalous Lagrangian before Loop Mixing}
\label{AA}

\indent \indent
The non--anomalous Lagrangian of the Hidden Local Symmetry Model can
be written~:
 \be
{\cal L}_{HLS}=({\cal L}_{A}+{\cal L}_{V}) ={\cal L}_{VMD} +{\cal L}_{\tau}
\label{AA1}
\ee
in order to split it up into convenient pieces. 
Removing the pseudoscalar field kinetic energy term, which is canonical, 
one has~:
{\small 
\be
\begin{array}{ll}
\hspace{-2.cm} ~& \displaystyle
{\cal L}_{VMD} =  \\[0.5cm]
\hspace{-2.5cm} ~& \displaystyle
+ ie \left[ 1 -\frac{a}{2} (1+ \Sigma_V +\frac{\Delta_V}{3})\right] A \cdot \pi^- \parsym \pi^+ 
-i e \frac{a}{6 z_A}[1 -z_V+\Sigma_V -\Delta_V +\frac{\Delta_A}{2}(1-z_V)]
 A \cdot K^0 \parsym \overline{K}^0 
\\[0.5cm]
\hspace{-2.cm} ~& \displaystyle
+ ie \left[ 1 -\frac{a}{2z_A} + 
\frac{a}{6z_A}(1-z_V -2  \Sigma_V -2  \Delta_V +\frac{\Delta_A}{2}(2+z_V))\right]
 A \cdot K^- \parsym K^+
\\[0.5cm]
\hspace{-2.cm} ~& \displaystyle +\frac{1}{2} \left [
m_{\rho^0}^2  (\rho^0)^2 + m_{\omg}^2 \omg^2 + m_{\phi}^2 \phi^2 \right]
\displaystyle +\frac{1}{9} a f^2_\pi e^2 (5+z_V +5 \Sigma_V+3\Delta_V) A^2 
\\[0.5cm]
\hspace{-2.cm} ~& \displaystyle -e
\left [f_{\rho\gamma} \rho^0 + f_{\omg \gamma}\omg -f_{\phi \gamma}\phi\right] \cdot A
 + \frac{i a g}{2}(1+ \Sigma_V) \left[
[\rho^0+ \Delta_V(1-h_V) \omg\right] \cdot\pi^- \parsym \pi^+ 
\\[0.5cm]
\hspace{-2.cm} ~& \displaystyle
+\frac{i a g}{4 z_A} \left[
(1+ \Sigma_V+h_V \Delta_V-\frac{\Delta_A}{2})~\rho^0+
(1+ \Sigma_V+(1-h_V)\Delta_V-\frac{\Delta_A}{2})~\omg
-\sqrt{2} z_V (1-\frac{\Delta_A}{2})~\phi
\right]K^- \parsym K^+
\\[0.5cm]
\hspace{-2.cm} ~& \displaystyle
+\frac{i a g}{4 z_A} \left[
(1+ \Sigma_V-h_V \Delta_V+\frac{\Delta_A}{2})~\rho^0-
(1+ \Sigma_V-(1-h_V)\Delta_V+\frac{\Delta_A}{2})~\omg
+\sqrt{2} z_V (1+\frac{\Delta_A}{2})~\phi
\right] K^0 \parsym \overline{K}^0
\end{array}
\label{AA2}
\ee
} 
\noindent in terms of the first step  renormalized vector fields\footnote{
In order to avoid heavy notations, the subscript $R_1$, which
actually affects each of the vector fields  in Eq. (\ref{AA2})
has been removed.}. The pseudoscalar fields shown here are
renormalized (it is the origin of the $z_A$ and $\Delta_A$
terms). Of course, we have only kept the lowest order symmetry breaking
contributions.

Some parameters have been introduced in Eq. (\ref{AA2})
for convenience; these are ($m^2=a g^2 f_\pi^2$)~:
{\small
\be 
\hspace{-1.cm}
\left \{
\begin{array}{c}
\displaystyle m_{\rho^0}^2= m_{\omg}^2 =\displaystyle m^2 
\left[ 1+ \Sigma_V \right]~~~,~~m_{\phi}^2 =\displaystyle m^2 z_V \\[0.5cm]
f_{\rho\gamma}=\displaystyle a g  f_\pi^2 \left[ 
1+ \Sigma_V +h_V\frac{\Delta_V }{3} \right]~,~~
f_{\omg \gamma}=\frac{a g  f_\pi^2}{3}\left[ 
1+ \Sigma_V +3 (1-h_V)\Delta_V \right]~,~~
f_{\phi \gamma}= -a g  f_\pi^2 \frac{\sqrt{2}}{3}z_V
\end{array}
\right.
\label{AA3}
\ee
} 

On the other hand, using~:
\be
\displaystyle 
m_{\rho^\pm}^2=m^2\left[ 1+ \Sigma_V \right]~~~,~~~
f_{\rho W} =  a g  f_\pi^2 \left[ 1+ \Sigma_V \right]
\label{AA4}
\ee
one has at lowest order in the breaking parameters~:
{\small
\be
\begin{array}{ll}
{\cal L}_{\tau}& = \displaystyle - \frac{i V_{ud}g_2}{2} W^+ \cdot 
\left [ (1 -\frac{a (1+ \Sigma_V )}{2} ) \pi^- \parsym \pi^0
+\frac{1}{\sqrt{2}} [1 -\frac{a}{2z_A}(1+ \Sigma_V )] K^0 \parsym K^-
\right]
\\[0.5cm]
\hspace{-3.cm} ~& \displaystyle  + m_{\rho^\pm}^2 \rho^+ \cdot \rho^- 
-\frac{g_2V_{ud}}{2}  f_{\rho W} W^+ \cdot \rho^-
+ \frac{i a g }{2} (1+ \Sigma_V)
\rho^- \cdot \left [\pi^+ \parsym \pi^0 +\frac{1}{z_A\sqrt{2}}\overline{K}^0 \parsym K^+  
\right ]  \\[0.5cm]

\displaystyle 
\hspace{-3.cm} ~& \displaystyle + \frac{f_\pi^2 g_2^2}{4} \left \{ 
\left[\displaystyle [(1+\frac{\Delta_A}{2}) z_A
+a   \sqrt{z_V} (1+ \frac{\Sigma_V +\Delta_V}{2})] |V_{us}|^2 + 
[1+a(1+ \Sigma_V )] |V_{ud}|^2\right]
\right \}W^+ \cdot W^-
\end{array}
\label{AA5}
\ee
} 
where one has limited oneself to write down only the terms relevant for
our purpose. The (classical) photon and $W$ mass terms \cite{HLSRef,Heath}
are not considered  and have been given only for completeness. 
However, it is worth remarking that the photon mass term does not prevent 
the photon pole to reside at $s=0$ as required \cite{Heath1998}, at leading
order. 

Our breaking scheme generates new couplings
for the charged $\rho$ mesons~:
\be
\left \{
\begin{array}{ll}
{\cal L}_{\tau}^\pm = \displaystyle - \frac{i V_{ud}g_2}{2} W^+ \cdot 
\left [  
\frac{i a g }{2} (1+ \Sigma_V)
\rho^- \cdot \left [g_{\rho \eta \pi}\pi^+ \parsym \eta  +
g_{\eta^\prime \pi}\pi^+ \parsym \eta^\prime  \right ] + {\rm herm.~conj.}
\right ] \\[0.5cm]
g_{\rho \eta \pi}=\displaystyle -\left [ \epsilon + \frac{\Delta_A}{2\sqrt{3}}\cos\theta_P
- \frac{\Delta_A}{\sqrt{6}}\sin\theta_P \right ] \\[0.5cm]
g_{\rho \eta^\prime \pi}=\displaystyle -\left [ \epsilon^\prime + 
\frac{\Delta_A}{2\sqrt{3}}\sin\theta_P
+ \frac{\Delta_A}{\sqrt{6}}\cos\theta_P \right ] 
\end{array}
\right .
\label{AA6}
\ee
because of the field redefinition given by Eqs. (\ref{eq19}), (\ref{eq22}) and
(\ref{eq23}). Therefore, the broken HLS model predicts decay modes
$\tau \ra \pi (\eta/\eta^\prime) \nu$ of small intensity absent from the original
Lagrangian.

\section{Elements of the $\delta M^2$ Matrix}
\label{BB}
\indent \indent  The perturbation $\delta M^2$ to the full mass matrix
$M^2$ is defined in Eq. (\ref{Eq32}). Keeping only the leading terms in
isospin breaking parameters, its matrix elements are~:
\be
\hspace{-1.5cm}
\left \{
\begin{array}{llll}
\displaystyle 
\epsilon_\rho=\left[ \frac{g_{\rho K \overline{K}}}{z_A} \right]^2 (1+ 2\Sigma_V)
\left[ \epsilon_2(s)+(2 h_V \Delta_V - \Delta_A)\epsilon_1(s)
\right] \\[0.5cm]
\displaystyle 
\epsilon_\omg=\left[ \frac{g_{\rho K \overline{K}}}{z_A} \right]^2 (1+ 2\Sigma_V)
\left[ \epsilon_2(s)+(2 (1-h_V) \Delta_V-\Delta_A)\epsilon_1(s) \right] \\[0.5cm]
\displaystyle 
\epsilon_\phi= 2\left[ \frac{g_{\rho K \overline{K}}}{z_A} \right]^2 z_V^2
\left[ \epsilon_2(s) - \Delta_A \epsilon_1(s) \right]
 \\[0.5cm]
\displaystyle 
\epsilon_{\rho \omg} = \left[ \frac{g_{\rho K \overline{K}}}{z_A} \right]^2
(1+ 2\Sigma_V)
\left[\epsilon_1(s) +(\Delta_V-\Delta_A) \epsilon_2(s)  \right]\\[0.5cm]
\displaystyle 
\epsilon_{\rho \phi} =-\sqrt{2}\left[ \frac{g_{\rho K \overline{K}}}{z_A} \right]^2
 z_V (1+ \Sigma_V)
\left[\epsilon_1(s) +(h_V \Delta_V-\Delta_A) \epsilon_2(s) \right] \\[0.5cm]
\displaystyle 
\epsilon_{\omg \phi} =-\sqrt{2}\left[ \frac{g_{\rho K \overline{K}}}{z_A} \right]^2
 z_V (1+ \Sigma_V )
\left[ \epsilon_2(s) +((1-h_V) \Delta_V-\Delta_A) \epsilon_1(s) \right] 
\end{array}
 \right. 
\label{BB1}
\ee
The functions $\epsilon_1(s)$ and $\epsilon_2(s)$ and the constant
$g_{\rho K \overline{K}}$ have been already defined in the main text by
Eqs. (\ref{Eq33}). We have also defined~:
\be
\displaystyle g_{\rho K \overline{K}}=\frac{ag}{4}
\label{BB2}
\ee

\section{Lagrangian Pieces with Renormalized Vector Fields}
\label{CC}
\indent \indent  
Coupling to a pion pair comes from the two Lagrangian pieces\footnote{\label{qq} Throughout
this Section, one takes profit of introducing irrelevant second--order 
terms in breaking parameters in order to write down expressions in the most 
concise way}~:
\be
\begin{array}{lll}
\displaystyle 
{\cal L}_{V \pi \pi} = \frac{iag}{2} \left[1+\Sigma_V\right]~
\left \{~\rho^0_R +\left[(1-h_V)\Delta_V-\alpha(s)\right]~\omg_R 
+\beta(s)~\phi_R
\right\}~\cdot \pi^- \parsym \pi^+ \\[0.5cm]
\displaystyle 
{\cal L}_{A \pi \pi} = ie \left[ 1 -\frac{a}{2} (1+ \Sigma_V
+\frac{\Delta_V}{3})\right] A 
\cdot \pi^- \parsym \pi^+
\end{array}
\label{CC1}
\ee
which exhibit the couplings to a pion pair depending on mixing angles.

Similarly, the  Lagrangian pieces relevant for couplings to $K^+K^-$ are
given by~:
{\small
\be
\hspace{-1.cm}
\begin{array}{lll}
\displaystyle 
{\cal L}_{V K^+K^-}  =\displaystyle \frac{iag}{4z_A} \left[1+\Sigma_V -\frac{\Delta_A}{2} \right] 
\times \left \{~ \left[ 1+ h_V  \Delta_V +\alpha(s) + \sqrt{2} z_V  \beta(s) \right] \rho^0_R +
\right. \\[0.5cm]
 \displaystyle \left .
\hspace{1.cm}+\left[1+ (1-h_V)\Delta_V-\alpha(s) + \sqrt{2} z_V \gamma(s) \right] ~\omg_R 
-\left[\sqrt{2} z_V (1-\Sigma_V) -\beta(s) - \gamma(s) \right]~\phi_R
\right\}~\cdot K^- \parsym K^+ \\[0.5cm]
\displaystyle 
{\cal L}_{A K^+K^-} =  ie \displaystyle 
\left[ 1 -\frac{a}{6z_A}
[2+z_V +2 \Sigma_V +2 \Delta_V -\frac{\Delta_A}{2} (2+z_V) ]\right] A 
\cdot K^- \parsym K^+
\end{array}
\label{CC2}
\ee
}
and by~:
{\small
\be
\hspace{-1.cm}
\begin{array}{lll}
\displaystyle 
{\cal L}_{V K^0 \overline{K}^0}   
=\displaystyle \frac{iag}{4z_A} \left[1+\Sigma_V +\frac{\Delta_A}{2} \right] 
\times \left \{~ \left[ 1- h_V  \Delta_V -\alpha(s) - \sqrt{2} z_V  \beta(s) \right] \rho^0_R +
\right. \\[0.5cm]
\hspace{1.cm} \displaystyle \left .
 -\left[1- (1-h_V)\Delta_V+\alpha(s) + \sqrt{2} z_V \gamma(s) \right] ~\omg_R 
+\left[\sqrt{2} z_V (1-\Sigma_V) +\beta(s) - \gamma(s) \right]~\phi_R
\right\}~\cdot K^0 \parsym \overline{K}^0 \\[0.5cm]
\displaystyle 
{\cal L}_{A K^0 \overline{K}^0} = - ie \displaystyle \frac{a}{6z_A}
\left[ 1-z_V + \Sigma_V -\Delta_V +\frac{\Delta_A}{2} (1-z_V)\right] A 
\cdot K^0 \parsym \overline{K}^0
\end{array}
\label{CC3}
\ee
}
\noindent
for $K^0 \overline{K}^0$ couplings. Setting $b=a(z_V-1)/6$ and $\mu=z_V \sqrt{2}$,
the  $s$--dependent loop transition functions $\Pi_{V \gamma}$
are~:
{\small
\be
\hspace{-1.cm}
\begin{array}{lll}
\Pi_{\rho \gamma} =&  \displaystyle [1-\frac{a}{2}(1+\Sigma_V+\frac{\Delta_V}{3})] \frac{\Pi_{\pi\pi}^\gamma(s)}{g_{\rho \pi \pi}}
+ (z_A-\frac{a}{2} -b) \frac{\epsilon_S(s)}{g_{\rho \pi \pi}} 
+b \frac{\epsilon_D(s)}{g_{\rho \pi \pi}}\\[0.5cm]
\Pi_{\omg \gamma} =&  \displaystyle [1-\frac{a}{2}(1+\Sigma_V+\frac{\Delta_V}{3})]
\left[(1-h_V)\Delta_V-\alpha(s)\right] \frac{\Pi_{\pi\pi}^\gamma(s)}{g_{\rho \pi \pi}}
+ (z_A-\frac{a}{2} -b) \frac{\epsilon_S(s)}{g_{\rho \pi \pi}} 
-b \frac{\epsilon_D(s)}{g_{\rho \pi \pi}}\\[0.5cm]
\Pi_{\phi \gamma} =&  \displaystyle [1-\frac{a}{2}(1+\Sigma_V+\frac{\Delta_V}{3})]
\beta(s)  \frac{\Pi_{\pi\pi}^\gamma(s)}{g_{\rho \pi \pi}}
- (z_A-\frac{a}{2} -b) \mu \frac{\epsilon_S(s)}{g_{\rho \pi \pi}} 
+b \mu \frac{\epsilon_D(s)}{g_{\rho \pi \pi}}
\end{array}
\label{CC4}  
\ee
}
where~:
\be
\hspace{-1.cm}
\begin{array}{lll}
\displaystyle \epsilon_S(s)=\epsilon_2(s)+\epsilon_1(s)&{\rm and}&
\epsilon_D(s)=\epsilon_2(s)-\epsilon_1(s)
\end{array}
\label{CC5}  
\ee
The expressions in Eqs. (\ref{CC4}) are
very close to their partner in \cite{taupaper} or ~\cite{ExtMod1}, 
as only first--order perturbation terms are meaningful.
 \section{The Anomalous Lagrangian Pieces}
\label{DD}
\indent \indent  The full Anomalous Lagrangian can be written~:
\be
{\cal L}_{anomalous}={\cal L}_{VVP}+{\cal L}_{AVP}+{\cal L}_{AAP}+{\cal L}_{VPPP}+{\cal L}_{APPP}
\label{DD1}
\ee
where $A$ denotes the electromagnetic field. It incorporates the Wess--Zumino--Witten terms
and the FKTUY Lagrangian \cite{FKTUY}. The Lagrangian pieces occuring in Eq. (\ref{DD1})
are\footnote{For clarity, the new constant parameters are denoted exactly as they are
defined in  \cite{HLSRef}.} \cite{HLSRef}~:
\be
\left \{
\begin{array}{lll}
{\cal L}_{VVP}=& \displaystyle - \frac{N_c g^2}{4 \pi^2 f_\pi} ~c_3
 \epsilon^{\mu \nu \alpha \beta}{\rm Tr}[ \partial_\mu V_\nu \partial_\alpha V_\beta P] \\[0.5cm]
 {\cal L}_{AVP}=& \displaystyle - \frac{N_c ge }{8 \pi^2 f_\pi} ~(c_4- c_3)
 \epsilon^{\mu \nu \alpha \beta}\partial_\mu A_\nu
 {\rm Tr}[ \{ \partial_\alpha V_\beta , Q \} P] \\[0.5cm]
 {\cal L}_{AAP}=& \displaystyle - \frac{N_c e^2 }{4 \pi^2 f_\pi} ~(1- c_4)
 \epsilon^{\mu \nu \alpha \beta}\partial_\mu A_\nu \partial_\alpha A_\beta{\rm Tr}[Q^2 P]\\[0.5cm]
 {\cal L}_{VPPP}=& \displaystyle - i \frac{N_c g }{4 \pi^2 f_\pi^3} (c_1-c_2-c_3)
   \epsilon^{\mu \nu \alpha \beta}{\rm Tr}[V_\mu \partial_\nu P \partial_\alpha P \partial_\beta P]\\[0.5cm]
  {\cal L}_{APPP}=& \displaystyle - i \frac{N_c e}{3 \pi^2 f_\pi^3} [1- \frac{3}{4}(c_1-c_2+c_4)]  
  \epsilon^{\mu \nu \alpha \beta} A_\mu{\rm Tr}[ Q \partial_\nu P \partial_\alpha P \partial_\beta P]
\end{array}
\right .
\label{DD2}
\ee
where the $c_i$ are parameters not fixed by the model. $N_c$ is the number of colors fixed to 3.
The $V$ and $P$ field matrices are the bare ones. 
 \section{The $V_{R_1}P\gamma$ Coupling Constants}
\label{EE}
\indent \indent  In order to express the $V_{R_1}P\gamma$ couplings, it is appropriate to define
the angle $\delta_P=\theta_P-\theta_0$ ($\tan{\theta_0}=1/\sqrt{2}$)~:
\be
\left \{
\begin{array}{ll}
\displaystyle \sin{\theta_P}=\frac{1}{\sqrt{3}}
(\cos{\delta_P}+\sqrt{2}\sin{\delta_P})\\[0.5cm]
\displaystyle \cos{\theta_P}=\frac{1}{\sqrt{3}}
(\sqrt{2} \cos{\delta_P}-\sin{\delta_P})
\end{array}
\right.
\label{EE1}
\ee
and some parameter expressions which reflect the various ways, nonet symmetry breaking
in the PS sector occurs~:
\be
\left \{
\begin{array}{ll}
\displaystyle 
x=1-\frac{3 z_A^2}{2 z^2_A+1} v\\[0.5cm]
\displaystyle 
x^\prime=1-\frac{3 z_A}{2 z^2_A+1} v\\[0.5cm]
\displaystyle 
x^\second=1-\frac{3}{2 z^2_A+1} v
\end{array}
\right.
\label{EE2}
\ee
where $v$ is the nonet symmetry breaking parameter  defined 
in Eq. (\ref{eq22}).
Finally, we also have defined $G=-e g (c_3+c_4)/(8 \pi^2 f_\pi) $.  
The $\rho_{R_1}^{0}P\gamma$ coupling constants are~:
{\small 
\be
\left \{
\begin{array}{ll}
\displaystyle
g_{\rho^{0}\pi^0\gamma}^0 = \frac{G}{2}~
\left[
1-3\left[\frac{\Delta_A}{2} +(1-h_V)\Delta_V \right]
 -3 \epsilon  \sin{\delta_P}+3\epsilon^\prime \cos{\delta_P}
\right]\\[0.5cm]
\displaystyle
g_{\rho^{0}\eta\gamma}^0 = \frac{G}{2}~
\left[
\sqrt{2}(1-x^\prime) \cos{\delta_P}-(2 x+1) \sin{\delta_P}
+\left[\frac{\Delta_A}{2} +(1-h_V)\Delta_V \right]\sin{\delta_P}
-\epsilon
\right]\\[0.5cm]
\displaystyle
g_{\rho^{0}\eta^\prime\gamma}^0 = \frac{G}{2}~
\left[
(2 x+1)\cos{\delta_P} + \sqrt{2}(1-x^\prime) \sin{\delta_P}
-\left[\frac{\Delta_A}{2} +(1-h_V)\Delta_V \right]\cos{\delta_P}-\epsilon^\prime
\right]\\[0.5cm]
\displaystyle
g_{\rho^{\pm}\pi^\mp\gamma}=\frac{G}{2}
\end{array}
\right.
\label{EE3}
\ee
}
In the $\omg_{R_1}P\gamma$ sector,  one has~:

{\small 
\be
\left \{
\begin{array}{ll}
\displaystyle
g_{\omg \pi^0\gamma}^0 = \frac{3G}{2}~
\left[
1-\frac{1}{3}\left[\frac{\Delta_A}{2} +h_V\Delta_V \right]
-\frac{\epsilon}{3} \sin{\delta_P}
+\frac{\epsilon^\prime}{3} \cos{\delta_P}
\right]\\[0.5cm]
\displaystyle
g_{\omg \eta\gamma}^0 = \frac{G}{6}~
\left[
\sqrt{2}(1-x^\prime) \cos{\delta_P}-(2 x+1) \sin{\delta_P}
+9 \left[\frac{\Delta_A}{2} +h_V\Delta_V \right]\sin{\delta_P}-9 \epsilon
\right]\\[0.5cm]
\displaystyle
g_{\omg \eta^\prime\gamma}^0 = \frac{G}{6}~
\left[
(2 x+1) \cos{\delta_P}  +\sqrt{2}(1-x^\prime) \sin{\delta_P}
-9 \left[\frac{\Delta_A}{2} +h_V\Delta_V \right]\cos{\delta_P}-9 \epsilon^\prime
\right]\\[0.5cm]
\end{array}
\right.
\label{EE4}
\ee
}
and, finally,  the $\phi_{R_1}P\gamma$ sector provides much simpler expressions~:
{\small 
\be
\left \{
\begin{array}{ll}
\displaystyle
g_{\phi \pi^0\gamma}^0 = \frac{G}{2}~
\left[
 \frac{2 \epsilon}{z_A} \cos{\delta_P}
+\frac{2\epsilon^\prime}{z_A} \sin{\delta_P}
\right]\\[0.5cm]
\displaystyle
g_{\phi\eta\gamma}^0 = \frac{G}{3 z_A}~
\left[
(2+ x^\second) \cos{\delta_P}-\sqrt{2}(1-x^\prime)  \sin{\delta_P}
\right]\\[0.5cm]
\displaystyle
g_{\phi\eta^\prime\gamma}^0 = \frac{G}{3 z_A}~
\left[
\sqrt{2}(1-x^\prime)  \cos{\delta_P}+ (2+ x^\second) \sin{\delta_P}
\right]\\[0.5cm]
\end{array}
\right.
\label{EE5}
\ee
}

Finally, the $K^*$ sector is described by~:
{\small 
\be
\left \{
\begin{array}{ll}
\displaystyle
g_{K^{*\pm} K^\pm\gamma} = \frac{G}{2} \sqrt{\frac{z_T }{z_A}}
\left[
2-\frac{1}{z_T}
\right] \left( 1-\frac{\Delta_A}{4}
\right)\\[0.5cm]
\displaystyle
g_{K^{*0} K^0\gamma} = -\frac{G}{2} \sqrt{\frac{z_T }{z_A}}
\left[
1+\frac{1}{z_T}\right] \left( 1+\frac{\Delta_A}{4}
\right)
\end{array}
\right.
\label{EE6}
\ee
}
where $z_T$ is another breaking parameter \cite{taupaper,ExtMod1} not discussed here.

 \section{The $VPPP$ Coupling Constants}
\label{FF}
\indent \indent  The $VPPP$ coupling constants in the $P_0 \pi^- \pi^+$
($P_0=\pi^0,~\eta,~\etp$) have been defined for the $R_1$ renormalized fields
in Eq. [\ref{eq62}). With an obvious naming, they are obtained 
by multiplying each of~:
  \be
\left \{
\begin{array}{ll}
\displaystyle  
g_{\rho \pi^0}^0=-\frac{1}{4} \left [\frac{\Delta_A}{2} +3(1-h_V)\Delta_V
- \frac{\cos{\theta}_P }{\sqrt{3}}
(\epsilon_1+\sqrt{2}\epsilon_2)-\frac{\sin{\theta}_P }{\sqrt{3}}
(\epsilon_2-\sqrt{2}\epsilon_1)
\right ]\\[0.5cm]
\displaystyle 
g_{\rho \eta}^0= \frac{1}{4\sqrt{3}} \left [
\left[ 1+ 2z_A\frac{1-z_A}{2z_A^2+1}~v \right ] \cos{\theta_P }
-\sqrt{2}
\left[ 1-z_A\frac{2z_A+1}{2z_A^2+1}~v
\right]\sin{\theta_P} 
\right]\\[0.5cm]
\displaystyle 
g_{\rho \eta^\prime}^0= \frac{1}{4\sqrt{3}} \left[
\sqrt{2}
\left[ 1-z_A\frac{2z_A+1}{2z_A^2+1}~v
\right] 
\cos{\theta_P} 
+
\left[ 1+ 2z_A\frac{1-z_A}{2z_A^2+1}~v \right]\sin{\theta_P}
\right]\\[0.5cm]
\displaystyle  
g_{\omg \pi^0}^0= \frac{3}{4} \\[0.5cm]
\displaystyle 
g_{\omg \eta}^0= -\frac{\sqrt{3} }{12} \left [\cos{\theta}_P
-\sqrt{2}\sin{\theta}  \right ]  \left [ h_V \Delta_V +\frac{3\Delta_A}{2}\right ] 
-\frac{3}{4} \epsilon\\[0.5cm]
\displaystyle 
g_{\omg\eta^\prime}^0= -\frac{\sqrt{3} }{12} \left [\sqrt{2}\cos{\theta}_P
+\sin{\theta}  \right ]  \left [h_V \Delta_V+\frac{3\Delta_A}{2}\right ] 
-\frac{3}{4} \epsilon^\prime 
\\[0.5cm]
\displaystyle 
g_{\phi \pi}^0= 0\\[0.5cm]
\end{array}
\right.
\label{FF1}
\ee
by $D=-3 g (c_1-c_2-c_3)/(4 \pi^2 f_\pi^3)$, which depends on the FKTUY parameters
$c_1-c_2$ and $c_3$ not constrained by the model. 
Only the leading correction terms have been retained. 
\section{The $VP\gamma$ Couplings for Renormalized Vector Fields}
\label{GG}
\indent \indent  Let us define the quantities~:
\be
\displaystyle
k_{[V_{R_1} P_0 \gamma]}^0=\frac{1}{G N_c}g_{V_{R_1} P_0 \gamma}^0~~~~,~~(G=-\frac{egc_3}{4 \pi^2 f_\pi})
\label{GG1}
\ee
for each $V_{R_1}=\rho_{R_1},~\omg_{R_1},~\Phi_{R_1}$ and $P_0=\pi^0,~\eta,~\eta^\prime$.
The $g_{V_{R_1} P_0 \gamma}^0$ can be found in Appendix \ref{EE} in Eqs. (\ref{EE3}), 
(\ref{EE4}), (\ref{EE5}).
The functions $H^{P_0}_{V_i}$ occuring in Eq. (\ref{eq66})  
provide the  couplings of the physical vector fields to a photon
and a neutral meson. They are given by~:
\be
\displaystyle
\left \{
\begin{array}{ll}
H^{P_0}_{\rho_{R}}=k_{[\rho_{R_1} P_0 \gamma]}^0 + \alpha(s) k_{[\omg_{R_1} P_0 \gamma]}^0
-\beta(s) k_{[\Phi_{R_1} P_0 \gamma]}^0\\[0.5cm]
H^{P_0}_{\omg_{R}}=k_{[\omg_{R_1} P_0 \gamma]}^0-\alpha(s) k_{[\rho_{R_1} P_0 \gamma]}^0
-\gamma(s) k_{[\Phi_{R_1} P_0 \gamma]}^0\\[0.5cm]
H^{P_0}_{\Phi_{R}}=k_{[\Phi_{R_1} P_0 \gamma]}^0+\beta(s) k_{[\rho_{R_1} P_0 \gamma]}^0
+\gamma(s) k_{[\omg_{R_1} P_0 \gamma]}^0
\end{array}
\right.
\label{GG2}
\ee

These  definitions help in writing the cross sections in a way similar to those in \cite{ExtMod1}.
When expanded,  the $H^{P_0}_{V_i}$ functions may contain contributions of order greater than 1
in some of the breaking parameters. These higher--order contributions are irrelevant and
can be dropped out.

\section{The Functions $N_i(s)$ in $e^+e^-\ra \pi^0 \pi^+ \pi^- $ Annihilations}
\label{HH} 
\indent \indent 
The amplitude for the transition $\gamma^*\ra \pi^0 \pi^+ \pi^- $ is much simply  
expressed in terms of the following complex functions~:
{\small
 \be 
\hspace{-1.cm}
\left \{
\begin{array}{lll}
\displaystyle
N_0(s)=
\frac{2\epsilon_1(\cos{\theta_P}-\sqrt{2} \sin{\theta_P})+
2\epsilon_2(\sqrt{2} \cos{\theta_P}+\sin{\theta_P}) -\Delta_A \sqrt{3}
}{6\sqrt{3}} \frac{1}{D_{\rho^0}(s)} 
\\[0.7cm]
\displaystyle
N_1(s)=\frac{F_{\omg \gamma}(s)}{D_\omg(s)}+
\left[ \alpha(s) -(1-h_V)\Delta_V \right]
\frac{F_{\rho \gamma}(s)}{D_{\rho^0}(s)}+
\gamma(s)\frac{F_{\phi \gamma}(s)}{D_{\phi}(s)}\\[0.6cm]
\displaystyle
N_2(s)=\frac{1}{D_{\rho^0}(s_{+-})}+
\frac{1}{D_{\rho^+}(s_{0-})}+
\frac{1}{D_{\rho^-}(s_{0+})}\\[0.6cm]
\displaystyle
N_3(s)=\left[\alpha(s_{+-}) -(1-h_V)\Delta_V  \right]
\left[ \frac{1}{D_{\rho^0}(s_{+-})} - \frac{1}{D_{\omg}(s_{+-})}
\right]\\[0.5cm]
\displaystyle
N_4(s)=\left[ \frac{\epsilon_1(\cos{\theta_P}-\sqrt{2} \sin{\theta_P})+
\epsilon_2(\sqrt{2} \cos{\theta_P}+\sin{\theta_P}) 
}{\sqrt{3}} -\left[\frac{\Delta_A}{2}-(1-h_V)\Delta_V \right ]
\right] \frac{1}{D_{\rho^0}(s_{+-})} 
\\[0.5cm]
\displaystyle
N_5(s)=\frac{\gamma(s_{+-}) - \gamma(s)}{2D_{\rho^0}(s_{+-})}
\\[0.5cm]
\displaystyle 
N_6(s)=\frac{2\epsilon_1(\cos{\theta_P}-\sqrt{2} \sin{\theta_P})+
2\epsilon_2(\sqrt{2} \cos{\theta_P}+\sin{\theta_P}) -\Delta_A \sqrt{3}
}{6\sqrt{3}} \frac{1}{D_{\rho^0}(s_{+-})} 
\end{array}
\right.
\label{HH1}
\ee
}
 $s_{+-}$, 
$s_{0-}$ and $s_{0+}$ are the invariant mass squared of the corresponding pion
pairs from the final state. $s$ is the off--shell photon invariant mass squared.
All other parameters and functions have been defined in the body of the text.

The connection with the Kuraev--Siligadze $(x,y)$ parametrization \cite{Kuraev} is defined
by($m_0=m_{\pi^0}$, $m_\pi=m_{\pi^\pm}$)~:
\be
\left \{
\begin{array}{lll}
\displaystyle
s_{+-}=& s(2x+2y-1)+m_0^2\\[0.5cm]
\displaystyle
s_{+0}=& s(1-2y )+m_{\pi}^2\\[0.5cm]
\displaystyle
s_{-0}=& s(1-2x)+m_{\pi}^2
 \end{array}
 \right .
 \label{HH2}
\ee
The integration limits can be found in \cite{Kuraev}; they are also reminded in
\cite{ExtMod1}. The Kuraev--Siligadze kernel is~:
 \be
 \displaystyle
G(x,y)=  4 
(x^2-\frac{m_\pi^2}{s})(y^2-\frac{m_\pi^2}{s})
-\left (1 -2x -2y +2 x y + \frac{2m_{\pi}^2 - m_0^2}{s}
\right)^2
 \label{HH3}
\ee

                    \bibliographystyle{h-physrev}
                     \bibliography{vmd3}
 
\clearpage

\end{document}